\documentclass{aa}
\usepackage{natbib}
\bibpunct{(}{)}{;}{a}{}{,} 
\usepackage{graphicx}
\usepackage[varg]{txfonts}
\newcommand{\thetaone}{$\theta^1$\,Ori}

\newcommand{\thetaonec}{$\theta^1$\,Ori\,C}
\newcommand{\thetaoned}{$\theta^1$\,Ori\,D}

\newcommand{\thetaonef}{$\theta^1$\,Ori\,F}
\newcommand{\thetatwo}{$\theta^2$\,Ori}
\newcommand{\thetatwoa}{$\theta^2$\,Ori\,A}
\newcommand{\thetatwob}{$\theta^2$\,Ori\,B}
\newcommand{\thetatwoc}{$\theta^2$\,Ori\,C}
\newcommand{\degree}{\mbox{$^{\circ}$}}               
\newcommand{\micron}{\mbox{\,${\mu}$m}}               
\newcommand{\Msolar}{\mbox{\,$M_{\odot}$\/}}          
\newcommand{\Mjup}{\mbox{\,$M_{\rm Jup}$\/}}          
\newcommand{\HII}{\mbox{H\,{\footnotesize II}}}       
\newcommand{\FeII}{\mbox{[Fe\,{\footnotesize II]}}}   
\newcommand{\OIII}{\mbox{[O\,{\footnotesize III]}}}   
\newcommand{\HeI}{\mbox{He\,{\footnotesize I}}}       
\newcommand{\SiI}{\mbox{Si\,{\footnotesize I}}}       
\newcommand{\water}{\mbox{H$_2$O}}                    
\newcommand{\methane}{\mbox{CH$_4$}}                  
\newcommand{\magnit}[2]{\mbox{$\mbox{\rm #1}^{\mbox{\rm\tiny m}}%
     \!\!\!.\!\,\, \mbox{\rm #2}$}}                   
\newcommand{\oversim}[2]{\lower0.5ex\vbox{\baselineskip=0pt\lineskip=0.2ex
     \ialign{$\mathsurround=0pt #1\hfil##\hfil$\crcr#2\crcr\sim\crcr}}}
\newcommand{\eg}{\mbox{\hbox{e.g.,}}}             
\newcommand{\idest}{\mbox{\hbox{\it i.e.,}}}          

\newcommand{\kmpers}{\mbox{\,km\,s$^{-1}$}}           
\begin{document}
\title{A JWST survey of the Trapezium Cluster \& inner Orion Nebula}
\subtitle{I\@. Observations \& overview}
%
%
\author{M.~J.~McCaughrean\inst{1} \and S.~G.~Pearson\inst{1}
}
%
%
\institute{$^1$European Space Agency, ESTEC, Keplerlaan 1, 2201 AZ Noordwijk, The Netherlands}
%
\date{Received 2 October 2023; accepted ...}
%
%
\abstract{%
We present a near-infrared survey of the Trapezium Cluster and inner Orion Nebula using
the NASA/ESA/CSA James Webb Space Telescope. The survey with the NIRCam instrument covers 
$10.9\times 7.5$ arcminutes ($\sim 1.25\times 0.85$\,pc) in twelve wide-, medium-, and narrow-band
filters from 1--5\micron{} and is diffraction-limited at all wavelengths, providing a maximum
spatial resolution of 0.063 arcsec at 2\micron{}, corresponding to $\sim 25$\,au at Orion. 
The suite of filters chosen was designed to address a number of scientific questions including 
the form of the extreme low-mass end of 
the initial mass function into the planetary-mass range to 1\Mjup{} and below; the nature
of ionised and non-ionised circumstellar disks and associated proplyds in the infrared with 
a similar resolution to prior HST studies; to examine the large fragmented outflow from
the embedded BN-KL region at very high resolution and fidelity; and to search for new jets 
and outflows from young stars in the Trapezium Cluster and the Orion Molecular Cloud~1 behind.
In this paper, we present a description of the design of the observational programme, 
explaining the rationale for the filter set chosen and the telescope and detector modes used to 
make the survey; the reduction of the data using the JWST pipeline and other tools; the creation
of large colour mosaics covering the region; and an overview of the discoveries made in the
colour images and in the individual filter mosaics. Highlights include the discovery of
large numbers of free-floating planetary-mass candidates with masses as low as 0.6\Mjup,
a significant fraction of which are in wide binaries; new emission phenomena associated with
the explosive outflow from the BN-KL region; and a mysterious ``dark absorber'' associated with
a number of disparate features in the region, but which is seen exclusively in the F115W
filter. Further papers will examine those discoveries and others in more detail.
}
\maketitle

\section{Introduction} \label{sec:introduction}
The constellation of Orion the Hunter is one of the best-known in the northern winter and 
southern summer skies and its asterism is instantly recognisable. In more detail, the handle
of the sword hanging below Orion's belt looks nebulous to the naked eye and unsurprisingly drew 
the attention of the earliest telescopic observers from 1610 onwards, including Galileo, 
Fabri de Peiresc, Cysat, and Huygens, later notably joined by Parsons and Herschel. 
The so-called Orion Nebula was seen at its core to have a small group of stars which became 
known as the Trapezium, and the combined system became the forty second in Messier's famous list. 

Astrophysical studies of the region arguably started with the first photographs 
\citep{draper1880} and spectroscopy \citep{huggins1865}, revealing a mix of stars and gas 
which Huggins speculated could indicate ``a more advanced state towards the formation of
a number of separate bodies, such as exist in our sun and in the stars''. The Orion 
Nebula became recognised as one of the nearest regions to Earth with massive stars that
were likely only a few million years old which illuminate their surroundings,
and continued to draw the attention of observers as telescope and instrument technology
improved. Detailed photography led to the recognition of a significant number of fainter 
stars in a roughly 0.3 parsec region around the Trapezium, the so-called Trapezium Cluster 
\citep{trumpler31, baade37}, while early infrared studies of the region revealed a number
of bright sources without optical counterparts embedded in the Orion Molecular Cloud 1 (OMC-1)
core behind the optical nebula \citep{becklin67, kleinmann67}, sources often 
collectively referred to as BN-KL, painting a picture of ongoing star formation in the region. 

Extensive imaging and spectroscopic studies of the nebular gas and dust have been made over
the decades leading to a basic model of the Orion Nebula as an ionised blister on the face 
of the giant molecular cloud OMC-1, albeit with significant geometric, ionisation, and 
velocity structure \citep[see][for reviews]{peimbert82, odell01b, odell01a}. Conversely, the 
presence of the bright nebula made detailed studies of the lower-mass young stellar population 
challenging until the advent of red-sensitive digital detectors and the use of narrow-band
filters to suppress the nebular emission lines, making it possible to 
detect cool low-mass stars through the reddening of the region \citep{herbig86}. Later
detailed optical imaging photometry and spectroscopy elucidating the mass function and 
star formation history of the wider Orion Nebula Cluster (ONC) and the inner Trapezium
Cluster, both part of the part of the Orion OB1 association \citep{hillenbrand97}.
Long time-baseline photographic surveys also enabled the study of proper motions of both 
stars and nebular features \citep{jones88, vanaltena88, walker88}.

Near-infrared mapping \citep{lonsdale82}, raster scanning \citep{hyland84}, and then true 
imaging with infrared array detectors revealed the full extent of the stellar and brown dwarf 
population around the Trapezium 
\citep{mccaughrean88, zinnecker93, mccaughrean94, hillenbrand00, lucas00, lucas01, muench02},
with subsequent deeper infrared and spectroscopy with the new generation of 8--10\,m class 
telescopes reaching well into the planetary-mass domain, discovering and characterising 
free-floating objects with masses as low as 3--5\Mjup{} 
\citep{mccaughrean02, lada04, slesnick04, lucas05, meeus05, lucas06, riddick07, weights08}. 
More recent infrared imaging surveys have also extended well beyond the inner Trapezium and 
Orion Nebula Clusters, helping place them in the wider context of star formation across the 
Orion molecular clouds \citep{megeath12, drass16, meingast16, grossschedl19}, while 
observations at high-energy wavelengths have provided complementary information on the 
properties of the stars and their immediate circumstellar environments \citep[\eg][]{getman05}. 

The same advancements in infrared technology brought increased insight into to the sources 
embedded in the molecular cloud behind the Orion Nebula and Trapezium Cluster, OMC-1. 
The region around BN-KL was gradually resolved into a number of point sources, some only
seen at longer thermal-infrared wavelengths \citep{rieke73, werner83, wynnwilliams84,
minchin91, dougados93, gezari98, robberto05}, accompanied by substantial extended emission. 
In addition to reflection nebulosity and thermal continuum emission, the detection of
significant luminosity in emission lines of molecular hydrogen \citep{gautier76, beckwith78}
led to the discovery of a large, explosive  outflow system comprising many ``fingers'' of 
emission emanating from the core of BN-KL and extending over 90\degree{} to both
the N and W, as well as to the S and E \citep{taylor84, burton91, allen93, 
mccaughrean97, kaifu00, bally11,
bally15}. The fingers are also well traced in other molecular lines 
\citep[\eg][]{zapata09, bally17}, and some extend into the Orion Nebula where they are seen 
as Herbig-Haro objects \citep{axon84, bally00}. Embedded infrared point sources and molecular 
outflows are also seen associated with a secondary cloud core, OMC-1S, roughly 0.2\,pc
south of BN-KL, another region of considerable interest \citep{bally00, doi02, zapata04,
smith04, getman05, zapata05, robberto05, zapata06, henney07, rivilla13}.

Another key to the advancement of optical and near-infrared studies of the Orion Nebula
and Trapezium Cluster has been the Hubble Space Telescope, launched in April 1990 on STS-31. 
The first observations of the cluster were made using WF/PC and were impacted by the spherical 
aberration of the improperly-polished 2.4\,m primary mirror \citep{prosser94, odell93}, 
limiting the spatial resolution, image quality, and point source sensitivity. 

Following the installation of WFPC-2 and its corrective optics during Servicing Mission~1 
(SM1) in 1993, however, the inner Orion Nebula and its associated population of young stars 
could finally be seen with the full diffraction-limited resolution of the telescope. 
One major finding was the confirmation of the so-called `proplyds' ---
photoevaporating, externally-ionised circumstellar disks \citep{odell94, bally00} ---
confirming the essential predictions of prior ground-based optical and radio observations,
as well as the aberrated HST images, in spectacular visual detail 
\citep{laques79, churchwell87, garay87, odell93}. A population
of circumstellar disks seen in silhouette against the bright background emission of the
Orion Nebula was also discovered \citep{odell94, odell96, mccaughrean96, bally00},
and many new jets and outflows emanating from the young stars of the region were 
identified in these surveys \citep{bally00, odellhenney08}.

Near-infrared observations of the stellar and substellar population in the inner region and some 
key objects became possible with the installation of NICMOS during SM2 in 1997 
\citep{luhman00, chen98, throop01, andersen11}, 
while further servicing missions and instrument upgrades to HST including ACS (SM3B) and 
later WFC3 (SM4) lead to public surveys in the optical and near-infrared covering a much 
larger region, out to $\sim 30\times 30$ arcmin or $3.4\times 3.4$\,pc 
\citep{dario09, robberto13}. These wide-field HST surveys yielded larger catalogues 
of ionised and silhouette disks \citep{ricci08}, and an extension of the sub-stellar and 
planetary-mass population detected by HST \citep{dario10, dario12}, ultimately to roughly the 
same $\sim 3$\Mjup{} limit seen in prior ground-based observations \citep{robberto20}.

By way of an introduction to our new JWST survey of the inner Orion Nebula and Trapezium
Cluster, this is by necessity a very incomplete survey of the vast literature on optical
and infrared studies of the nebula, the background molecular cloud, and their associated 
stellar populations, and the broad and important work done at high-energy, 
far-infrared, millimetre, and radio wavelengths to understand the region is only briefly 
mentioned. The reader is referred to the reviews of \citet{odell01b, odell01a, 
bally08, odell08, muench08} among others for more detail, for how these observations 
have helped inform our understanding of \HII{} regions, PDR's, and molecular cloud cores 
on one hand, and of the properties and evolution of the stellar and sub-stellar content 
in young star-forming regions on the other. 

The remainder of this paper is organised as follows:
Section~\ref{sec:jwst} gives a broad overview of JWST and its instrumentation, in particular
NIRCam, as a background to the design of our Orion imaging survey observations;
Section~\ref{sec:programme} describes the key scientific questions we wanted to
address and how the observational programme was designed to meet them;
Section~\ref{sec:datareduction} describes the data reduction, photometry, and
creation of the large colour composite images presented here; and
Sections~\ref{sec:orionnebula}--\ref{sec:galaxies} give top-level summaries of
some of the features seen in the images and discoveries made from them.
More complete analyses and context covering individual topics will be presented in future 
(\eg{} Pearson \& McCaughrean 2023, submitted, and others in preparation).

Finally, throughout this paper, we shall assume the distance of $390\pm 2$\,pc to the
Orion Nebula and Trapezium Cluster recently derived from a {\it Gaia\/} EDR3-based study 
by \citet{maizapellaniz22}, which is excellent agreement with the {\it Gaia\/} DR2 value of
$389\pm 3$\,pc of \citet{kounkel18}. These {\it Gaia\/} studies have resulted in a significant 
revision of the previous radio parallax distance of 414\,pc \citep{menten07, kim08} most
commonly used in recent literature on the Orion Nebula. 

\section{JWST and its instruments} \label{sec:jwst}
The NASA/ESA/CSA JWST is a cryogenic infrared observatory built under the responsibility
of prime contractor Northrop-Grumman with Ball Aerospace providing the Optical Telescope
Element (OTE) and the Integrated Science Instrument Module (ISIM), under the management of 
NASA Goddard Space Flight Center, Greenbelt, USA \citep{gardner23}. 
The observatory was launched towards its 
home in orbit around the Sun-Earth L2 point on an Arianespace Ariane 5 (VA-256) from the
Centre Spatiale Guyanais in Kourou, French Guiana, at 12:20UTC on Christmas Day (25 December) 
in 2021, and is operated by the Space Telescope Science Institute, Baltimore, USA\@.  

The primary mirror comprises 18 hexagonal gold-coated beryllium segments and spans a diameter 
of approximately 6.5\,m. The primary mirror and other optical components of JWST are figured,
polished, and aligned to yield diffraction-limited resolution at all wavelengths above 1.1\micron,
better than the design specification of $\geq 2$\micron{} \citep{rigby23}. Located above 
Earth's atmosphere and passively cooled behind its very large deployed sunshield to 
$\sim 40$\,K, the full near- and mid-infrared wavelength range is accessible without the 
atmospheric absorption bands that characterise ground-based infrared astronomy and the 
background flux is reduced far below that experienced by ground-based telescopes due to 
atmospheric OH airglow, and atmospheric and telescope thermal emission, with JWST's 
near-infrared background level set by the zodiacal light in the solar system.

Via the secondary mirror and so-called aft optics, which includes the fine steering mirror,
the light collected from astronomical targets is sent to 
the focal plane in the ISIM and into the entrance apertures of the four science instruments 
provided by the three space agencies and ESA member states. Three of those instruments collect 
and analyse light in the far-red optical and near-infrared from 0.6--5.3\micron{} (NIRCam, 
NIRISS, and NIRSpec), while MIRI covers the mid-infrared range from 4.9--27.9\micron. 

For the current near-infrared imaging survey of the inner Orion Nebula, only NIRCam was used.
Built by The University of Arizona and Lockheed Martin, NIRCam is the workhorse near-infrared 
imager of JWST and provides a wide range of observational modes spanning scientific imaging, 
coronography, and low-dispersion spectroscopy for sensing the wavefront of the JWST OTE
\citep{rieke23}. 

Given the critical function provided by the latter mode, considerable redundancy is built in 
to NIRCam, including the provision of two completely separate camera modules (side A and side B) 
which can be used simultaneously to observe two adjacent $\sim 2.2\times 2.2$ arcmin fields 
separated by a 44 arcsec intra-module gap. 

Because JWST is diffraction-limited, the angular resolution is linearly dependent on the
wavelength and thus some optimisation of the pixel scale is needed across the 0.6--5.0\micron{}
wavelength range of NIRCam. As a result, within each module the light is split with a dichroic
between Short Wavelength (henceforth SW) and Long Wavelength (LW) channels covering 
0.6--2.3\micron{} and 2.4--5.0\micron, respectively, both of which observe simultaneously. 

The SW channel has a nominal pixel scale of 0.031 arcsec, designed to Nyquist sample
the 0.063 arcsec diffraction-limited resolution of JWST at 2\micron, while the LW channel
has a pixel scale of 0.063 arcsec to sample the 0.127 arcsec resolution limit of the 
telescope at 4\micron. Thus to cover the $2.2\times 2.2$ arcmin field of a given module,
the LW channel uses a single $2048\times 2048$ pixel H2RG array provided by Teledyne 
Technologies, with the mercury-cadmium-telluride (HgCdTe) detector layer optimised with 
a wavelength cut-off at 5\micron, while the SW channel uses four $2048\times 2048$ 
pixel H2RG array with the HgCdTe material optimised to cut-off at 2.5\micron. In the latter 
case, the detectors are separated by gaps of 4--5 arcsec.

The multi-modular design of NIRCam, with a total of ten H2RG detectors across the two modules 
and two channels simultaneously observing at two wavelengths across overlapping but 
non-contiguous fields-of-view has considerable consequences for the design of an observational 
programme to survey a region larger than the instrument field-of-view, as well as for the 
subsequent data reduction and analysis. Further constraints and considerations arise due to the 
minimum read-out and thus integration time of 10.737 seconds when reading out a full NIRCam 
detector, especially in regions like the Orion Nebula and Trapezium Cluster, has many 
stars at brightnesses up to magnitude 4 and above, as well as very bright structured nebular 
emission. In addition, the L2 orbit of JWST and the relatively small semiconductor bandgap 
of the HgCdTe detectors can yield a significant flux of cosmic ray events. 

The Astronomers Proposal Tool (APT) used to design JWST observations provides a number of modes
for setting the integration time and non-destructively reading out the NIRCam detectors,
for choosing the filters used simultaneously in the SW and LW channels, for rejecting 
cosmic rays, and for orienting and moving the telescope and/or the fine-steering mirror to 
cover the inter-detector gaps, the inter-module gaps, and to make a mosaic covering the desired 
area of the sky. It is also recommended to take multiple images at any given location with small
dithers on pixel and sub-pixel levels to further help eliminate bad pixels and fully sample 
the resolution of the telescope. A further consideration is the overall amount of data being
generated as part of the daily JWST downlink allowance, which may also influence the choice
of observing mode and scheduling constraints. This is particularly important for NIRCam given
the very large data volume that can be generated by reading out ten $2048\times 2048$ detectors
every 10.7 seconds.

This survey lies squarely in one of the four key pillars identified in the science case
for JWST \citep{gardner06}, namely the birth of stars and protoplanetary systems. It makes 
use of several of the core technical capabilities of the observatory, namely the use of a 
large collecting area and diffraction-limited spatial resolution to detect very faint, 
very low-mass young brown dwarfs and planetary-mass objects, as well as to examine detailed 
structures in disks, jets, and outflows. A cryogenic space telescope gives access to key 
parts of the electromagnetic spectrum either blocked by Earth's atmosphere or rendered 
challenging by high sky and thermal background flux.

And although the current survey only uses one of JWST's four scientific instruments,
NIRCam, follow-up spectroscopic observations of the lowest-mass brown dwarfs and planetary-mass 
objects detected here will be conducted with NIRSpec in JWST Cycle~2 
(Programme 2770)\footnote{%
\url{https://www.stsci.edu/jwst/science-execution/program-information?id=2770}
}. 
There is of course an excellent case for studies in Orion using MIRI for both imaging and imaging
spectroscopy of the deeply embedded star formation associated with OMC-1 and OMC-1S, as well 
as the driving sources of lower-mass outflows newly revealed below. Conversely, it is worth 
noting the slitless spectroscopic mode of NIRISS is likely to be significantly compromised 
by the bright nebulosity of the region, while the originally-planned tuneable filter imager 
mode would have been ideal for identifying and classifying young low-mass objects
via their atmospheric absorption bands, the method used here employing NIRCam imaging instead.
%

This information helps set the scene for the way we designed our observations for the Orion
Nebula, as described in the following section.

\section{Programme design} \label{sec:programme}
There were three key scientific goals behind this survey which informed the choice of
NIRCam filters used:
\begin{enumerate}
\item The discovery and initial characterisation of candidate objects at the low-mass end of the
brown dwarf regime and into the planetary-mass domain, potentially at 1\Mjup{} and below. At
an age of $\sim 1$\,Myr, objects between 1––13\Mjup{} are expected to have effective 
temperatures of $\sim 890$--2520\,K \citep{phillips20} and thus have atmospheres similar 
to older late M, L, and T field dwarfs showing absorption due to water and, at the lower masses 
and temperatures, methane, allowing them to be identified against a
potential population of more distant field stars seen through the molecular cloud. The cooler
objects will also radiate most of their bolometric luminosity at longer wavelengths in the 
NIRCam bands. Other issues are the possibility of infrared excess emission from
warm circumstellar material and reddening due to the dust in the parent molecular cloud. 
These considerations led to the selection of wide-band filters to measure the overall spectral 
energy distribution from 1--5\micron{} (F115W, F277W, F444W) and groups of medium-band 
filters in and straddling the atmospheric water and methane absorption features 
(F140M, F162M, F182M for \water{} and F300M, F335M, F360M for \methane{}).
\item The measurement of the sizes of circumstellar disks seen as silhouettes against the
bright background of the nebula at near-infrared wavelengths, for comparison with their sizes
in HST images at visible wavelengths, in order to characterise the dust properties. Commensurate
with the discovery images made with HST in the H$\alpha$ n=3--2 line of ionised hydrogen at 
656.3\,nm, the best option with JWST is the Paschen-$\alpha$ n=4--3 line at 1875\,nm or
1.875\micron{} (F187N). The brightness of the nebula in this line, along with the narrowness
of the F187N filter, ensures the maximum contrast of dark silhouettes against the background
and minimises the continuum flux from the parent stars. In the Pa$\alpha$ line, the 
diffraction-limited spatial resolution of the 6.5\,m JWST is 59 milliarcsec, almost 
identical to the 56 milliarcsec resolution of the 2.4\,m HST in the H$\alpha$ line.
\item Detailed imaging of the explosive outflow ``fingers'' from the BN-KL region in OMC-1 behind 
the Trapezium Cluster to examine structure, excitation, and extinction, along with making proper
motion measurements with respect to extant high-resolution ground- and space-based imaging. 
Also, the intention was to survey for other outflows from young stars and embedded sources
in the region. The key emission line tracer for the BN-KL outflow is molecular hydrogen, 
with ground-based observations generally focussing on the v=1--0 S(1) ro-vibrational line
at 2.12\micron{} (F212N), but as a cryogenic telescope above Earth's atmosphere, JWST also
offers narrow-band filters centred on longer-wavelength lines such as the 1--0 O(5) line
at 3.23\micron{} (F323N) and the 0--0 S(9) line at 4.69\micron{} (F470N). For this survey,
we used F212N and F470N, while knowing that the 3.23\micron{} line also appears in the F335M
filter. Similarly, another key tracer of the BN-KL outflow is the forbidden \FeII{} line
at 1.64\micron, which predominantly appears near the tips of the various fingers. JWST does
have a corresponding narrow-band filter (F164N), but here we rely on that line also appearing 
in the F162M filter, while another \FeII{} line at 1.257\micron{} appears in the 
F115W filter, as do other key \HII{} region and outflow tracers including the \HeI{} line 
at 1.083\micron{} and Paschen-$\beta$ at 1.282\micron. Finally, where flows emerge into 
the H\,II region, they can be ionised by the UV radiation field of the central OB stars and 
will thus be detectable in the F187N filter, among others.
\end{enumerate}

It is worth noting that beyond the core scientific goals elucidated above, the same filter set
should permit a wide range of other purposes. As just one example, the F335M filter also covers
the bright polycyclic aromatic hydrocarbon (PAH) emission feature at 3.3\micron, as is very
evident in the data described below, while the F444W filter covers not only the 4.69\micron{}
line of H$_2$, but also many lines of CO (1--0) R- and P-branch emission, which is known to
be a complementary tracer of young outflows probing different excitation regions
\citep[see, \eg][]{ray23}. Thus our survey should have considerable legacy potential.

As NIRCam allows for simultaneous observations in the SW and LW channels, we chose the following
six pairs of filter for our survey: F187N + F470N, F212N + F300M, F140M + F335M, F162M + F360M,
and F182M + F277W\@. Central wavelengths, bandwidths, and cut-on/cut-off wavelengths are
given in Table~\ref{tab:filters}, along with the nominal integration times outside 
overlapping and gapped regions, point-source detection limits in low-nebulosity
regions, and the utility of the filter in the context of this survey. 

\begin{table*}[p]
\begin{center}
\begin{tabular}{lcccccccp{6cm}}
\hline
Filter  & Pivot $\lambda$ & Bandwidth & Blue cut-on $\lambda$ & Red cut-off $\lambda$ 
        & Integration time & Saturation & Limiting & Purpose \\
        & (\micron)       & (\micron) & (\micron)             & (\micron)             
        & (sec)           & magnitude   & magnitude           & \\ \hline
F115W	& 1.154	& 0.225	& 1.013	& 1.282	& 515.365 & 18.1 & 29.2 & 
          Continuum, \HeI, \FeII, Pa-$\beta$ \\
F140M	& 1.404	& 0.142	& 1.331	& 1.479	& 773.064 & 17.4 & 28.3 & 
          \water{} absorption set \\
F162M	& 1.626	& 0.168	& 1.542	& 1.713	& 773.064 & 16.8 & 27.5 & 
          \water{} absorption set, \FeII{} \\
F182M	& 1.845	& 0.238	& 1.722	& 1.968	& 773.064 & 16.6 & 26.3 & 
          \water{} absorption set \\
F187N	& 1.874	& 0.024	& 1.863	& 1.885	& 773.064 & 15.7 & 25.5 & 
          H$^+$ n=4--3 Pa-$\alpha$ \\
F212N	& 2.120	& 0.027	& 2.109	& 2.134	& 773.064 & 16.2 & 24.9 & 
          H$_2$ v=1--0 S(1) \\
F277W	& 2.786	& 0.672	& 2.423	& 3.132	& 773.064 & 16.4 & 23.6 & 
          Continuum \\
F300M	& 2.996	& 0.318	& 2.831	& 3.157	& 773.064 & 16.3 & 23.8 & 
          \methane{} absorption set \\
F335M	& 3.365	& 0.347	& 3.177	& 3.537	& 773.064 & 15.8 & 23.3 & 
          \methane{} absorption set, PAH, H$_2$ v=1--0 O(5) \\
F360M	& 3.621	& 0.372	& 3.426	& 3.814	& 773.064 & 15.6 & 22.7 & 
          \methane{} absorption set \\
F444W	& 4.421	& 1.024	& 3.881	& 4.982	& 515.265 & 14.9 & 22.6 & 
          Continuum, CO (1--0) \\
F470N	& 4.707	& 0.051	& 4.683	& 4.733	& 773.064 & 14.3 & 22.0 & 
          H$_2$ v=0--0 S(9) \\ \hline
\end{tabular}
\end{center}
\caption[]{Parameters of the NIRCam filters used for the Cycle 1 GTO Trapezium Cluster and 
Orion Nebula survey and the nominal integration times used for each. Corresponding limiting 
magnitudes for detection at the faint end and saturation at the bright end are given for
point sources in Vega magnitudes, although in practical terms both are variable across 
the survey given the often bright and complex nebulosity. Approximate completeness limits 
lie 3--3.5 magnitudes brighter. A brief description of the utility of each filter in this work 
is also given. The filters are also illustrated in
Figure~\ref{fig:spectralseries} and detailed transmission profiles are available at:
\url{https://jwst-docs.stsci.edu/jwst-near-infrared-camera/nircam-instrumentation/nircam-filters}
}
\label{tab:filters}
\end{table*}

\subsection{Region surveyed}
While the Orion Nebula is very large, with the so-called Extended Orion Nebula or EON covering
roughly 0.5\degree{} or 3.5\,pc in diameter, the relatively limited observing time available to 
a JWST Science Working Group Interdisciplinary Scientist meant that only section of the region 
could be surveyed at this point. 

Nevertheless, the great bulk of the stellar and substellar population of the Trapezium Cluster 
at the heart of the Orion Nebula is concentrated in and around the $\sim 5$\,arcmin (0.6\,pc) 
diameter Huygens Region, the brightest part of the nebula centred on the Trapezium OB stars. 
Extending a little further then covers the Huygens Region and the associated Bright Bar and 
Dark Bay regions to the south-east and east/north-east of the Trapezium, respectively, 
along with most of the Trapezium Cluster, but also covering the BN-KL region and the full 
extent of the outflow fingers, and the OMC-1 ridge with the OMC-1S and other embedded sources 
to the west. The latter ridge also extends to the north towards OMC-2 and contains a plethora 
of young embedded stars, but will need to be covered in a future survey.

In addition, the quantised nature of the NIRCam field-of-view and the need to come up with a
scheme that fills the inter-module and inter-detector gaps while maximising the efficiency of
the survey, plus the quantised nature of the NIRCam read-out patterns introduces a series of 
``steps'' in the time needed as the survey expands both spatially and in terms of sensitivity. 
After experimenting with various possibilities and trading off the additional science versus
the time required, we decided to cover a contiguous region that is $\sim 11$\,arcmin in right 
ascension that is centred E-W along the OMC-1 ridgeline through BN-KL and OMC-1S, and 
$\sim 7.5$\,arcmin in declination that is centred N-S just north of the Trapezium, to cover 
the full extent of the BN-KL outflow to the north and most of the Bright Bar to the south.
For reference, the nominal central JWST pointing coordinate for the survey was:
05h 35m 14.1140s, $-05$\degree{} 23\arcmin{} 14.45\arcsec[] (J2000.0).

Covering this region required a mosaic of NIRCam pointings arranged in 2 columns and 5 rows, with 
the telescope at a position angle (V3PA) of 270\degree{}, meaning that the columns and rows
extended in $\delta$ and $\alpha$, respectively. The overlap between the 5 rows was set at only
2\% to ensure maximum E-W coverage while maintaining a small overlap between rows; the spacing
between the columns was set to 50\% to maximise N-S coverage while filling the inter-module 
gap\footnote{%
The following parameters were used for the main NIRCam mosaic in APT: 
Rows = 5, Columns = 2, Row overlap = 2\%, Column overlap = 58\%, Row shift = $-1.0$, 
Column shift = 0.5, V3PA = 270\degree.
}.

This scheme was used for five out of the six filter pairs described above. For the F115W + F444W
pair, we created a slightly wider mosaic with a 30\% overlap between the rows, \idest{}
a $2\times 7$ position arrangement\footnote{%
Similarly, the following APT parameters were used for the F115W + F444W mosaic: 
Rows = 7, Columns = 2, Row overlap = 30\%, Column overlap = 58\%, Row shift = $-1.0$, 
Column shift = 0.5, V3PA = 270\degree.
}.
The aim here was to deliver good sampling of the stellar field with more overlap, to ensure 
that a very accurate astrometric registration could be achieved could be built across the 
various sections of the mosaic, to enable further proper motion studies, for example. As 
described below, this proved unnecessary for the present work, as a carefully-adjusted 
selection of {\it Gaia\/} stars delivered a mosaic that is accurate at the $<0.1$ arcsec level, 
but the F115W and F444W mosaics were still used scientifically. 

\begin{figure*}[p]
\centering
\includegraphics[width=\textwidth]{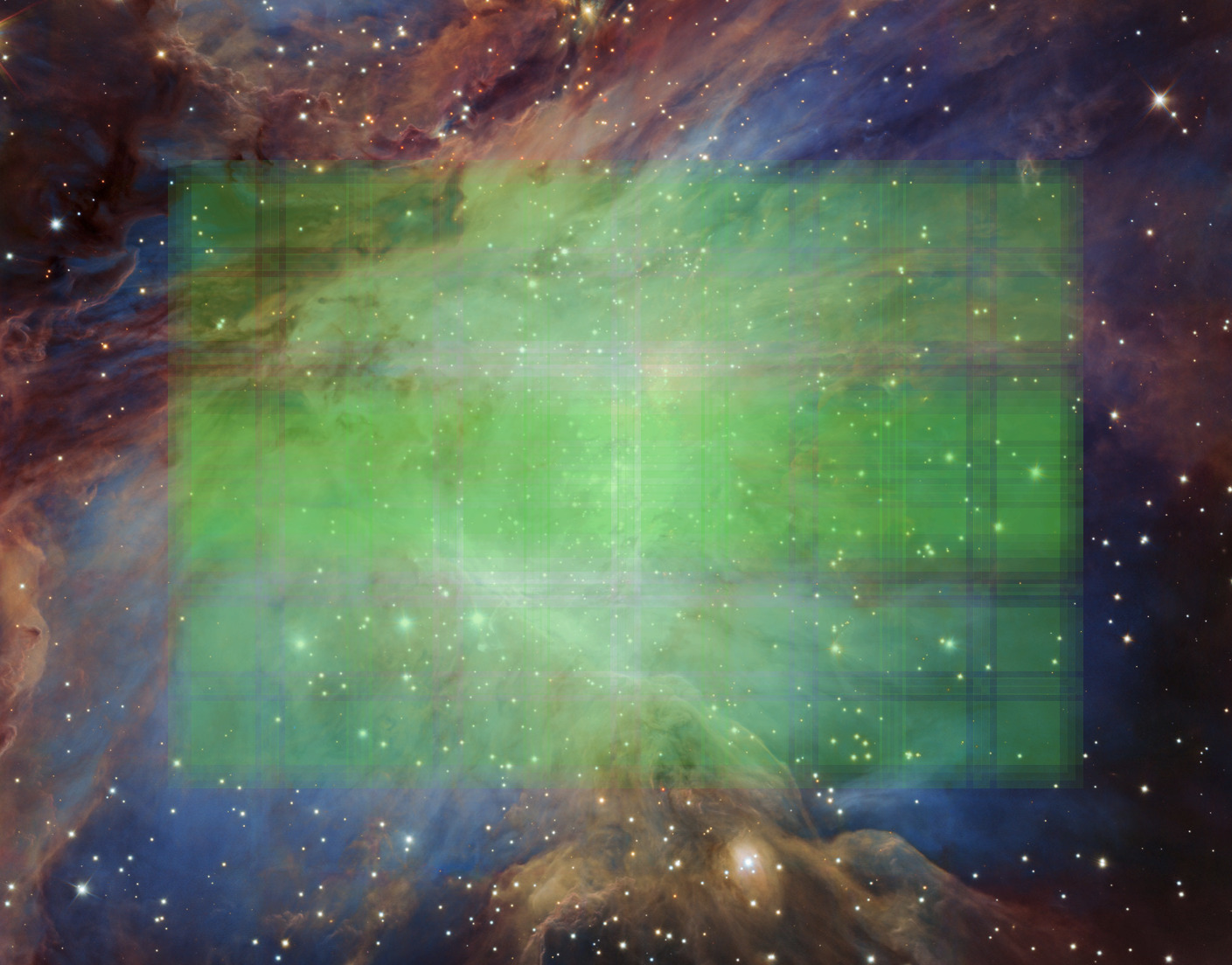} 
\caption{A schematic representation of the main region surveyed and the NIRCam SW detector 
footprints including mosaicing and dithering over a ground-based near-infrared image of the 
Orion Nebula and Trapezium Cluster from \citep{drass16}. The main region surveyed with 
JWST shown here spans $11\times 7.5$ arcmin, while the F115W + F444W mosaics extend slightly 
wider in right ascension. 
}
\label{fig:surveyregion}
\end{figure*}

Finally, to cover the small inter-detector gaps, remove bad pixels, and provide some better
sampling of the PSF, we used the INTRAMODULEX dithering pattern with 6 dithers for the main
survey and 4 dithers for the F115W + F444W pair. No sub-pixel dithering was used.
Figure~\ref{fig:surveyregion} shows a schematic overview of the coverage of NIRCam detector
footprints over the region surveyed. 

\subsection{Integration times}
In deciding the integration time to be used, several constraints came into play: 
\begin{itemize}
\item The predicted fluxes of the faintest targets being sought, \idest{} planetary-mass 
objects at 1\Mjup{} and less, at an age of $\sim 1$\,Myr, and at the distance of the Orion 
Nebula, in each of the key filters; 
\item The impact on the sensitivity of the observations given the high and variable brightness 
of the nebular background on one hand, and the variable reddening seen towards objects at
different locations and depths into the background molecular cloud on the other; 
\item The extreme range of stellar and nebular brightnesses in the region and the desire
to avoid saturation as far as possible on targets of interest at the bright end;
\item The total amount of time available, balancing the depth at any given location, the 
area to be surveyed and thus the chances of finding potentially rare objects, and the filters 
to be used to maximise the scientific return;
\item The data downlink budget. 
\end{itemize}

While a range of models have been developed over the years converting the interior and
atmospheric properties of brown dwarfs and planetary-mass objects as a function of their
evolution into predicted observational fluxes 
\citep[\eg][]{dantona94, chabrier00, baraffe03, marley07, baraffe15, marley21, chabrier23},
the variable nebular brightness and reddening are harder to account for in a
strictly quantitative way, especially in filters without good precursor imaging (\eg{} F335M). 
Therefore we used our pre-existing VLT near-infrared (J, H, \& Ks band) imaging of the 
Orion Nebula and Trapezium Cluster to make approximate adjustments between the predictions
of the exposure time calculators (including the JWST ETC) and actual detections of objects
down to $\sim 3$\Mjup{} in Orion, accounting for the increased spatial resolution of JWST
(a crucial advantage in pulling faint point sources above the nebula background), the
lower sky background seen by JWST, plus rough extrapolations for the likely nebular brightness,
the underlying dust reddening law, and so on.

Very crudely, indications were that $\leq$1\Mjup{} objects would be detectable through moderate
extinction in 10--15 minutes of integration time in the key medium-band filters, with less time
required to detect the same point sources in the wide-band filters and reduced sensitivity in 
the narrow-band filters, although they were primarily intended for extended targets. 

However, single exposures of this duration would clearly be impossible in the region, leading to
saturation for many stars and substellar objects, as well as on the nebula background over 
a wide region, plus excessive cosmic ray hits. At the other extreme, using the NIRCam RAPID 
read-out mode would deliver a new image every 10.737 seconds, the minimum full-frame time, but
would quickly overwhelm the JWST Solid State Recorder and the available downlink budget.

Balancing these constraints, we chose to use the sample-up-the-ramp non-destructive read-out
modes of NIRCam that measures the charge collected on a pixel every 10.737 seconds until the
integration is terminated and the pixel reset. The brightness of the flux arriving at
each pixel is calculated from the slope fitted to the sequence of read-outs. With at least
three frames, anomalies in the slope can be detected and thus the impact of cosmic rays 
mitigated. Similarly, if a pixel saturates after several read-outs, a change in slope is
seen and the brightness of the source determined by fitting only the unsaturated samples.

To further reduce the data volume, we used the SHALLOW2 read-out mode, which co-adds 
the first two frames separated by 10.737 seconds (a so-called group), skips three frames, 
co-adds the next two frames as another group, skips three frames, and so on. 
In this way, with the first two frames combined into a single group, the minimum integration 
time would be 21.474 seconds: anything that saturates in that time would be lost. The next 
group would only be complete after 75.159 seconds, meaning many other objects could become 
saturated in that time.

However, all of the co-added read-out modes in NIRCam also save and downlink the initial 
frame, the so-called {\tt Frame\,0}, which yields an effective minimum integration time of 10.737 
seconds and thus significantly expands the dynamic range. Anything that is bright enough to 
saturate in less than 10.737 seconds is lost, but anything that saturates after that is 
essentially recoverable. 

In the end, for the main survey, we used SHALLOW2 mode with three groups (NGRPS = 3), 
yielding a maximum on-chip integration time of 128.844 seconds. Rather than use a longer 
sequence of groups and risk saturating more pixels (thus essentially wasting observing time), 
we elected to dither then start a new integration. For the main survey, we used six dithers
and thus the effective integration time at one pointing is 773.064 seconds.

For the F115W + F444W astrometric observations, we used the BRIGHT2 mode, which co-adds 
sequential pairs of frames without skipping any, and also saves {\tt Frame\,0}. With NGRPS = 6,
this also yields 128.844 seconds on-chip and with four dithers, a total integration time
of 515.365 seconds 


Given our mosaicing scheme as described above, the bulk of the region surveyed then has the
full integration time of 773 seconds (or 515 seconds for F115W + F444W), with the exception of 
a $\sim 1.5$\,arcmin strip stretching across the whole 11 arcmin E-W region where the integration 
is doubled thanks to the scheme used to fill the inter-module gap. Conversely, there are 
narrow strips all over the survey where the integration time is reduced due to the inter-detector 
gaps, as is also the case along the edges of the mosaic and most notably in the four corners. 

Summing over all individual integrations, filter pairs, dithers, and mosaic positions, the
total on-source science time for the survey was 12.76 hours.

For the main survey, 480 images were taken for each SW filter (10 mosaic positions $\times$ 
6 dithers $\times$ 8 detectors) and 120 for each LW filter (10 mosaic positions
$\times$ 6 dithers $\times$ 2 detectors), yielding 3000 images overall. The F115W + F444W 
survey yielded a further 560 images (14 mosaic positions $\times$ 4 dithers $\times$ 
10 detectors), for a combined total of 3560 images.

\subsection{Scheduling}
Given the distance between the $2\times 5$ pointings of the survey ($2\times 7$ 
for F115W + F444W), each of the 10 (14) pointings became a separate JWST visit, and despite 
the use of the SHALLOW2 read-out mode and its reduced data rate, the volume of data
produced meant that the visits could not all be scheduled consecutively and were in fact 
executed over a week. In principle, that could be problematic for photometry of young stars 
such as those in Orion, since they are well-known to be variable on similar timescales. 
However, as ten of the twelve filters, including the key medium-band ones, are observed
in a single visit, the maximum time between photometry for a single object is nominally
only $\leq 3$\,hrs.

Specifying a fixed telescope orientation (V3PA = 270\degree{}) for all of the visits to
ensure the maximum area was covered in the allocated time without leaving gaps in the mosaic
necessarily constrained the schedulability of the programme. On top of which, as Orion
lies within 30\degree{} of the ecliptic, the windows are quite narrow, just a three weeks in
September-October each year, 3 months before the ``classical'' northern midwinter ground-based 
observing season for Orion (JWST does not point in the anti-Sun direction but roughly 
orthogonally to the Sun-Earth-L2 line).


In the end, the observations for this programme\footnote{%
\url{https://www.stsci.edu/jwst/science-execution/program-information?id=1256}
}
were made between 13:39:59UTC on 26 
September 2022 and 03:24:08UTC on 2 October 2022\footnote{%
More detailed timings for the individual visits can be found at: \\
\url{https://www.stsci.edu/jwst/science-execution/program-information?id=1256}}.
Including all telescope pointings, acquisitions, and movements, the nominal total amount of
wall clock time as calculated by APT and charged to this programme was 34.33 hours. Combined
with the total amount of on-source science time from above, that yields an efficiency of 37\%.

\section{Data reduction} \label{sec:datareduction}
\subsection{Customising the JWST pipline} 
To reduce the observations, we retrieved the Stage~0 data products for JWST Programme~ID 1256
from the Barbara A. Mikulski Archive for Space Telescopes (MAST)%
\footnote{\url{http://dx.doi.org/10.17909/vjys-x251}}.

We then ran the Stage~1, 2, and~3 reduction steps using a custom version of the JWST 1.11.3 
pipeline \citep{bushouse23} and Calibration Reference Data System mapping 
{\tt jwst pmap\_1100}. Stage 1 was run using the optional step argument 
{\tt det1.ramp\_fit.suppress\_one\_group = False}. Stage~2 was run using the default 
reduction pipeline. A custom version of the Stage~3 pipeline was used to align the individual 
images to {\it Gaia\/} Data Release~3 \citep[GDR3,][]{gaia16, gaia23} and combine the images into 
the final full mosaics, as described below.

When downloading the data from MAST, it was noted that the World Coordinate System (WCS) of 
Visit~2 for the ten filters in the main survey and Visit~7 of F115W + F444W were found to be 
in error by $\sim 15$ arcsec, resulting in major discontinuities in the mosaics. The error
was adjusted by manually adding an offset to the WCS data stored in data model in the 
{\tt asdf} tree in the header of each {\tt \_cal.fits} file. This approximate correction 
did not not take into account distortion effects, but significantly reduced the search 
radius needed for subsequent finer alignment.

Despite having taken the F115W + F444W with an overlapping mosaic pattern to ensure a good
astrometric calibration, we decided to start by aligning the F470N data to GDR3, as the
images in this filter had the largest overlap between the faintest {\it Gaia\/} stars and 
unsaturated JWST sources. We compiled an absolute reference catalogue of $\sim 650$ 
high-quality GDR3 sources that excluded flagged binaries, close pairs, extended galaxies, 
and knots of nebulosity, the latter being a major source of contamination in this region.

For each of the Stage~2 {\tt \_cal.fits} images, we compiled an individual source catalogue. 
The $x,y$ coordinates of the centre of the corresponding GDR3 sources were determined 
using a non-pipeline recentring routine. Each source was also weighted depending on the 
quality of the fit and whether it was found to be saturated in the {\tt \_cal.fits} data. 
The Stage~3 {\tt TweakReg} routine was then run on each of the {\tt \_cal.fits} individually. 
The absolute reference catalogue was passed to the {\tt TweakReg} routine using the 
{\tt tweakreg.abs\_refcat = path-to-file} step argument. The source catalogues were 
saved as {\tt .ecsv} files and were passed to the {\tt TweakReg} routine by updating the 
{\tt asn} with the file path. This process was repeated for each {\tt \_cal.fits} file 
individually as the pipeline defaults to expanding the absolute reference catalogue, 
which causes alignment errors. The individually aligned files were then resampled into a 
full combined F470N mosaic using step arguments: 
{\tt tweakreg.skip = True}, {\tt skymatch.skip = True}, {\tt resample.fillval = 'nan'}.

From this F470N mosaic, a new absolute reference catalogue of $\sim 1500$ sources was constructed. 
There was significantly more overlap of non-saturated sources between this catalogue and those
constructed for the other 11 filters than there was between those filters and GDR3, resulting in
an improved alignment. The F470N catalogue was used to repeat the above process for the 
remaining 11 filters, in each aligning the individual {\tt \_cal.fits} files to the F470N 
absolute reference catalogue before combining and resampling them into full mosaics.

MacBook Pro~M1 laptops were used for the pipeline processing. While the full LW filter mosaics
could be combined in one go, the SW filter mosaics were too big and thus were split into two
sections of 3 rows and 2 columns, one comprising the easternmost 3 rows and the other the
3 westernmost rows. These could be readily combined as there was a full row of overlap, but
this was only done at the final colour composite mosaic stage as described below.

\subsection{Source detection, cataloguing, and aperture photometry} 
Sources were detected in the Level~3 mosaics produced by Stage~3 of the pipeline. First, the 
two-dimensional background of each image was estimated and subtracted using the 
DAOPHOT {\tt MMM} algorithm as implemented in {\tt Astropy} \citep{bradley23, astropy22}
using a $30 \times 30$ pixel box and a $5 \times 5$ pixel filter. 
We used the {\tt MMMBackground} algorithm to divide the input data into a grid of 
$30\times 30$ pixels boxes and then used its mode estimator of the form 
$(3\times {\rm median}) - (2\times {\rm mean})$ to calculate the background level of each box, 
thus creating a low-resolution background map. This image was then median filtered to suppress 
local under or over estimations, with a window of size of $5 \times 5$ pixels. The final 
background map was calculated by interpolating the low-resolution background map.

Sources were then identified using DAOStarFinder with a threshold of $2\sigma$ and a model
PSF for each of the 12 JWST filters employed \citep{perrin14}. Sources that were detected 
in $\geq 3$ filters were then added to a preliminary source catalogue which was checked by 
eye against the images to remove spurious sources, such as bad pixels, knots of nebulosity,
diffraction spikes, and persistence spots that had been erroneously flagged as point 
sources. The by-eye examination was also used to visually classify other sources including
proplyds, outflows, and galaxies. The final catalogue contains 3090 sources.


Aperture photometry was performed using {\tt Photutils} package \citep{bradley23}
in {\tt Astropy} \citep{astropy22}. We used the {\tt aperture\_photometry} routine to obtain
fluxes for all of the sources in our catalogue, using apertures of 2.5 and 4.5 pixels radius
for the sources while the background was measured in an annulus with inner and outer radii of 
5 and 10 pixels, respectively, using a sigma-clipped median. The {\tt PIXAR\_SR} header 
keyword was used to convert from surface brightness (MJy\,sr$^{-1}$) to point source flux 
(Jy) and then to Vega magnitudes using the zeropoints provided by the Spanish Virtual 
Observatory (SVO) filter profile service \citep{rodrigo20}. To convert the aperture 
magnitudes to total magnitudes, we used the aperture corrections provided by the JWST 
reference files for the respective filter, interpolated to the corresponding aperture radius. 

By examining the differences in magnitude measured in the 2.5 and 4.5 pixel radius apertures,
we were able to distinguish between point and extended sources. Point sources should have the
same magnitudes through both apertures after the corrections above, while extended sources
such as galaxies and nebular knots will appear brighter in the larger aperture. Sources
where the median difference between the apertures across all filters in which they were
detected exceeded 0.1 magnitudes were classified as extended. Sources with a neighbour within
1 arcsec were excluded from this automated classification and checked manually. 

Given the large collecting aperture of JWST, the high spatial resolution, and the
relatively long minimum on-chip integration time, point sources saturate at fairly faint
brightnesses. Across the 12 filters, this also depends on wavelength and bandwidth, but
for example the saturation limit lies at \magnit{18}{1} for F115W, \magnit{16}{6} for F182M,
\magnit{15}{8} for F335M, and \magnit{14}{3} for F470N (all Vega magnitudes). 
Very roughly, this corresponds to masses higher than $\sim 0.6$--1.4\Msolar{} in the 
narrow-band filters, $\sim 0.1$--0.2\Msolar{} in the medium-band filters, and 
$\sim 0.07$--0.15\Msolar{} in the wide-band filters, assuming zero extinction. 

A summary of the limiting magnitudes and saturation limits for each filter is given 
in Table~\ref{tab:filters}. 

\begin{figure*}[p]
\centering
\includegraphics[width=\textwidth]{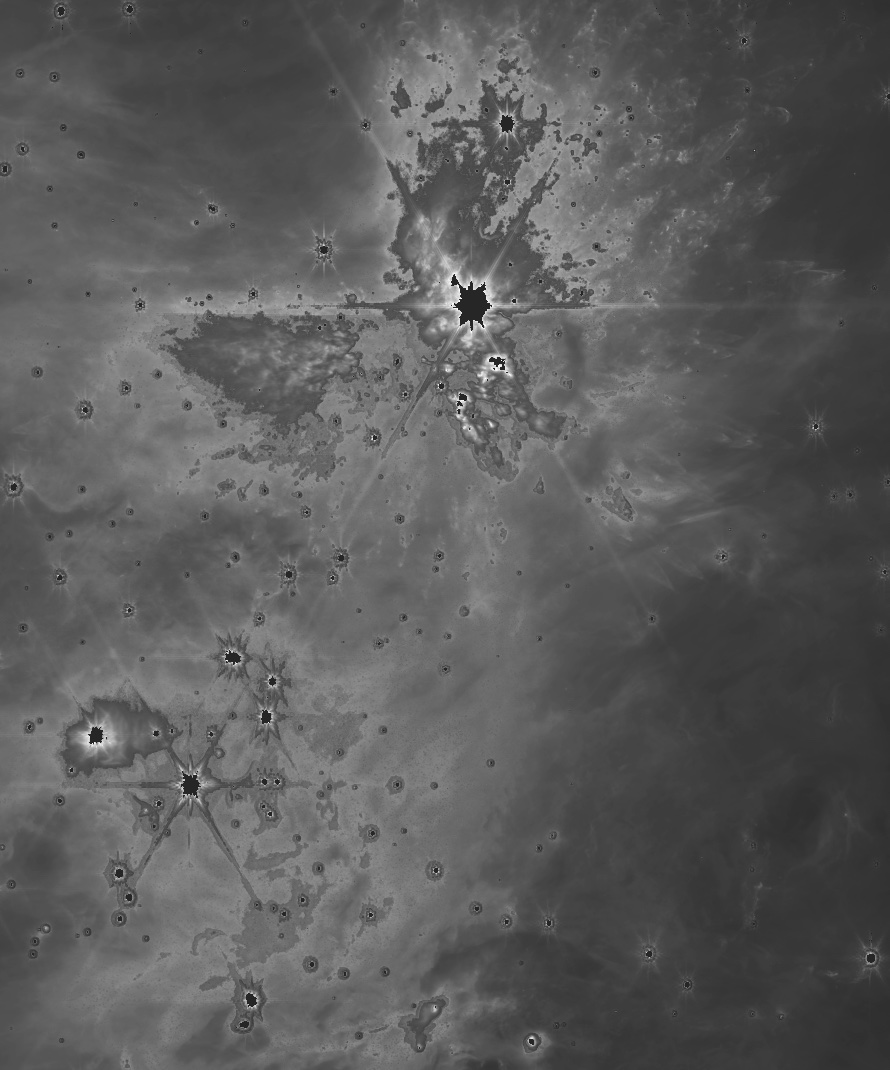} 
\caption{An illustration of the intensity wrapping issue encountered in bright regions in
Orion prior to the pipeline fix in July 2023. The image shows a section of the F277W mosaic
with the Trapezium stars in the lower left and BN-KL in the top centre. Fully saturated
pixels in the centres of the bright stars are rendered as {\tt NaN} or zero by the pipeline, but 
flux below that is seen to be wrapped due to a bug when switching from full slope fitting
to the use of {\tt Frame\,0}. In fact, in more detail, two separate wrappings are seen, indicating
a more subtle issue. The fact that unsaturated detail can be seen in the wrapped regions
suggested that this could be recovered as it now has been.}
\label{fig:intensitywrapping}
\end{figure*}

One particular issue which delayed some aspects of the data analysis was a bug in the
pipeline implementation of the {\tt Frame\,0} dynamic range extension which was only rectified in 
July 2023. Prior to this, particularly bright regions close to saturation were seen to
be ``wrapped'' in intensity, with a sudden drop in brightness and then a subsequent increase:
see Figure~\ref{fig:intensitywrapping}. It was evident that there was more dynamic range to 
be had and once the pipeline had been fixed, this problem was eliminated.

\subsection{Scaling, cleaning, and combination for colour mosaics}
The flux- and astrometrically-calibrated FITS mosaics for each of the twelve filters are the 
primary source for the scientific analysis described here and in future papers in this series. 
However, the same mosaics can be adjusted to compress the dynamic range, aligned, and combined
into colour composites for outreach purposes, but also to elucidate key aspects of the science
and even make discoveries.

From this point, all processing of the FITS files output by the pipeline was performed
using IRAF \citep{tody86, tody93}.

\subsubsection{Cropping} \label{sec:cropping}
The twelve SW mosaics (two overlapping sections for each filter) and six LW mosaics were 
carefully examined and cropped to sizes which lacked any missing data. In particular, the 
use of only four primary dithers for the F115W and F444W imaging resulted in a small periodic 
data loss along the upper and lower
edges of the SW F115W mosaics, due to incomplete coverage of the inter-detector gaps. Excluding
all rows affected by these periodic indents results in a loss of $\sim 2.5$\% in declination
coverage in the F115W mosaic compared to all of the other SW channels mosaics and it was thus
not included in the final SW colour composite. It nevertheless remains very important for
the science described below.

\subsubsection{Dynamic range compression}
The dynamic range in the images between the central OB stars and brightest nebulosity on 
one hand and the for western corners and Dark Bay on the other is very high and this was
compressed by taking the common logarithm of the data values, making sure to adjust the lower
tail of the background sky histogram to be above zero beforehand to avoid clipping.

\subsubsection{Saturation}
Even after the successful inclusion of the {\tt Frame\,0} processing to extend the dynamic range,
the cores of many stars and even some nebulosity in the wide-band filters remained saturated.
The default setting of the pipeline is to set saturated pixels to {\tt NaN} and thus to enable
further processing outside the scientific environment, these pixels needed to be set to a high
positive value in order to avoid them appearing as black craters in the middle of bright stars.

For each mosaic individually, the brightest valid pixels were identified and that value was
then substituted for every instance of {\tt NaN}. Given the high base level of nebulosity and
the way that the NIRCam read-out scheme and pipeline use slope fitting to establish brightnesses,
this is not entirely valid across the whole mosaic, but it is good enough for cosmetic
purposes. Future PSF-fitting photometry to the wings of such stars may make it possible 
to ``fill'' the saturated cores with more appropriate brightness extensions. 

\subsubsection{Residual noise and other non-astronomical features}
Imaging in the infrared in a region with many bright stars inevitably leads to undesirable
effects, not least persistence in the HgCdTe detectors. Careful examination of every mosaic
revealed a plethora of residual features that had not picked up by the pipeline when
aligning, stacking, and combining the many images, including:

\begin{description}
\item [\bf Persistence:] it is well-known that HgCdTe detectors exhibit a persistence 
effect in which illumination in one image can fill traps in the detector material which
later decay and cause residual signals in subsequent ones for several minutes. This effect 
is particularly noticeable when very bright stars leave ghost images of themselves in 
later images taken at different pointings \citep{smith08}.
Of course, the Trapezium Cluster is full of extremely bright stars and imprinted persistence
patterns were a major concern ahead of the observations. Fortunately, the overall bright 
nebulosity appears to have ameliorated the impact, restricting persistence to just a few
specific cases.
First, many of the bright stars in the darker SW part of the
survey were seen to have ghost images at around 8 arcsec (and occasionally at 18 arcsec also)
exactly at 45\degree{} to the NW, and in some cases, a faint trail could be seen 
connected the ghost to the star distinct from the JWST diffraction spikes at 30\degree. 
These are evidently residuals and may be linked to dither pattern steps, although they
would then be expected to the NE, SW, and SE as well. In some cases, the residuals had
been removed by the pipeline, but in many others not.
Second, there were also some very major persistence `tracks' closer to an E-W orientation
and only partly repaired by the pipeline. These may be related to the initial pointing at
the Trapezium Cluster as it was acquired for each visit. 
Third, some major patches of nebulosity were seen apparently randomly strewn across the mosaics
These were only seen in a couple of filters and were heavily mottled, unlike any astrophysical
source elsewhere in the images, and they had a broadly common appearance. We concluded that these
were also persistence artefacts, perhaps from the very bright nebular regions around the 
Trapezium and BN-KL\@. 
\item [\bf 1/f noise:] the NIRCam detectors are known to exhibit 1/f or pink noise 
\citep{rauscher11} manifesting itself as row-to-row offsets and thus striping. The visibility
of these stripes is a function of the background flux: the lower it is, the stronger they become. 
Therefore the striping is more prevalent in the SW channel images, as each pixel subtends just 
one quarter of the solid angle of a single LW pixel, in narrow-band images like F212N,
and towards the edges of the survey where the nebular background is low. Given our chosen 
telescope orientation at V3PA = 270\degree, those stripes are vertical in declination. 
\item [\bf Snowballs:] these are large circular transient features with slightly elevated fluxes
above the background that occur in images made with HgCdTe arrays. It is not known whether
these features are a result of radioactive alpha decay due to impurities in the detectors or
cosmic rays \citep{cillis18}.
While multiple dithers allow them to be substantially removed, often a faint caldera-like 
structure remains visible with a slight rise in intensity above the background to a 
``crater rim'', with the flux dropping back to background levels inside the crater.
Residual snowballs are particularly evident in regions with low background near the
edges of our survey.
\item [\bf Cosmic rays:] the majority of cosmic rays are eliminated by the sample-up-the-ramp
non-destructive read-out modes used by NIRCam and the image stacking process, but some escape 
and need to be removed. Care needs to be taken though as single pixel cosmic ray events
can become blurred out into multiple pixel events in the process of resampling and drizzling,
meaning that they can be mistaken for very faint point sources or vice versa.
\item [\bf Fake cosmic rays:] at the time of writing, there is a bug in the JWST pipeline that
appears to inject random hot pixels into one of the H2RG detectors (A2) of the SW channel
during the blotting component of the {\tt OutlierDetection} routine. This leads to periodic 
regions of the mosaics that appear to be peppered with large numbers of cosmic rays. While 
this has been reported to the pipeline developers, no fix has been implemented as yet. As
with real cosmic rays, these can be blurred out and mimic faint astrophysical point sources,
so care must be taken when cleaning them.
\item [\bf Optical ghosts:] in a very small number of cases next to the brightest isolated stars
like \thetatwoa{} and \thetatwob{} there appear to be bright ghosts which resemble the 18 segment
JWST primary. Similarly, \thetatwoa{} lies close to the edge of a detector in one of the
mosaic/dither positions and optical ghosting is evident which leads to a poor intensity match
when stacked with other images.
\item [\bf Other artefacts:] we do not see any clear evidence for other NIRCam 
scattered light artefacts\footnote{%
\url{https://jwst-docs.stsci.edu/jwst-near-infrared-camera/nircam-instrument-features-and-caveats}
},
including the so-called ``claws'', ``wisps'', ``dragon's breath''
or ``ginkgo leaves'', despite the fact that the latter two effects in particular are due 
to bright stars just outside the NIRCam field-of-view and that the Trapezium Cluster is 
full of very bright stars. It is 
possible, however, that these low-level artefacts are lost in the complex nebulosity of 
the region and/or that they have been removed by dithering. It is also possible
that some of the artefacts seen around \thetatwoa{} and \thetatwob{} in some filters are
related to Type 1 dragon's breath. 
\end{description}

To remove these residual, post-pipeline artefacts for cosmetic reasons, the individual filter,
logarithmically-scaled FITS mosaics were first converted to 16-bit PNG files using the
{\tt ImageMagick} package. Cleaning was then done using a combination of GIMP with the 
G'MIC-Qt package installed, Adobe Lightroom, and Adobe Photoshop, as follows.

The first step was removing 1/f noise from the images where it was noticeable: in practice that
meant all SW channel images, plus F300M and F470N\@. The G'MIC-Qt ``Banding Denoise'' 
algorithm was used in tiled mode, with its various parameters specifically tuned for
each image. Particular care was taken to avoid overly aggressive application of the
algorithm so as not to introduce additional artefacts (\eg{} negative `trails' above and
below small extended sources like galaxies and `shadows' above and below features in
bright horizontal diffraction spikes) or to eliminate real structures like the
short vertical diffraction spikes due to the secondary support structure. 

In general, this step removed all trace of the 1/f noise
with the exception of the F212N images, where the striping was very strong in the darker
parts of the surveyed region. In this case, wide residual bands were visible and these
were reduced using very slight adjustments to the background brightness levels in and out of the 
bands using a brush mask in Lightroom.

As described above, residual real and fake cosmic rays in the data were not always easily 
distinguishable from faint point sources as they were blurred out in the resampling and
drizzling steps in the pipeline. Experiments were made with automated cleaning in IRAF,
but no parameters could be found which removed most cosmic rays without also removing faint
sources. 

Therefore the spot healing tool of Lightroom was used to manually clean the residual cosmic rays, 
both real and fake, along with snowballs and persistence features, including faint spots,
trails, and large-scale features. The user clicks on a feature using an adjustable radius 
circle, the tool finds a nearby region with similar noise properties, and generates a seamless 
`patch' to the region selected. Largely this works very cleanly, but has to be taken that the 
tool does not choose a region that contains another source, otherwise it becomes cloned. Whenever 
there was a question about the reality of a faint point source or other feature, images at other 
wavelengths were checked: there are very few if any real astrophysical sources which only 
appear in a single filter. 

This cleaning approach meant a careful and methodical manual inspection of all twelve 
SW mosaics and all six LW mosaics at high magnification, a process that took a total of 
fourteen days work in August 2023\footnote{%
Credit is due to the musicians and curators of the various Japanese ambient and Shibuya-kei 
playlists on Spotify which helped the first author remain mostly sane during this otherwise
soul-destroying period.}.

Finally, there were a few obvious large-scale patches in some of the mosaics due to intensity 
steps at the edges of individual images which had not been appropriately matched by the pipeline. 
These features were eliminated by the application of regional masks and slight exposure 
adjustments in Adobe Photoshop. Similarly, some very abrupt steps in intensity in a few 
wavelengths around \thetatwoa{} and \thetatwob{} were removed via cloning and masking of other 
sections of PSF of the stars, taking care not eliminate nearby fainter point sources. 

The cleaned mosaics were primarily prepared for combination into colour composites, but can
also be used for certain scientific projects, as the overall intensity scaling can be 
inverted and the original flux-calibrated intensities recovered. However, for obvious reasons,
any potential discoveries made in these mosaics would need to checked against the 
original Stage 2 or Stage 3 pipeline products.

\subsection{Colour composites} \label{sec:colourcomposites}
\begin{figure*}[p]
\centering
\includegraphics[width=\textwidth]{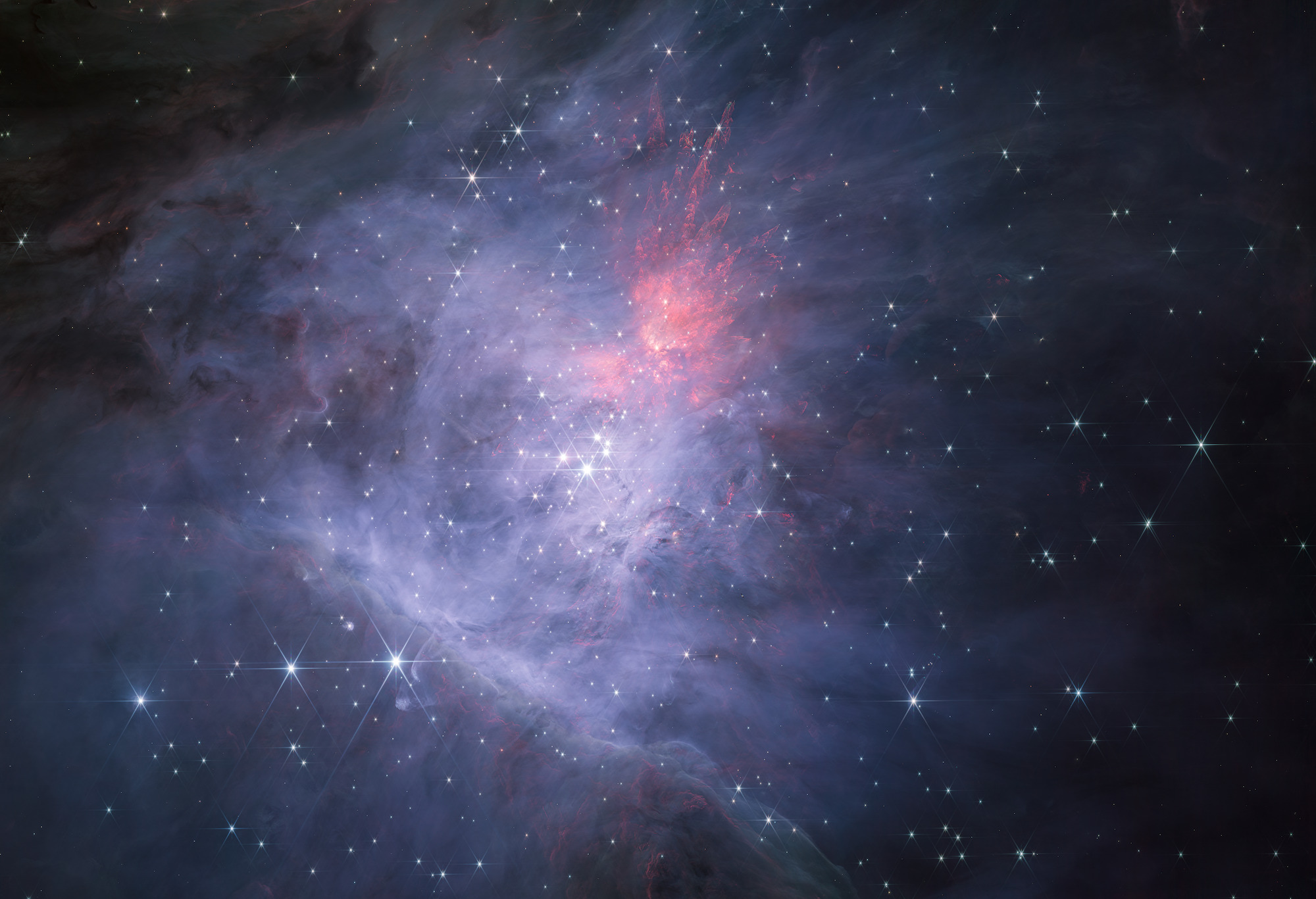} 
\caption{%
The Short Wavelength colour composite covering the full JWST NIRCam survey of the inner 
Orion Nebula and Trapezium Cluster, comprising a combination of five SW channel filters 
(F140M, F162M, F182M, F187N, F212N) as described in the text. A total of 2,400 individual
H2RG images have been combined to make this mosaic.
The image is centred at 05h 35m 14.10s, $-05$\degree{} 23' 13.2'' (J2000.0), covers 
$655.8 \times 448.2$ arcsec at 31.2275 mas/pixel, $10.93 \times 7.47$ arcmin, 
or $1.24\times 0.85$\,pc assuming a distance of 390\,pc, and is oriented close to 
N is up, E left, with a rotation angle of +0.4477\degree{} E of N\@. 
The version shown here is greatly reduced from the $21000\times 14351$ pixel original.
}
\label{fig:swcomposite}
\end{figure*}

\begin{figure*}[p]
\centering
\includegraphics[width=\textwidth]{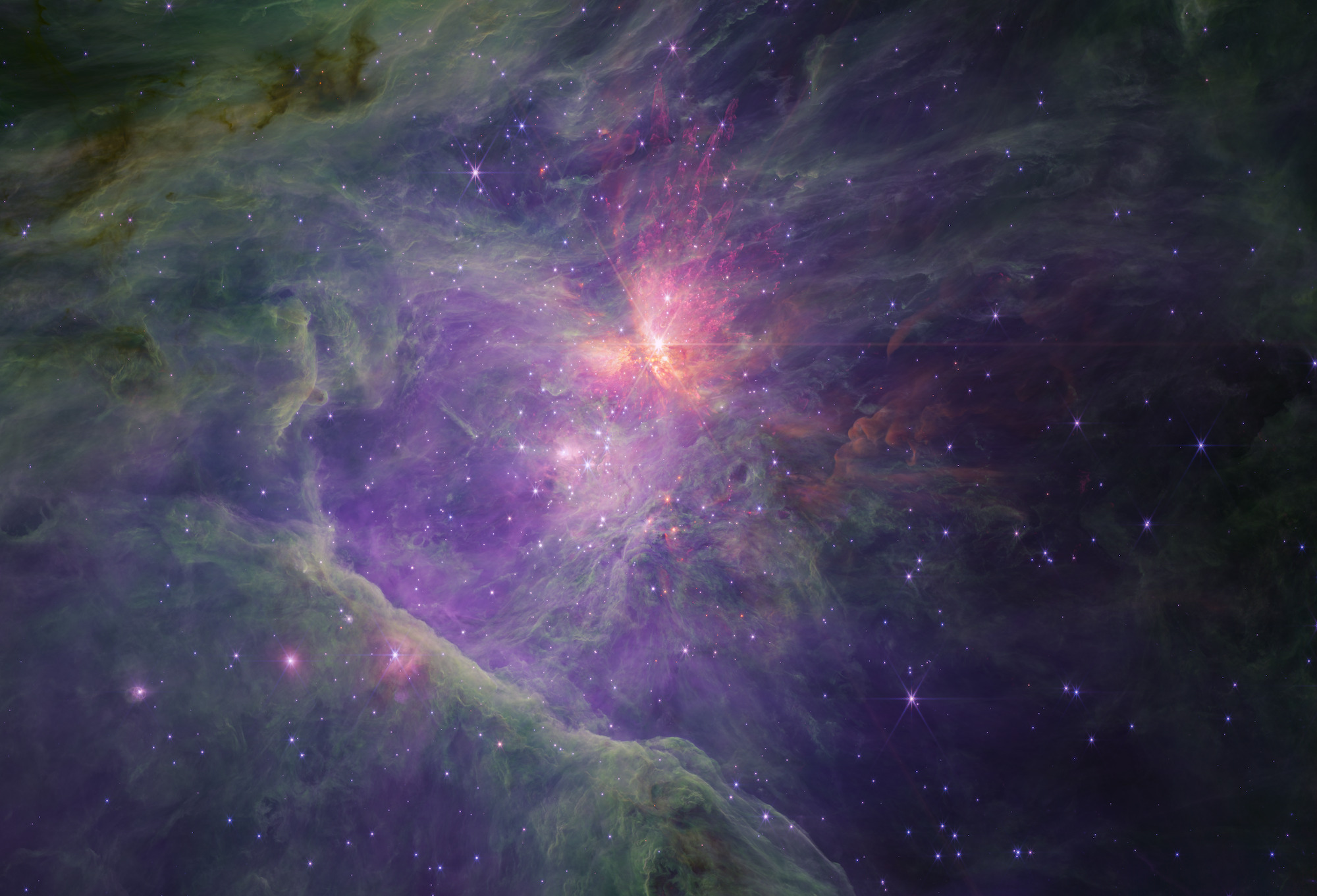} 
\caption{%
The Long Wavelength colour composite covering the full JWST NIRCam survey of the inner 
Orion Nebula and Trapezium Cluster, comprising a combination of six LW channel filters 
(F277W, F300M, F335M, F360M, F444W, F470N) as described in the text. A total of 712
individual H2RG images have been combined to make this mosaic.
The image is centred at 05h 35m 14.10s, $-05$\degree{} 23' 14.5'' (J2000.0), covers 
$657.2 \times 447.25$ arcsec at 62.9108 mas/pixel, $10.95 \times 7.45$ arcmin, 
or $1.24\times 0.85$\,pc assuming a distance of 390\,pc, and is oriented close to 
N is up, E left, with a rotation angle of $-0.024$\degree{} E of N\@. 
The version shown here is greatly reduced from the $10446\times 7109$ pixel original.
}
\label{fig:lwcomposite}
\end{figure*}

The cleaned, logarithmically-scaled mosaics for the various filters were then used to create
two colour composites, one each for the SW channel and LW channel, as shown in 
Figures~\ref{fig:swcomposite} and~\ref{fig:lwcomposite}, respectively.

For the LW channel composite, the six mosaics were opened in Photoshop and aligned as layers
in a single image. Despite the same {\it Gaia\/} stars having been used to define the astrometric
reference frame for all wavelengths, it turned out that in addition to slight pixel shifts
between the mosaics, very small scale adjustments were also needed in some cases to
ensure that stars lined up across the whole region. Using the F277W mosaic as a base, the F300M
and F360M mosaics aligned with just $x,y$ pixel shifts, while the F335M, F444W, and F470N
mosaics need to be rescaled by 1--3 pixels in $x$ and $y$ to line up properly, so by
roughly 0.02\% across the whole $\sim 10,500$ pixel mosaic. No rotations were needed to
align the layers.

For the SW channel, two separate composites were made using the E and W overlapping sections
initially. As with the LW composite, some slight adjustments to the scale on the order of 
0.01--0.02\% were needed to ensure the various wavelengths lined up: using F140M as the fiducial, 
F162M lined up with just an $x,y$ shift, while the remaining wavelengths all needed small, 
different scale adjustments. As with the LW composite, no rotations were needed between 
wavelengths.  The E and W sections of the SW mosaics were then combined after the following stage.

With the different wavelength images aligned, each layer could be assigned colours and their
intensities adjusted to create an aesthetically pleasing and as far as possible, scientifically
meaningful colour composite.

For the LW composite, all six filters were used and colours assigned in order of wavelength, 
with RGB values and notional colour names\footnote{%
Colour names from: \url{https://colornamer.robertcooper.me}
}
as follows: 
F277W (52,0,255) electric ultramarine;
F300M (0,0,255) blue;
F335M (0,255,0) green;
F360M (255,255,0) yellow;
F444W (255,122,0) heatwave;
F470N (255,0,0) red.

For the SW composite, only five filters were used, with F115W left out because of the gaps
along its northern and southern edges (see Section~\ref{sec:cropping}), which would have yielded
a smaller mosaic. Colours were assigned in order of wavelength with one exception: F187N,
the Pa-$\alpha$ line should be yellow or orange in principle, but was set to a medium grey-blue
instead. The rationale for doing so is to make a visual reference to the colour scheme often
used for ground-based near-infrared images of the Orion Nebula 
\citep[\eg][]{ukirt89, kaifu00, mccaughrean02, tamura06, drass16}
with the J-band as blue, H-band as green, and
K-band as red. As Pa-$\alpha$ is not visible from the ground, the diffuse ionised gas of the
region is represented by the Pa-$\beta$ line of hydrogen instead at 1.282\micron, thus
in the J-band and blue. As we do not include the F115W filter and thus Pa-$\beta$ in the colour
composite, this is reasonable substitution. The full set of filter RGB values and 
notional colour names is:
F140M (0,0,255) blue; 
F162M (0,255,0) green;
F182M (148,132,0) medusa green;
F187N (85,95,233) flickering sea;
F212N (255,255,0) red.

In practice, careful inspection of the colour composites revealed a small number of outlier
objects seen in one filter only, likely due to the misidentification of cosmic rays as real
sources or the inadvertent cloning a star while cleaning, along with other artificial features, 
so the individual wavelength mosaics were iteratively recleaned and recombined several times 
before the final composites were arrived at. 

Once the two colour composites were finalised, they were imported into Lightroom for final
processing involving slight adjustments to the global intensity mapping and relatively 
small local adjustments to contrast using the Texture and Clarity settings to bring out 
some of the more delicate features in the region. 

The final colour composite images are shown in reduced form here: Figure~\ref{fig:swcomposite}
shows the Short Wavelength (SW) channel mosaic, combining the F140M, F162M, F182M, F187N, and
F212N images as described above, while Figure~\ref{fig:lwcomposite} shows the Long Wavelength
(LW) channel mosaic combining the F277W, F300M, F335M, F360M, F444W, and F470N images as
described. Note that the full images do not cover precisely the same area of sky, mainly 
because the LW composite is rotated by 0.5\degree{} clockwise with respect to the SW 
composite due to a slight difference in the orientations between the SW and LW channels 
of NIRCam. 

In their full original resolution (31.2275 mas/pixel for the SW image and 62.9108 mas/pixel
for the LW image), the mosaics span $21000 \times 14351$ pixels and $10446\times 7109$ pixels, 
respectively. Both are available for download at full size and explorable at full resolution
using pan-and-zoom technology at the following web locations:
\begin{itemize}
\item \url{https://www.esa.int/Science\_Exploration/Space\_Science/Webb/Webb\_s\_wide-angle\_view\_of\_the\_Orion\_Nebula\_is\_released\_in\_ESASky}
\item \url{https://www.esa.int/ESA\_Multimedia/Images/2023/09/Orion\_Nebula\_in\_NIRCam\_short-wavelength\_channel}
\item \url{https://www.esa.int/ESA\_Multimedia/Images/2023/09/Orion\_Nebula\_in\_NIRCam\_long-wavelength\_channel}
\item \url{https://sky.esa.int/?jwst\_image=webb\_orionnebula\_shortwave}
\item \url{https://sky.esa.int/?jwst\_image=webb\_orionnebula\_longwave}
\end{itemize}
The latter links are to ESA's ESASky portal, which also allows the overlay of other images and
catalogue information, so should be of scientific use also
\citep[\url{http://sky.esa.int},][]{giordano18}. 

\section{The Orion Nebula} \label{sec:orionnebula}
In the introduction to this paper (Section~\ref{sec:introduction}), we gave a broad but
necessarily incomplete review of observational studies of the Orion Nebula and associated
stellar populations, the latter being a main focus of our JWST survey. However, the nebula
itself is also of great interest, with its extended emission and absorption due ionised, 
atomic, and molecular gas, plus obscuration by and emission from dust, as reviewed by
\citep{peimbert82, odell01b, odell01a, odell08}. 

The JWST images provide excellent high-spatial resolution and high-fidelity imaging of the inner
nebula in ionised, atomic, and molecular gas as well as dust, in tracers and at wavelengths
that are complementary to the extensive HST visible, Spitzer mid- and long-wavelength infrared,
interferometic millimetre, and radio observations. In some cases, comparisons should enable
a better understanding of the excitation and extinction conditions in the nebula (\eg{}
comparing ionised hydrogen emission in H$\alpha$ with Pa-$\alpha$, molecular hydrogen
emission at 2.12 and 4.69\micron), detailed morphology across tracers of different
temperature and density conditions (\eg{} in the Bright Bar PDR and around the Dark Bay),
and dynamics by measuring proper motions seen over the almost 30 year HST--JWST observational
time baseline. Here we only wish to draw attention to some of the general features of the 
nebula as seen in the JWST images –-- future studies using these data will go far deeper.

The SW colour composite looks broadly similar to ground-based three-colour near-infrared 
images made using the broad-band J-, H-, and K-band filters as blue, green, and red, respectively, 
with a general blue-purple haze due to ionised hydrogen emission. Of course, as described 
in Section~\ref{sec:colourcomposites}, this partly by design, with the assignment of the 
Pa-$\alpha$ line to blue tones, in place of Pa-$\beta$ which lies in the J-band filter.
In the JWST data, Pa-$\beta$ lies in the F115W filter, which is not used in the colour
composite. However, there is clearly far more small-scale detail seen in the JWST SW images 
due to the higher-spatial resolution, which also serves to yield significantly smaller stellar 
point-spread functions, thus further emphasising the nebula emission. 

The LW colour composite by comparison looks very different, stars taking another
step further back. The image is dominated by purple ionised emission in the central region 
around the Trapezium stars, with browner emission due to dust even closer
in, as described in Section~\ref{sec:trapezium}. There is a significant green in the image
due to PAH emission at 3.3\micron{} in the F335M filter. This is particularly bright along
the Bright Bar PDR described below, but also up into the Dark Bay to the E and NE, and with the 
central region too. The latter takes on an almost 3D appearance, suggestive of the back wall
of the ionised blister around the \thetaone{} stars. There are also holes and bubbles in the
PAH emission, some apparently related to outflows from the BN-KL and OMC-1S regions embedded
behind the blister, as described in more detail in Sections~\ref{sec:bnkl} 
and~\ref{sec:jetsoutflows} below. Finally, there are some peculiar brown ``smooth clouds'' 
to the W of the Trapezium and another N of BN-KL, both also hinted at in the SW composite.

\subsection{The Bright Bar} \label{sec:brightbar}
The Bright Bar is photodissociation or photon-dominated region (PDR), running from NE to SW 
between the \thetaone{} and \thetatwo{} stars. In the SW composite, a very fine 
transition is seen between the predominantly blue-purple ionised gas, to greener and
then red emission, the latter dominated by 2.12\micron{} H$_2$. The latter has an almost
3D appearance of clouds being illuminated from afar, in strong visual agreement with the 
finding that external FUV radiation from the OB stars is responsible for the emission in 
the Bright Bar and that the H$_2$ emission is fluorescent, not shocked as seen elsewhere 
in the region \citep{tielens93}. In the LW image, a well-defined interface is seen between
the ionised gas in purple and the PAH emission in green --- the latter also has a yellow
tinge along the brightest part of the Bar perhaps, which may be due to a contribution 
from CO emission in the F444W filter.

Substantial work has been done to combine infrared and millimetre observations of the Bar to 
arrive at a comprehensive picture of an edge-on PDR showing the transition from ionised to 
atomic to molecular gas 
\citep[\eg][]{tauber94, youngowl00, pellegrini09, goicoechea16}, and a comprehensive
study of the PDR is the subject of a dedicated JWST near- and mid-infrared programme which 
will doubtless make use of these wider-field data \citep{berne22}, so further discussion is 
referred there. 

One interesting point in our images, however, is that the 2.12\micron{} H$_2$ emission in
the colour composite and even more clearly the F212N image appears to delineate the
surface of a tapering pillar with its narrow end near \thetatwoa, and from the illumination 
of the various clouds and clumps along the pillar, the latter star also appears to be making 
significant contributions to the UV radiation, not just the Trapezium. 

\subsection{Eastern pillars} \label{sec:pillars}
\begin{figure*}[p]
\centering
\includegraphics[width=\textwidth]{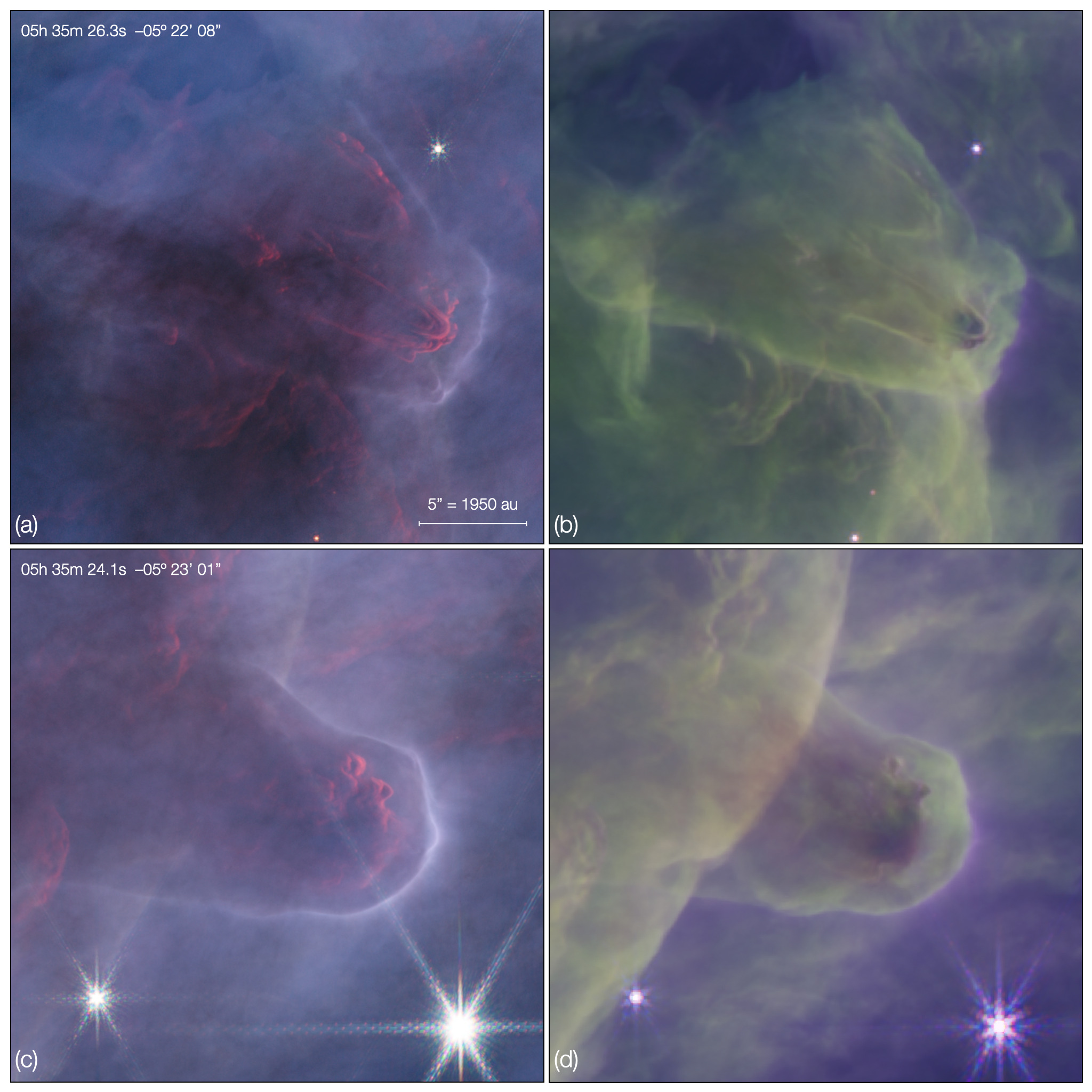} 
\caption{%
Close-up views of the two pillars to the E of the Trapezium extracted from the SW and
LW composites, illustrating the multiple layers revealed by the various JWST filters. In
the SW composites ((a) for the North Pillar, (b) for the South Pillar), relatively smooth, bright 
ionisation fronts are seen in blue $\sim 500$\,au ahead of clumpier structures outlined in red,
presumably fluorescent H$_2$ emission in F212N\@. In the equivalent LW composite images
((c) and (d)), the ionisation is seen in purple, albeit lacking a bright rim, while the
pillars themselves are shrouded in green, presumably PAH emission in F335M, turning to
a darker brown towards the denser regions of the clumps. Although superficially similar
in size and shape in the two composites, careful inspection reveals that the pillars
are slightly ``shrunken'' in the LW images, with the edge of the PAH emission lying 
$\sim 0.1$--0.2 arcsec or $\sim 40$--80\,au inside the ionisation front. There is little
evidence of F470N H$_2$ emission in the LW matching the F212N H$_2$ seen in the SW images.
Each image is $24.5\times 24.5$ arcsec or $\sim 0.05\times 0.05$\,pc assuming a distance of
390\,pc. The Trapezium and \thetaonec{} lie $\sim 0.25$\,pc to the WSW\@. N is up
and E left in all four panels.
}
\label{fig:pillars}
\end{figure*}

On a much smaller scale, but perhaps simpler geometrically and thus more accessible to 
modelling, there are two pillars seen to the E of the Trapezium, above the Bright Bar and
in the Dark Bay, which are also clearly being impacted by the 
intense UV radiation from the OB stars (Figure~\ref{fig:pillars}).

These pillars are large enough to be evident in ground-based infrared images 
\citep[\eg][]{mccaughrean02, drass16}, and the South Pillar is seen in the
HST Treasury survey at visible wavelengths as a half-hidden red rim \citep{robberto13},
but the JWST images reveal them as $\sim 7000$\,au or 0.03\,pc long protrusions from a wider
molecular cloud along the edge of the ionised region. 

Their appearance is particularly dramatic in the SW composite, with bright ionised rims seen
in blue-white facing towards the Trapezium and then roughly 1--1.5 arcsec (390--585\,au) behind 
that there are clumps and sub-pillars rimmed with the red of H$_2$ at 2.12\micron, presumably
UV-excited and fluorescing. The extended sub-pillars are particularly clear in the North Pillar 
and similar red-rimmed clouds are seen along the cloud from which the pillars protrude. 

In the LW composite, the pillars are again rimmed with ionised emission, this time
seen in purple with the interiors of the pillars glowing green in PAH emission. The clumps
and sub-pillars illuminated red in the SW composite now take on a browner, dustier 
appearance, implying that, as expected, these are higher density structures inside the
main pillars. These structures cannot be very dense down their entire
length, however, as the South Pillar appears to be transparent near the base, with
stronger PAH emission from a rim on the parent cloud apparently shining through. There
is no obvious equivalent to the strong red rims in the 4.69\micron{} line of H$_2$. 

In more detail, a careful comparison shows that the outer rim of the
green PAH emission seen in the LW composite lies slightly inside ionised rim seen in the
SW composite: the offset is on the order of 0.1--0.2 arcsec or $\sim 40$--80\,au. These
observations seem particularly amenable to modelling of the pillar structure, factoring
the radiation field of the Trapezium stars and plausible density profiles for the pillars.

\section{The Trapezium region} \label{sec:trapezium}
The massive Trapezium OB stars, \thetaone, lie at the heart of the Orion Nebula and provide
much of the ionising flux and wind that illuminates and shapes the nebula, along with the
\thetatwo{} stars below the Bright Bar. The region is much more complex than the nominal 
four stars of the Trapezium would suggest, with each of the massive stars being a hierarchical
multiple system \citep[\eg][]{preibisch99, grellman13, gravity18}, and a high density of
lower-mass stars being arrayed around them \citep{mccaughrean94}. In addition to their 
wide-reaching impact on the scale of the full Orion Nebula, the OB stars have a significant
impact on the stars local to them and the protoplanetary disks. The proplyds 
are disks that are being externally photoevaporated and ionised by the OB stars, yielding 
characteristic `tadpole-shaped' nebulae as material flowing away is shaped by the radiation 
and winds from the same stars. 
The objects were initially detected in ground-based optical emission-line imaging
by \citet{laques79} and also at radio and x-ray wavelengths \citep{churchwell87, garay87,
felli93}, but it was the post-refurbishment Hubble Space Telescope images of 
\citet{odell93, odell94, bally00} that revealed their full structure, with many proplyds 
around the Trapezium with tails pointing away from the OB stars. The OB stars also heat 
dust in the region, perhaps generated by the photoevaporated disks, resulting in the 
Ney-Allen nebula at thermal-infrared wavelengths, 
a series of shells and arcs curving away from \thetaonec, with a significant concentration of 
emission coming from around \thetaoned{} 
\citep{ney69, wynnwilliams74, mccaughrean91, 
hayward94a, hayward94b, smith04, smith05, robberto05}.

\begin{figure*}[p]
\centering
\includegraphics[height=21cm]{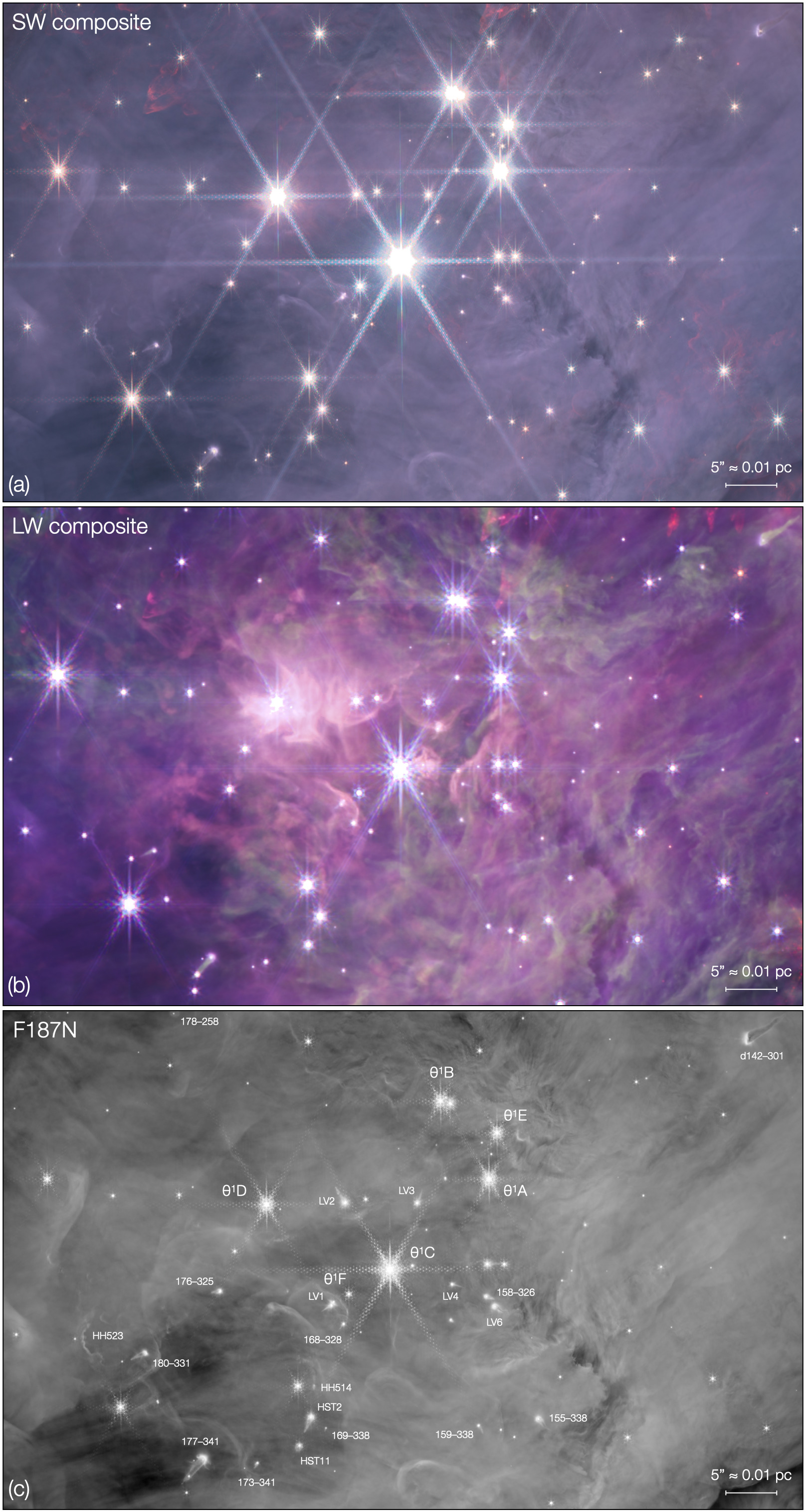} 
\caption{%
Subsections of the SW composite (Figure~\ref{fig:swcomposite}), the LW composite
(Figure~\ref{fig:lwcomposite}), and F187N image focussing on the Trapezium stars and the
region to the immediate south where a number of well-known proplyds are located. Some of
these are indicated in the F187N image using a mix of the nomenclatures established by
\citep{laques79, odell93, bally00, ricci08}, along with some Herbig-Haro objects and the 
names of the Trapezium stars. The global and local intensity and contrast have been 
adjusted with respect to the main composites to make key features clearer. 
The images are centred just N of \thetaonec{} at 05h 35m 16.46s, $-05$\degree{} 23' 21.8'' 
(J2000.0), are oriented close to N up, E left, and each spans $67.1\times 41.8$ arcsec
or $0.127\times 0.079$\,pc assuming a distance of 390\,pc. 
}
\label{fig:trapezium_triptych}
\end{figure*}

Figure~\ref{fig:trapezium_triptych} shows cut-outs from the SW and LW composites centred
on \thetaonec{} (panels (a) and (b), respectively), albeit with the dynamic range and contrast 
adjusted to see more detail. In addition, a similar cut-out from the Pa-$\alpha$ F187N mosaic 
is shown (panel (c)), as this suppresses the continuum emission from stars while better 
emphasising the ionised gas emission. The latter is annotated with the names of the 
\thetaone{} OB stars, as well as many of the ionised proplyds around them. 

The proplyds and stand-off ionisation fronts are also seen in the SW composite cut-out at lower
contrast, along with the first hint of red nebulosity around \thetaoned{} 
\citep[cf.][]{mccaughrean94}, and some of the southward fingers of shocked H$_2$ emission 
from the OMC-1 outflow discussed in more detail in Section~\ref{sec:omc1outflow}. Many of 
the proplyds are also seen in the LW cut-out and some show green tails, indicative of 
PAH emission in the material being swept back from the circumstellar disk by the OB stars.

The LW image also provides the highest-ever resolution image of the dust shells which comprise
the Ney-Allen nebula. 
The previously most-detailed images were taken on the 8-m diameter Gemini-S at
11.7\micron, yielding a maximum spatial resolution of 0.35 arcsec \citep{smith05}, 
while the resolution of the JWST images is 0.127 arcsec at 4\micron. Against that, the
stars are still moderately bright at 4\micron, while at 10\micron{} and beyond, the
dust arcs dominate entirely. As the temperature of
the dust in the region is around 300\,K \citep{hayward94a}, we are likely seeing a mix
of thermal emission, reflected light, and perhaps PAH emission associated with the dust. The
brown colour of the shells and arcs, indicating more emission at longer wavelengths, fading away
at $\sim 0.05$\,pc radius from \thetaonec{} to be replaced by green PAH emission, would tend
to support that. There is green PAH emission nearer \thetaonec{} also, but that may be a
projection effect, if the Trapezium stars are embedded in a 3D dust pocket. 

The brightest region in the JWST LW image coincides, as expected, with \thetaoned,
which \citet{robberto05} suggest may harbour a photoevaporating disk, releasing dust,
while \citet{smith05} favour scenario where \thetaoned{} is located closer to the background
ionisation wall and PDR on the Orion Nebula. Another prominent arc to the SE of \thetaonec{} 
is associated with IRS4 of \citet{wynnwilliams74}, which is in turn almost identical
in shape to the ionisation front in the Pa$-\alpha$ image between proplyd LV1 (aka 168--326), 
\thetaonef, and ultimately \thetaonec. The same arc is seen in the HST H$\alpha$ and
\OIII{} images. The other obvious dust arcs in the LW image can similarly be identified with 
the proplyds LV2 (167--317), LV3 (163--317), and LV4 (161--324), although not all show bright 
ionisation fronts. These coincidences confirm the model proposed by \citet{smith05} in which
the dust is being lit up at the point where material flowing away from the photoevaporating disks 
collides with the stellar wind of \thetaonec. More detailed modelling and follow-up NIRSpec 
and MIRI IFU spectroscopy should provide excellent probes of the excitation and radiation 
processes at work in this region.

\section{Brown dwarfs and planetary-mass objects in the Trapezium Cluster}
\label{sec:cluster}
One of the main scientific drivers for this JWST survey of the Trapezium Cluster was the search 
for candidate planetary-mass objects (PMOs) extending well below the $\sim 3$\Mjup{} achieved
in ground-based imaging surveys. The Orion Nebula is well out of the galactic plane and has 
a dense molecular cloud behind it, so background field star contamination should be limited, 
while the close distance to Orion also mitigates against substantial foreground contamination 
by field brown dwarfs. Indeed, over much of the survey region, there is no evidence for a 
large population of faint objects at the limit of sensitivity, suggesting both that the initial 
mass function does not continue to extremely low-masses in large numbers and conversely that 
the survey is not significantly polluted by the field. On top of which, it is possible to use 
the large variety of filters in NIRCam to provide an initial discrimination between young, 
low-mass objects in Orion and field contaminants on the basis of their spectral energy 
distributions. 

\begin{figure*}[p]
\centering
\includegraphics[width=\textwidth]{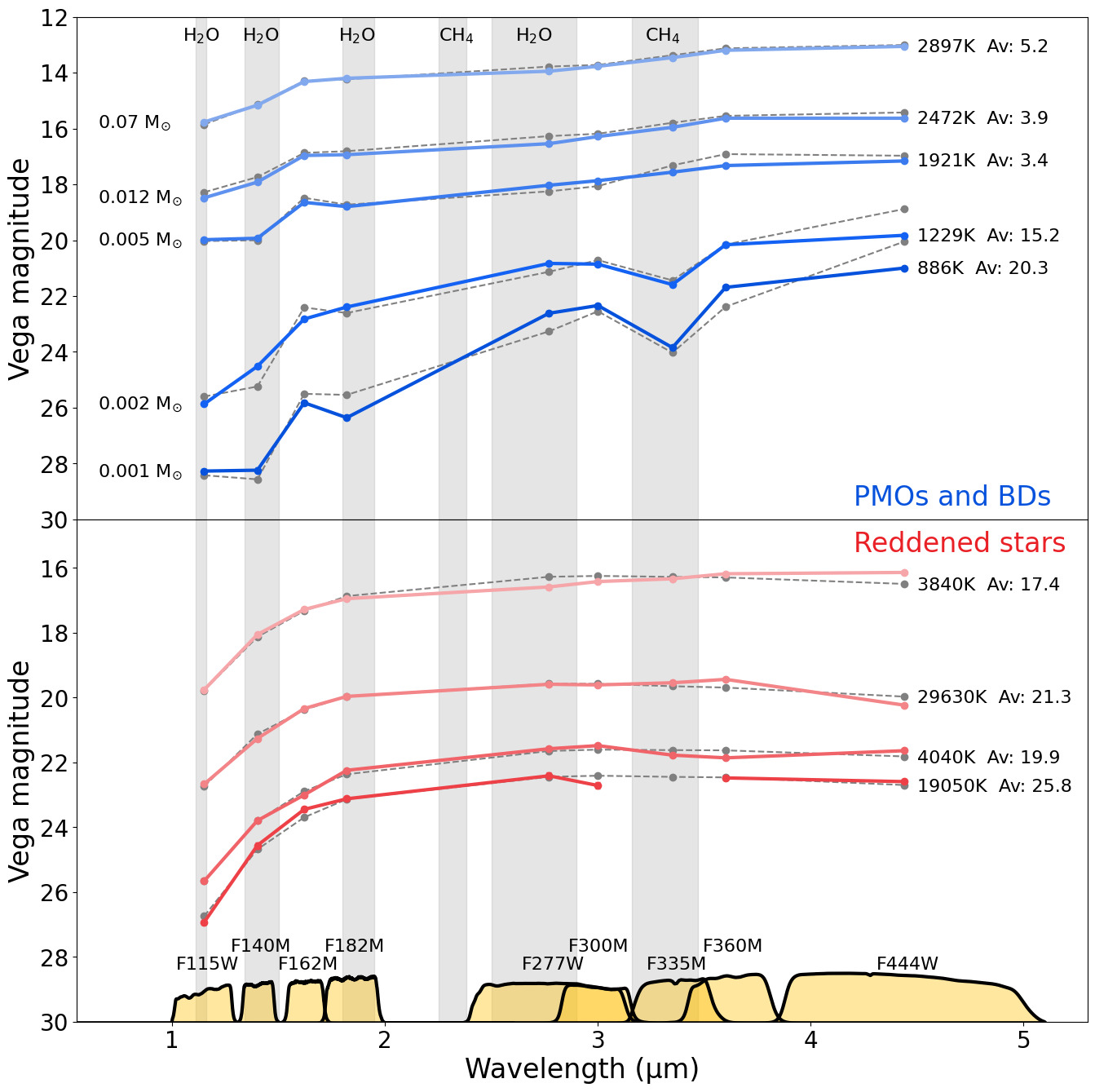}
\caption{%
A selection of point sources identified in our survey along with their measured magnitudes
and derived properties. The top panel shows sources (solid lines) which are well-matched 
to models (dashed lines) for young (1\,Myr low-mass sources at the distance of Orion, 
from the brown dwarf limit at $\sim 0.07$\Msolar/73\Mjup{} through the deuterium-burning limit 
at 0.012\Msolar/13\Mjup{} to Jupiter-mass objects at 0.001\Msolar/1\Mjup{}. Absorption
due to \water{} becomes evident in the SW filters as the effective temperature decreases 
into the brown dwarf regime and then at lower temperatures in the planetary-mass regime, 
\methane{} absorption is seen in the LW bands. The best fit effective temperatures and extinction
values are shown for each source. The lower panel shows a ``control'' sample of objects
lacking any of the absorption bands, marking them as background, reddened field stars at 
higher temperatures. For reference, the plot also shows the bandpasses of the 9 wide-
and medium-band filters used to measure the crude spectral energy distributions.
See Pearson \& McCaughrean (submitted, 2023) for more detail.
}
\label{fig:spectralseries}
\end{figure*}

Young ($\sim 1$\,Myr) PMOs with masses between 1 and 13\Mjup{} have effective temperatures 
of 890--2520\,K \citep{phillips20}, which means that an equivalent blackbody spectral energy 
distribution (SED) would peak between 1.15--3.3\micron, \idest{} covered by our JWST NIRCam 
wavelength range. In fact, much like the spectra of late M, L, and T field dwarfs spanning the 
same range of effective temperatures (albeit at higher surface gravities), the cool atmospheres 
of young PMOs are nothing like blackbodies and are dominated by broad bands of atmospheric 
\water{} and \methane{} absorption. These strong bands of 
molecular absorption confine the spectrum to a series of narrow peaks which can be identified 
using photometry through an appropriate set of medium- and wide-band filters, providing a robust 
method for distinguishing PMOs from more massive, distant, and reddened background objects. 
The molecular absorption bands of \water{} and \methane{} are shown as grey bars in 
Figure~\ref{fig:spectralseries}, while the nine medium- and wide-band NIRCam filters used to 
measure their strength are shown along the bottom of the plot. 

The power of utilising these absorption features for identifying PMOs is demonstrated in 
Figure~\ref{fig:spectralseries}. In the top panel, the blue lines show NIRCam photometry for 
a sample of our candidate brown dwarfs and PMOs with decreasing brightness. The dashed grey 
lines show the corresponding best-fitting models for 1\,Myr objects \citep{chabrier23, baraffe15},
assuming a distance of 390\,pc and allowing the reddening to vary during the fit. The
resulting best fit masses for this illustrative sample range from 0.07 to 0.001\Msolar{}
or approximately 70 to 1\Mjup.

The evolution of the \water{} and \methane{} bands as sampled by the nine filters is clearly
demonstrated in the sequence as a function of decreasing mass from brown dwarfs to PMOs.
The \water{} absorption features at 1.4 and 1.9\micron{} are present at the star to brown 
dwarf transition and strengthen with decreasing mass. At lower effective temperatures and
thus masses below 5\Mjup, \methane{} absorption at 3.3\micron{} becomes a defining 
characteristic of the SED, and continues to strengthen with decreasing temperature. 

For comparison, the red lines in the bottom panel show a `control sample' of candidate 
reddened background stars seen in our survey, with fitted effective temperatures of 
$\sim 3800$--30\,000\,K\@. In contrast to the brown dwarf and PMO candidates, the reddened 
background star candidates do not show molecular absorption feature, rather having much 
smoother SEDs that are well fit by a reddened blackbody model (dashed grey lines). This
demonstrates that our multi-filter technique is effective at weeding out background stars
and identifying a cleaner sample of genuine Orion brown dwarfs and PMOs, although definitively
confirming these candidates will require spectroscopy – we will be obtaining NIRSpec MSA prism
(R$\sim$100) spectroscopy for many of our candidates in JWST Cycle 2.

In this way, we have identified several hundred candidate PMOs in our survey region. A brief 
description of the methodology and top-level results are given here, but the reader is referred 
to Pearson \& McCaughrean (2023, submitted) for a full presentation of the results and analysis.

A total of 540 candidates have been identified with masses below 13\Mjup, of which 168 are
at 5\Mjup{} or less.  At higher masses, our PMO candidates recover the objects found in previous 
deep infrared surveys of the Trapezium Cluster and Orion Nebula Cluster
\citep{mccaughrean02, lada04, slesnick04, lucas05, meeus05, lucas06, riddick07, weights08,
drass16, robberto20}, but the higher spatial resolution and lower background of JWST allow
us to extend greater than $\sim 4$--5\,mag deeper in apparent magnitude and thus easily to
1\,Mjup{}: in extremum, we have discovered candidates at 0.6\Mjup{} or roughly 2 Saturn 
masses. Such masses are well below the classical minimum mass of 7--10\Mjup{} thought to 
arise from 3D opacity limited fragmentation in molecular clouds 
\citep[\eg][]{rees76, low76, silk77, boss88, whitworth18}, 
and even below the minimum mass of 2.6\Mjup{} that might occur under well-tuned conditions 
in a 2D shock-compressed layer \citep{boyd05, whitworth07}: this poses a problem if 
these objects formed directly from molecular clouds. 

It is always possible of course that some of the very low mass objects seen in our JWST survey
could have formed as planets in circumstellar disks around young stars before being ejected 
due to interactions in the disk \citep{parker12, vanelteren19, scholz22}. However, another
discovery made in our survey poses a conundrum for such a solution.
Namely that there is a significant population of low-mass PMOs down to 1\Mjup{} which are
found in $\geq 100$\,au binary systems. There are 40 such systems in our data and
two triple systems, yielding a wide binary fraction of 9\%. This is quite unexpected given
that the fraction for field and cluster brown dwarfs over similar separations is close to zero. 

\begin{figure*}[p]
\centering
\includegraphics[height=22cm]{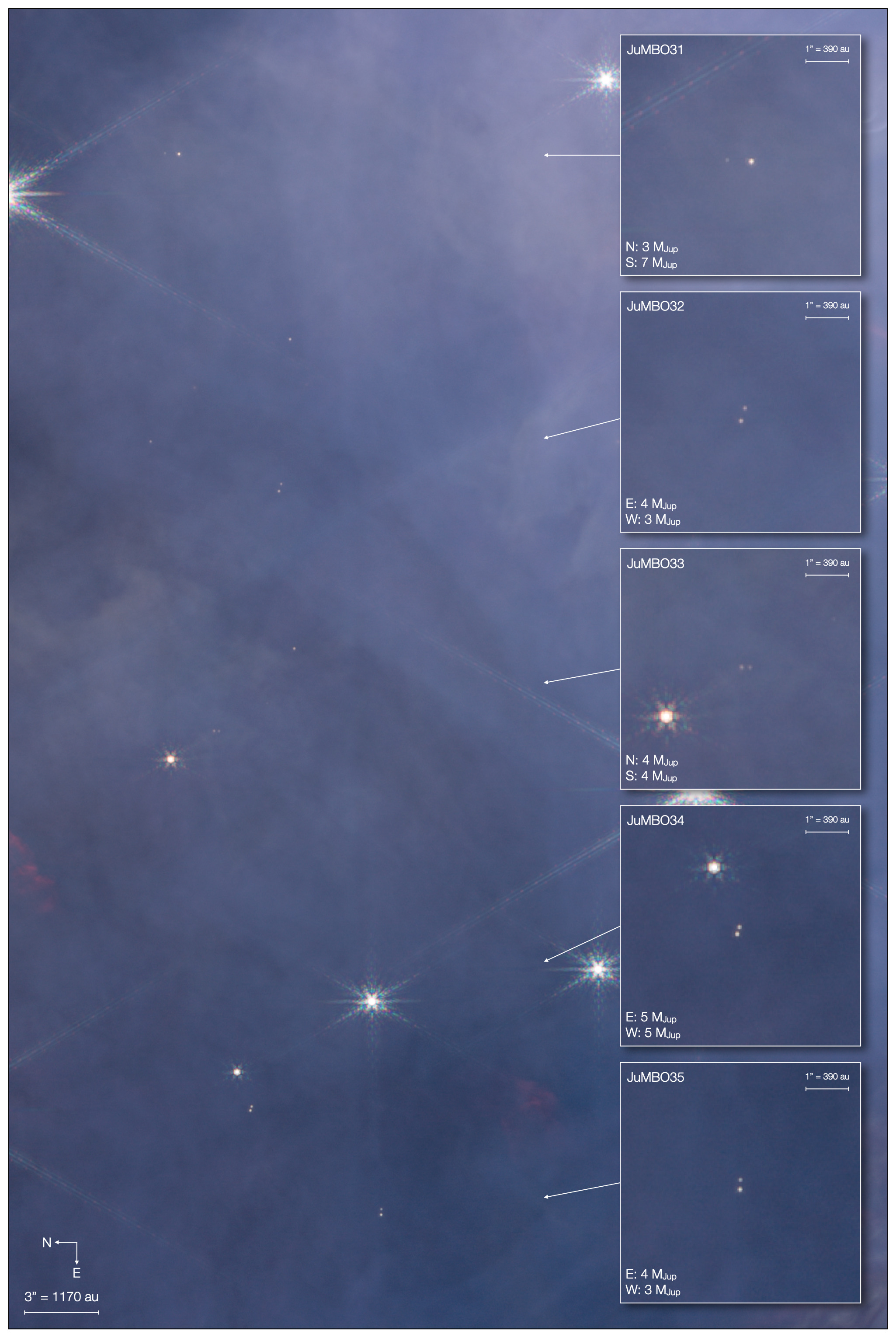}
\caption{%
A section of the SW composite located to the E of the Trapezium containing five
binary planetary-mass objects or JuMBOs, with cut-outs showing each pair in more detail.
The image has been rotated with N left and E down to show this E-W strip of JuMBOs more
effectively. It is centred at 05h 35m 27.0s, $-05$\degree{} 23' 27'' (J2000.0) and covers 
$52.3\times 35.3$ arcsec or $0.10\times 0.067$\,pc assuming a distance of 390\,pc.
}
\label{fig:jumbos_panels}
\end{figure*}

Figure~\ref{fig:jumbos_panels} shows a remarkable sample of five 
3--7\Mjup{} binaries in a small region to the east of the Trapezium. Given the generally 
rather low density of the cluster and field star populations at such separations, as is 
evident in the figure, these are clearly real pairs, not chance alignments. We have coined 
the acronym ``JuMBO'' for Jupiter-Mass Binary Object for these systems. 

The formation mechanism for JuMBOs is unknown. Can {\em pairs\/} of planets be ejected 
from a circumstellar disk and remain bound? Or perhaps new theoretical models may be needed
to explain not only how 1\Mjup{} objects can form in isolation, below the usual minimum
mass limits, but also that they can form as binaries. The particular conditions in 
the highly dynamic and energetic environment of the Orion Nebula can help 
\citep[see, \eg][]{whitworth07}, or perhaps unconventional mechanisms such as the 
fragmentation of low-mass starless disks must be invoked \citep[\eg][]{bodenheimer78}:
searching for similar objects in other young star-forming regions with varying
densities may help discriminate between potential models.

In any case, the larger number of such objects allows to examine the extreme low-mass end 
of the IMF in Orion, but also the planetary-mass binary fraction and other statistical 
properties, extending the many earlier studies on stellar and brown dwarf binaries in
Orion and other young star-forming regions 
\citep[\eg][]{petr98a, petr98b, scally99, scally05, reipurth07, petrgotzens08, kounkel16a, 
kounkel16b, duchene18, jerabkova19, tokovinin20, strampelli20, defurio22a, defurio22b}, 
and these issues are addressed in more detail by Pearson \& McCaughrean (2023, submitted).

\section{Circumstellar disks and proplyds} \label{sec:disksproplyds}
Virtually all of the circumstellar disks and ionised proplyds previously discovered by
\citep{odell94, mccaughrean96, bally00, ricci08} and others in the region covered by the
JWST survey are seen in many our images spanning the 1--5\micron{} range, apart from in a 
small number of cases where the central star has become so bright in the near-infrared as 
to swamp the silhouette and/or ionised emission. Equally, there are not many new discoveries
in this region, as the spatial resolution of JWST in the near-infrared is very close to that 
of HST at the key wavelengths around H$\alpha$.

That said, the new near-infrared images have an unprecedented combination of spatial resolution,
sensitivity, and fidelity at these wavelengths, as well as tracing a variety of ionised, 
fluorescent, and shocked species that can deliver new insights into the nature of young
protoplanetary disks in the region and how, in many cases, they are being impacted by their
proximity to intense sources of UV radiation and strong stellar winds.

\begin{figure*}[p]
\centering
\includegraphics[height=21.5cm]{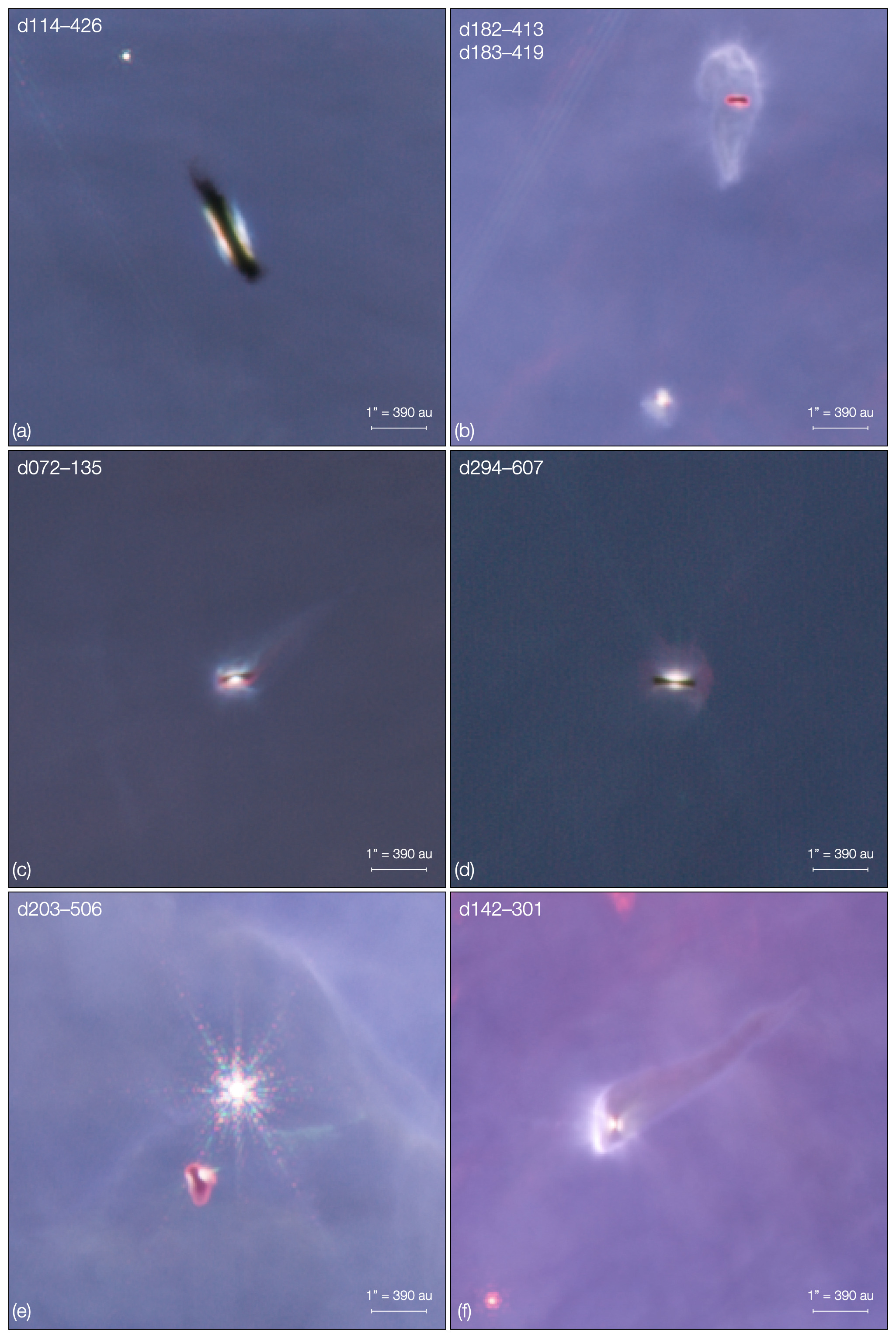} 
\caption{%
A selection of well-known silhouette disks in the Orion Nebula as imaged in the near-infrared 
with JWST\@. In this figure, a section of the main SW colour composite 
(Figure~\ref{fig:swcomposite}) has been extracted. Each panel is a 7.8 arcsec square and 
a 1 arcsec (390\,au) scale bar is shown. N is up and E left in all panels. The names in the top 
left corner encode the coordinate following the scheme of \citet{odell94} with the prefix 
`d' to indicate a disk: all objects have been previously 
catalogued from the various HST surveys \citep[\eg][]{bally00, ricci08}.
The local brightness and contrast has adjusted to maximise the visibility of 
relevant features, but the colour mix has not been altered. Thus as in the main SW colour 
composite, purple and blue reveal ionised gas and reflection nebulosity, red shows
H$_2$ emission at 2.12\micron, and green shows \FeII{} emission at 1.64\micron.
The features seen in the individual images are discussed in the text. 
}
\label{fig:silhouettes_swcolour}
\end{figure*}

\begin{figure*}[p]
\centering
\includegraphics[height=21.5cm]{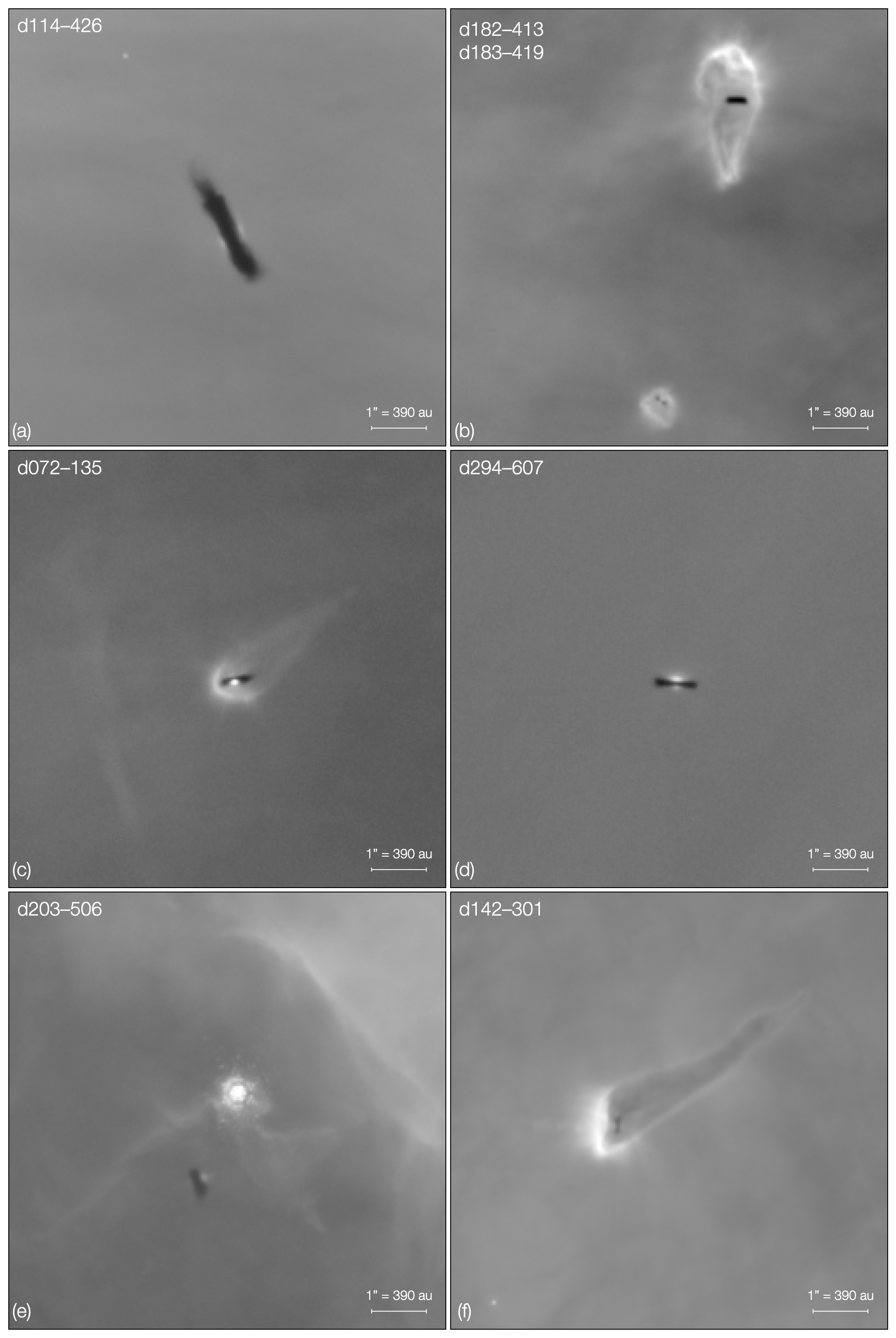} 
\caption{%
A selection well-known silhouette disks in the Orion Nebula as imaged in the near-infrared 
with JWST\@. In this figure, a section of the F187N image, dominated by Pa$\alpha$ emission
has been extracted and adjusted for brightness and contrast. For most systems, this 
maximises the visibility of the silhouette disk against the bright nebular background. 
Each panel is a 7.8 arcsec square and a 1 arcsec (390\,au) scale bar is shown. 
N is up and E left in all panels. The names in the top left corner encode the 
coordinate following the scheme of \citet{odell94} with the prefix 'd' to indicate
a disk: all objects have been previously 
catalogued from the various HST surveys \citep[\eg][]{bally00, ricci08}.
The features seen in the individual images are discussed in the text. 
}
\label{fig:silhouettes_f187n}
\end{figure*}

We have selected six previously known disks seen as silhouettes against the background 
nebula to illustrate the kinds of features that are seen in the JWST and which will be
studied in more detail in future papers. Figure~\ref{fig:silhouettes_swcolour} shows these
disks in the same combined short-wavelength colour scheme as in the main mosaic 
(Figure~\ref{fig:swcomposite}), while Figure~\ref{fig:silhouettes_f187n} shows the same
six disks as imaged in F187N, where the narrow filter isolates the Pa-$\alpha$ line and
reduces continuum contamination, this often yielding the best contrast between the 
silhouette disk and any other sources. These and many other disks and proplyds are 
also seen in many of the LW filters, albeit at lower spatial resolution and often with
even more contamination due to the central star. Thus we do not show those images in this
overview paper but do describe interesting features seen in them for these disks,
and it is clear that those data will also be important to studies of the
proplyds, including measurements of mass loss due to photoionisation and ablation. 
%

\begin{description}
\item [\bf d114--626:] This is by the largest silhouette disk known in the Orion Nebula, spanning
almost 1000\,au in diameter, including tendrils seen to the NE, and can be seen in ground-based
seeing limited images of the region. The disk is near edge-on and no trace of the central star
is seen directly: its presence is revealed by polar reflection nebulae above and below
the plane of the disk. In F187N, these are barely visible, showing the silhouette in best
detail. Preliminary measurements of the intensity profile across the horizontal plane 
suggest that it is the same size as measured in the H$\alpha$ line \citep{mccaughrean96},
confirming the results from earlier, lower-resolution Pa-$\alpha$ measurements made using
NICMOS on HST \citep{mccaughrean98, throop01}. This achromatic profile was taken as an 
indication of significant growth in the dust grains in the disk beyond typical interstellar 
sizes, but truncation of the disk by external forces likely also plays a role at least 
at the southwest end.

In the SW colour image, the extremities of the silhouette look substantially similar to the
F187N image and there is no strong evidence for colour shifts in the tenuous edges. Across the
middle, the polar hoods have grown much brighter thanks to the inclusion of wider filters, but 
they are also very wide and relatively linear. There are significant colour shifts moving
away from the midplane, from red to neutral, to blue on the E side, and from green to neutral
to blue on the W side. This might indicate that the W is tilted slightly towards the observer,
but caution is needed not to over-interpret colour shifts at the pixel level, as the SW colour
composite was registered globally, not locally. The star in the NE corner appears reasonably 
well registered, but a better job could be done with local sub-sampling and registration.
\item [\bf d182--413 (HST10) and d183--419 (HST17):] the former object is one of the best
known proplyds in Orion \citep{odell93} and is remarkable for the fact that its classic teardrop
is not oriented away from \thetaonec{} or \thetatwoa, the two dominant ionising sources in 
the region, and it may indeed be shaped by both \citep{shuping14}. Our new images confirm
the essential near-infrared emission-line features revealed by NICMOS on HST \citep{chen98} 
and the ground-based Keck laser AO observations by \cite{shuping14}, but much more clearly.
Namely a ionised outer shell with a 
brighter edge to the N, bright H$_2$ emission surrounding the dark edge-on silhouette disk, 
and PAH emission (as seen in F335M) from the disk (there is no evidence of a dark core
in the JWST PAH data as seen by \citet{shuping14}, but perhaps the 10-m diameter of Keck
wins over the 6.5-m diameter of JWST in this case) and along the E side of the body, 
There are bright ionised tendrils or spikes extending away from the main
body of the ionised shell, predominantly to the N, and there are bright neutral
spots on either side of the dark disk which are likely polar reflection nebulae. 
Further out, the southern knots of the HH517 outflow driven by d182--413 \citep{bally00} 
are clearly seen in the F187N data and there is an ionised knot roughly the same distance 
to the N as HH517~s3 is to the S which might be the counterflow.
d183--419 or HST17 lies just below d182--413 in projection and also shows a silhouette
disk with polar reflection nebulae. The ionised envelope looks quite different though
with more of a jellyfish shape, the top of the jellyfish pointing NW toward \thetaonec{} 
and the SE end much more open. There is tentative evidence for a spike emerging 
orthogonal to the disk to the NW, perhaps indicating a jet. The JWST F212N image also
shows a clear H$_2$ envelope surrounding the dark disk, as also noted by \citet{chen98}. 
\item [\bf d072--135:] this is a beautiful example of a  circumstellar disk undergoing 
photoevaporation combining many key features. It was first detected by \citep{bally00}
and imaged again in the visible by the HST Orion Treasury Programme \citep{ricci08, robberto13}, 
but is now revealed in much more detail by JWST\@. 
The disk is seen clearly as a silhouette with the central star and/or a polar reflection 
nebula on the S side, indicating the tilt of the disk. In the F187N image, there is a classic
teardrop-shaped ionised shell oriented toward \thetaonec{} with the rounded front side
very close to the edge of the disk and the tadpole tail behind. There are faint spikes 
of ionised emission pointing orthogonal to the shell, as seen for d182--413. Approximately
1000\,au ahead of the star there is a clear ionised bowshock. In the SW colour image,
much more is seen, with a layer of H$_2$ emission on the S side of the disk and
also extending into the tail of the tadpole. On the N side of the disk, there is
blue-green emission, indicating a mix of ionised hydrogen and \FeII. This also extends
into the tail, but separately from the H$_2$ emission. Finally, the tail of d072--135
is uniformly green in the LW composite, indicating PAH emission in the F335M filter.
\item [\bf d294--607:] another classic example of an edge-on circumstellar disk seen in
absorption against the background nebula, with polar reflection nebulae on either side,
a little brighter on the N side indicating a slight tilt. In the F187N image, there is
nothing else to be seen and since disk is relatively far from any other sources, 
an early preview of this image released on social media garnered some philosophical
attention \citep[\eg][]{koren22}. 
But the addition of the other wavelengths tells a slightly different story. d294--607 is not
just a passive protoplanetary system far from the madding crowd: first, it has a clear 
evidence for a faint jet in F162M, indicating \FeII{} emission, extending roughly an
arcsecond to the N, with a slight hint of a counterjet to the S\@. And the whole disk is
wrapped in mix of blue and red emission, indicating both ionised and molecular hydrogen,
with a well-defined edge to the WNW in the direction of \thetatwoa, which lies at a 
projected distance of $\sim 0.25$\,pc away. So even at this distance, the OB stars can
impact the evolution of protoplanetary systems.
\item [\bf d203--506:] this is a rather unusual system 
close to the Bright Bar and thus not that far 
from the Trapzium, but which seems to show no evidence for an ionised shell or rim. Rather, the
silhouette disk seen clearly in F187N is wrapped in a `shroud' of H$_2$ emission, albeit in
a peculiar shape that is extended on the NW side where also a bright white region is seen,
likely a polar reflection nebula, with a fainter counterpart on the SE side.
There is a narrow \FeII{} spike emerging to the NW and connected to a flask-shaped region of 
more \FeII{} emission starting roughly 500\,au to the NW, and there is the slightest hint of
green to the SE\@. Combined, these make up the HH520 jet. This whole system has recently 
been the subject of comprehensive studies with the VLT by \citep{haworth23} and JWST as part 
of the Early Release Science programme PDRs4All \citep{berne22, berne23}. Interestingly, 
the star to the NNW of d203--506 also hosts a proplyd, 203--504, with a classic
teardrop shape oriented towards \thetatwoa. There is only a slight hint of this proplyd in
the JWST data thanks to brightness of the central star, a reminder of the limitations of 
near-infrared imaging observations.


\item [\bf d142--301:] another of the very large proplyds with a silhouette disk embedded in an
ionised shell, discovered in the original HST surveys \citep{bally00}. The outer shell
has a blunt head parallel to the plane of the disk, rather than a rounded one such as seen 
in d072--135, and it has a very long, wiggling tail with a bright rim and dark core in 
F187N\@. There is clear evidence for an ionised jet extending perpendicular to the disk
to the E and well beyond the main ionisation front, and as with some of the other ionised
proplyds, there are spikes of ionised emission from the shell. In the SW colour image,
the tail turns brown-ish, suggesting extinction against the background nebula, while in
the LW image, it is uniformly green, again indicative of PAH emission (see the top-right
corner of Figure~\ref{fig:trapezium_triptych}b).
\end{description}

\section{The BN-KL region} \label{sec:bnkl}
The development of infrared astronomy through sensitive detectors and telescopes placed on
high, dry sites, made it possible to start surveying star-forming regions. One of the earliest
discoveries was of an bright near-infrared source to the NW of the Trapezium stars without
an optical counterpart: this became known as IRc1 or more commonly, BN for the discoverers
\citep{becklin67}. Longer wavelength observations made soon after revealed a more extended
region of mid- to far-infrared emission that is known as KL, also for its discoverers
\citep{kleinmann67}. 

The combined BN-KL region is a region of star formation embedded in
a molecular ridge, OMC-1, behind the Orion Nebula and Trapezium Cluster, and subsequent 
surveys at ever higher sensitivity and spatial resolution using the largest telescopes,
latest detector technology, and advanced imaging techniques in the near-, mid-, and 
far-infrared, and in the radio, have revealed a complex region containing multiple embedded 
sources, some point-like, some extended, with a prodigious combined infrared luminosity
\citep{rieke73, downes81, lonsdale82, minchin91, dougados93, menten95, gezari98, robberto05, 
sitarski13}.

\begin{figure*}[p]
\centering
\includegraphics[height=21cm]{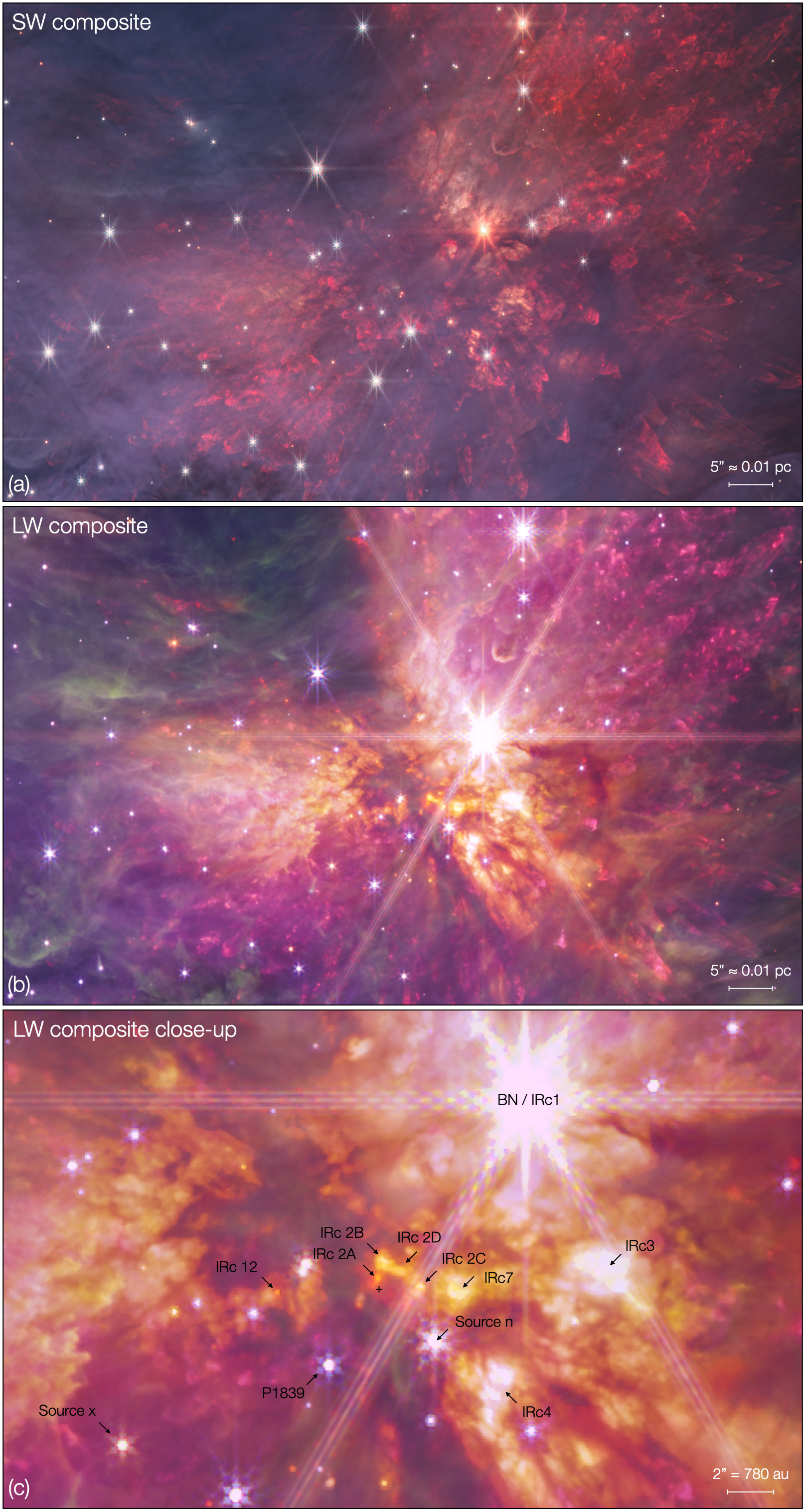} 
\caption{%
Subsections of the SW composite (Figure~\ref{fig:swcomposite}) and the LW composite
(Figure~\ref{fig:lwcomposite}) focussing on the BN-KL region. Panels (a) and (b) are 
centred at at 05h 35m 14.46s, $-05$\degree{} 22' 29.4'' (J2000.0), are oriented close to
N up, E left, and each spans $77.4\times 47.9$ arcsec or $0.146\times 0.090$\,pc assuming 
a distance of 390\,pc. Panel (c) zooms in further and covers $33.8\times 21.0$ arcsec
($\sim 13200 \times 8200$\,au): the well-known mid-infrared sources in the region are
identified (\citep{becklin67, kleinmann67, rieke73, downes81, dougados93, sitarski13}, while
the location of the deeply-embedded radio source `I' \citep{menten95} is marked with a 
cross just to the S of IRc2\,A without a label to avoid obscuring the surroundings. 
Source~x of \citep{lonsdale82} was recently demonstrated to be a high-proper motion
star moving at 55\kmpers{} to the SE and which may have been ejected from the region
around Source~I, with BN ejected in the opposite direction \citep{luhman17}. The 
global and local intensity and contrast have been adjusted with respect to the main composites 
to make key features clearer. There is some ``stepping'' in the colours and intensities
in some regions as several of the individual filter images hard saturate above a certain
brightness. 
}
\label{fig:bnkl_triptych}
\end{figure*}

Our JWST data cover the region from 1--5\micron{} and document the huge shift from the Orion 
Nebula region being dominated by the Trapezium OB stars at the shortest wavelengths
to the BN-KL region taking over in the background beyond 3\micron. Figure~\ref{fig:bnkl_triptych}
shows extracts from the SW and LW composites with the dynamic range and contrast adjusted to
reveal the key features in more detail, with a closer zoom into the LW composite to identify
the various sources in the heart of BN-KL\@. 
BN itself is detected at all JWST wavelengths, starting
as an innocuous faint point source with a small nebulosity to its NE in the F115W image, growing
brighter and illuminating what appears to be a large-scale turbulent reflection nebula 
extending NE and W, with fainter nebulosity to the NW, then to the SSW, and at longer
longer wavelengths to the E\@. This nebulosity is seen in yellow and orange colours both 
the SW and LW composites. 

Superimposed on that is a deeper red and crimson in both composites, with a rather different
turbulent: this is emission from shocked molecular hydrogen in the F212N (SW) and F335M, 
F444W, and F470N (LW) that extends into huge fingers of emission to the N to W quadrant
and also in the S to E quadrant above the Trapezium: this is the famous high-speed ``explosion''
from the BN-KL region revealed early infrared imaging observations of the region as described
by \citet{allen93}. We discuss this in more detail below in Section~\ref{sec:omc1outflow}.

In the LW composite, BN is dominant and there is a clearer separation between the 
yellow-orange reflection nebulosity and the crimson H$_2$ emission, with clear effects due 
to dust extinction in a belt extending NE--SW across the KL region to the SE of BN\@. 
Many of the sources in the inner region are bright and saturated in much of the LW data, but
it is still possible to make a clean identification of many of the well-known sources and
place them in context with the extended nebulosity. The presumed central engine of the 
region, Source~I \citep{menten95} remains invisible, as it was in the higher spatial resolution
Keck~AO images of \citet{sitarski13}. But the much greater fidelity of the JWST imaging and 
sensitivity to the extended nebula should help with the modelling of the region, with the
red ridge around the location of Source~I thought to represent one side of a dense
circumstellar disk surrounding it.

\section{The OMC-1 outflow} \label{sec:omc1outflow}
The BN-KL region has long been known to be associated with molecular emission, with two broad
peaks of near-infrared H$_2$ emission to the NW and SE of BN seen in early infrared maps 
\citep{gautier76, grasdalen76, beckwith78}, and broad, high-velocity CO emission in 
millimetre maps \citep{kuiper75, kwan76}. Modelling of the line intensities showed that 
shocks were involved, yielding an excitation temperature of around 2000\,K \citep{beckwith78}, 
as opposed to radiative excitation and fluorescence \citep{black76}. \citet{beckwith78} 
concluded that the constant luminosity of the infrared sources in the region were insufficient 
to power the emission and that ``a cataclysmic event such as supernova explosion'' must be 
responsible.

The presence of outflows to the NW of the Trapezium was first noted through the discovery of 
optical Herbig-Haro objects \citep{axon84}, while higher spatial resolution raster mapping 
in the v=1--0 S(1) line of H$_2$ at 2.12\micron{} suggested that that emission also broke 
up into a series of ``fingers'' or ``molecular jets'' \citep{taylor84}. Subsequent infrared 
imaging confirmed that these fingers were a series of bowshocks and wakes over a wide range 
of angles, trailing behind fast-moving ``shrapnel'' from an explosive event \citep{allen93}. 
The ``bullets'' were seen to be emitting in the higher-excitation \FeII{} line at 1.64\micron, 
but their origin remained unclear: they may be discrete objects flung from the source of the
explosion, or they may arise in a highly-energetic fragmenting stellar wind bubble more
distant from the source of the explosion \citep{stone95, mccaughrean97, dempsey20}. 

Proper motion analyses showed that the \FeII{} bullets and trailing H$_2$ wakes are expanding 
at speeds on the order of 200\kmpers, albeit some are expanding more slowly, which may be due 
to a projection effect, given the wide angle over which the fingers extend \citep{lee00, bally11}. 
These velocities and the spatial extent of the fingers from the source region suggest ages 
significantly less than 1000 years, although in more detail, the range of dynamical ages 
determined in this way is inconsistent with a single explosive event and ballistic motion. 
These differences can be reconciled when the interaction between the shrapnel and the surrounding 
medium is taken into account \citep{riveraortiz19}.

The fingers, bullets, and wakes have also been studied extensively in the near-infrared through 
polarimetry and spectroscopy \citep[\eg][]{burton91, minchin91, tedds95, tedds97, youngblood16}, 
as well as ever higher spatial resolution narrow-band H$_2$ imaging 
\citep{stolovy98, kaifu00, cunningham06, bally11, bally15}. Combined with similarly 
high-resolution molecular maps \citep{zapata09, bally17} and the link to optical Herbig-Haro 
objects thought to arise where some outflow fingers emerge from the molecular cloud and into 
the Orion Nebula \citep{graham03}, these studies have revealed the full outflow complex and 
its properties in much more detail. 

The current best scenario for its origin involves an interaction and/or merger between massive 
stars in the OMC-1 core about 500 years ago \citep{tan04, bally05, zapata09}, ejecting BN and 
Source~I at high velocity, the liberated energy powering a large-scale ``Hubble flow'' 
within 50 arcsec of the origin, and bullets and wakes extending at high speed further out 
to the N and NW, as well as to the S and SE \citep{bally17}. Source~x 
(see Figure~\ref{fig:bnkl_triptych}) is also moving at high velocity away from
the centre of the explosion and may also have been involved in the dynamical decay.

Our new JWST images of the outflow show it in unprecedented high resolution and high fidelity
across the 1--5\micron{} range. The large SW and LW composites (Figures~\ref{fig:swcomposite} 
and~\ref{fig:lwcomposite} show the full extent of the fingers, mostly traced in the H$_2$ lines 
of v=1--0 S(1) at 2.12\micron{} and v=0--0 S(9) at 4.69\micron{} in the F212N and F470N filters, 
respectively, but with additional emission in the wide- and medium-band filters, including the 
v=1--0 O(5) 3.23\micron{} H$_2$ line in F335M, \FeII{} at 1.64\micron{} in F162M, and the 
CO (1--0) R- and P-branch bands in F444W\@. There is also some ionised hydrogen emission at 
the very tips of the fingers due to Pa-$\alpha$ at 1.87\micron{} in F187N\@. Given this mix 
of filters and lines, H$_2$ emission is mostly red in the SW composite and more purple in 
the LW composite. \FeII{} shows up as green in the SW image, while the tips turn green-white 
once Pa-$\alpha$ emission is added, and CO shows up as yellow in the LW composite. 

The extent of the outflow to the N and NW is clear in both composites, but the counterflow 
to the S and SE can also be traced through fingers behind the Trapezium, as seen more clearly 
in the ALMA molecular images. The inner part of the flow to the NW, within $\sim 50$ arcsec of 
Source~I and thus coincident with the inner Hubble Flow of \citet{bally17}, appears more 
complex and turbulent, and in the LW composite, many of the purple H$_2$ shocks have more 
yellow tips, indicating additional CO emission, similar to that seen in the brightest
bowshocks of the much less energetic HH211 protostellar outflow in Perseus also recently 
studied with JWST \citep{ray23}. This can be seen more clearly in panel (b) of the blow-ups 
of the BN-KL region (Figure~\ref{fig:bnkl_triptych}).

\begin{figure*}[p]
\centering
\includegraphics[height=21cm]{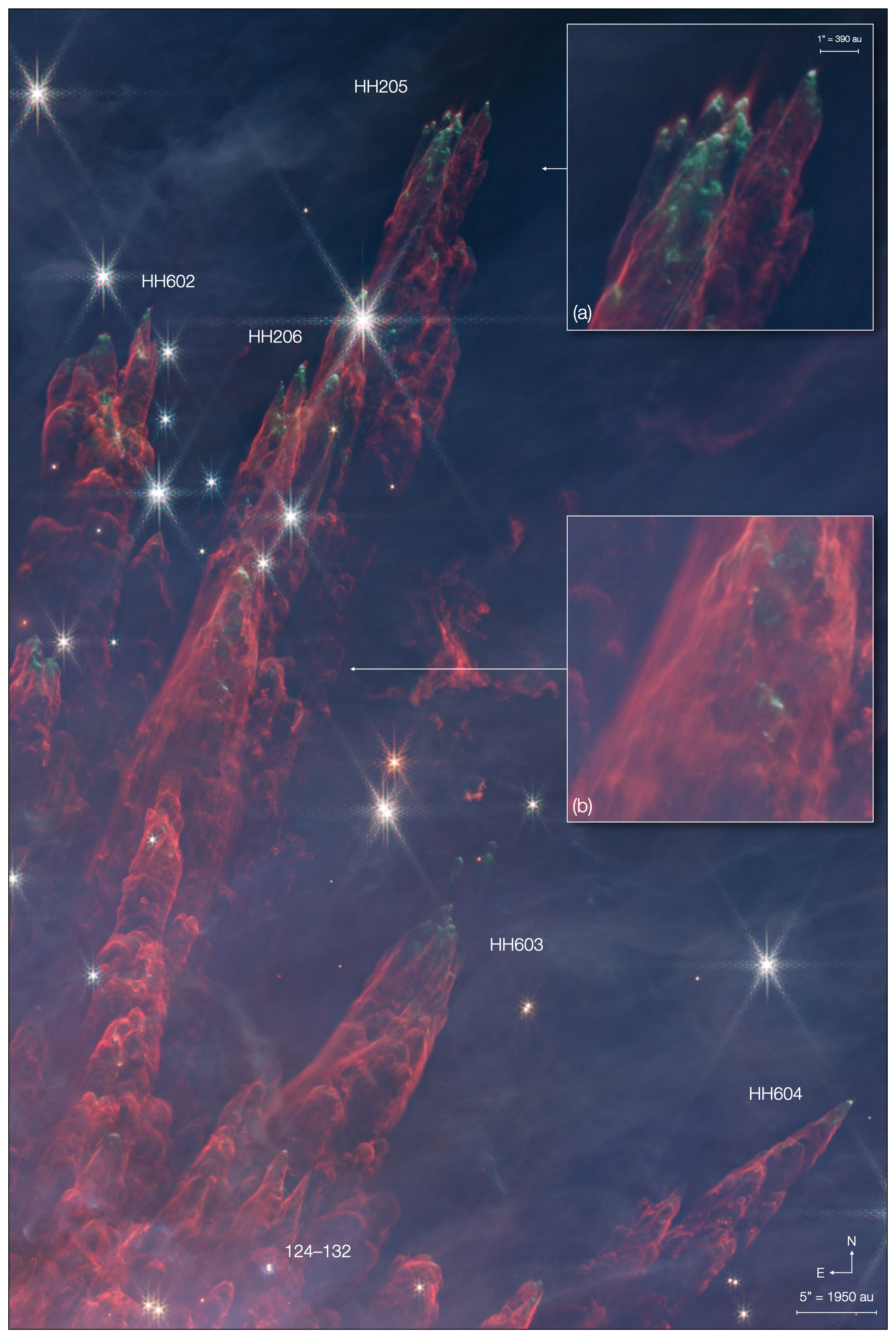}
\caption{%
A section of the SW composite located to the NNW of BN showing some of the fingers of 
emission expanding rapidly away from the the region around Source~I in the BN-KL region
(see Figure~\ref{fig:bnkl_triptych}). The fingers are predominantly red, indicating 
H$_2$ emission in the F212N filter, replaced by green and white emission near the tips 
of the fingers, indicating \FeII{} 1.644\micron{} emission in the F162M filter and a 
mix of emission also including ionised Pa-$\alpha$ in the F187N filter, respectively. 
Inset (a) shows a more detailed view of the tip of HH\,201 with the discovery that there is 
also additional H$_2$ emission wrapped around the \FeII{} emission and even extending into spikes 
pointing in the direction of motion of the finger. Inset (b) shows a region further down 
the same finger, revealing a mix of turbulent and apparently laminar flow along the
sides. Herbig-Haro object names are given for some of the fingertips and the proplyd
surrounding a binary system \citep[121--132,][]{robberto08} is marked.
The main image is centred at 05h 35m 11.6s, $-05$\degree{} 20' 54'' (J2000.0) and covers 
$54.9\times 82.2$ arcsec or $0.10\times 0.16$\,pc assuming a distance of 390\,pc.
The sub-panels are each $7.9\times 7.9$ arcsec.
}
\label{fig:fingers_panels}
\end{figure*}

By contrast, the extended fingers to the northwest are perhaps best seen in the SW composite, as 
shown in Figure~\ref{fig:fingers_panels}, not least because of the higher diffraction-limited 
resolution. The dominant hue is red due to the 2.12\micron{} H$_2$ line, but many of the 
bowshocks show a more yellow rim, likely indicative of other, lower intensity H$_2$ lines in 
the shorter wavelength filters. The tips of many of the fingers turn green, indicative of \FeII{} 
1.64\micron{} emission, and green-white at the tips as Pa-$\alpha$ 1.87\micron{} is added. 
A particularly notable case in HH603, where green \FeII{} knots appear to be detached and 
extended well beyond the initial finger, although it is also possible that two separate 
outflow components are seen superimposed along the line of sight. 

The decline in H$_2$ emission towards the tips is thought to be due to the dissociation of 
molecular hydrogen at the higher temperatures there, but unexpectedly and remarkably, the 
JWST images show red molecular hydrogen emission appearing {\em beyond\/} the \FeII{} in 
many cases. This H$_2$ emission appears to be wrapped around the tips and even extending in spikes 
of up to 200\,au in the direction of motion of the fingers. This most clearly seen in HH205 
(Figure~\ref{fig:fingers_panels}(a)), but HH206 and HH602 show the same effect, albeit with 
the H$_2$ spikes slightly off the corresponding expansion vectors.

One possible origin for this additional ``pre-shock'' H$_2$ emission may be fluorescence from
molecular hydrogen in the molecular cloud ahead of the shock, excited by intense emission 
from the tips of the bullets. One of the visible Herbig-Haro objects in the region, 
HH210 \citep{axon84, taylor86}, is moving at 425\kmpers{} and is known to emit soft 
x-rays \citep{grosso06}. HH210 is clearly seen in the full SW composite to the E of the largest 
H$_2$ fingers, but almost entirely in green, \idest{} in \FeII{} and Pa-$\alpha$: there is 
no associated H$_2$, as HH210 has fully emerged into the Orion Nebula. As \citet{dempsey20} 
note, it is possible that the bullets associated with the H$_2$ fingers also emit in x-rays, 
but the latter are sufficiently absorbed by the intervening molecular cloud to render them 
invisible.  Why the ``pre-shock'' H$_2$ forms spikes in the direction of expansion remains to be 
investigated.

Finally, it is worth briefly mentioning the structure of the fingers further down the H$_2$ 
wakes. These are mostly turbulent, breaking up into multiple arcs and bowshocks, consistent 
with the idea that the H$_2$ is shock excited. But the high resolution of JWST also reveals 
that in some places, the H$_2$ emission is much smoother, almost laminar in appearance 
(see Figure~\ref{fig:fingers_panels}(b)). This may yield further insight into the propagation 
of the bullets and their wakes through the molecular cloud and how the H$_2$ is excited. 

Beyond that, comparison of these new images with archival high-resolution AO observations of the 
fingers in H$_2$ \citep[\eg][]{bally15} should give us a more accurate picture of the velocity 
field on much smaller scales than previously possible (Bally, McCaughrean, \& Pearson 2023,
in preparation). Follow-up second-epoch JWST observations in future years will improve matters 
even further. 


\section{Other jets and outflows} \label{sec:jetsoutflows}
\begin{figure*}[p]
\centering
\includegraphics[width=11.3cm]{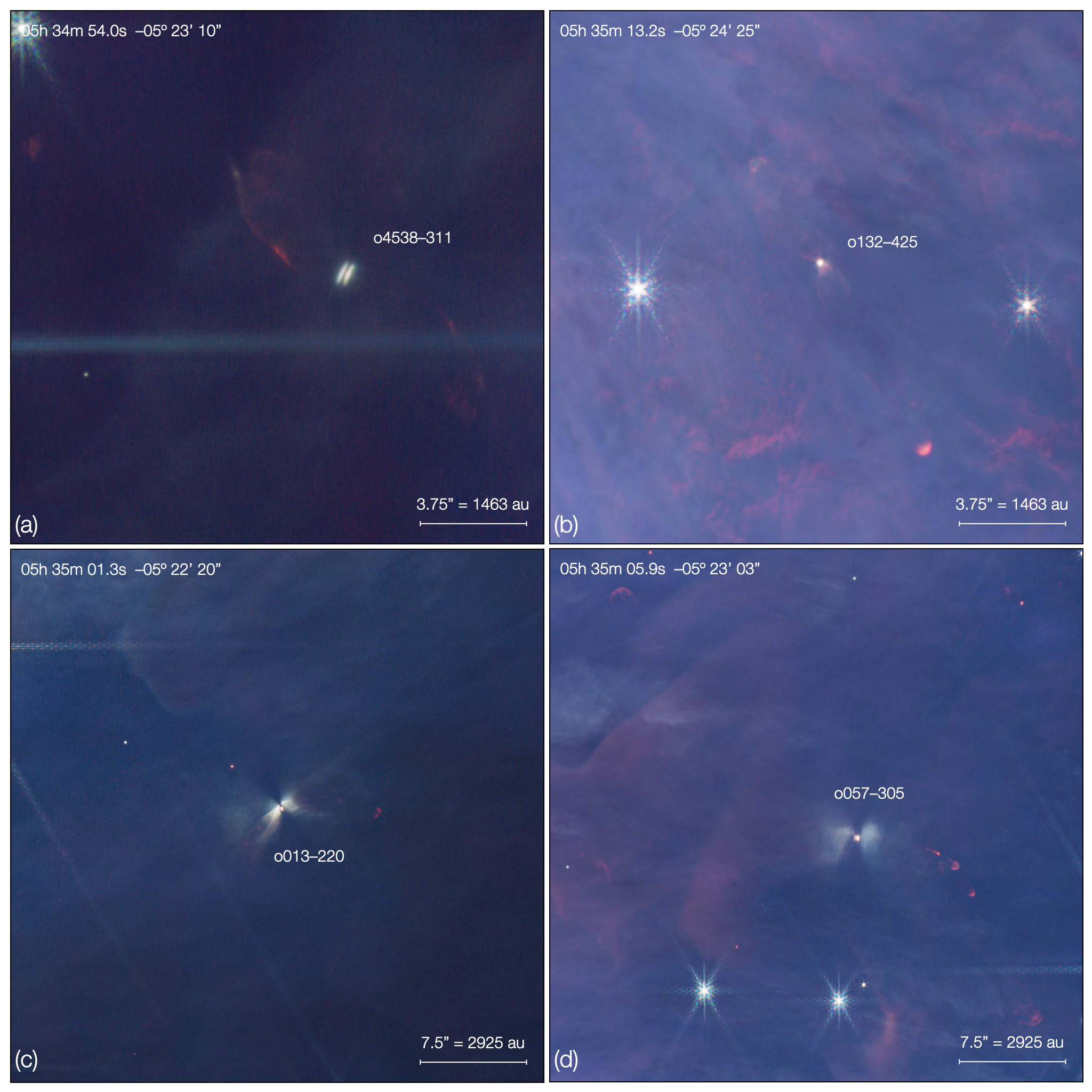} 
\includegraphics[width=11.3cm]{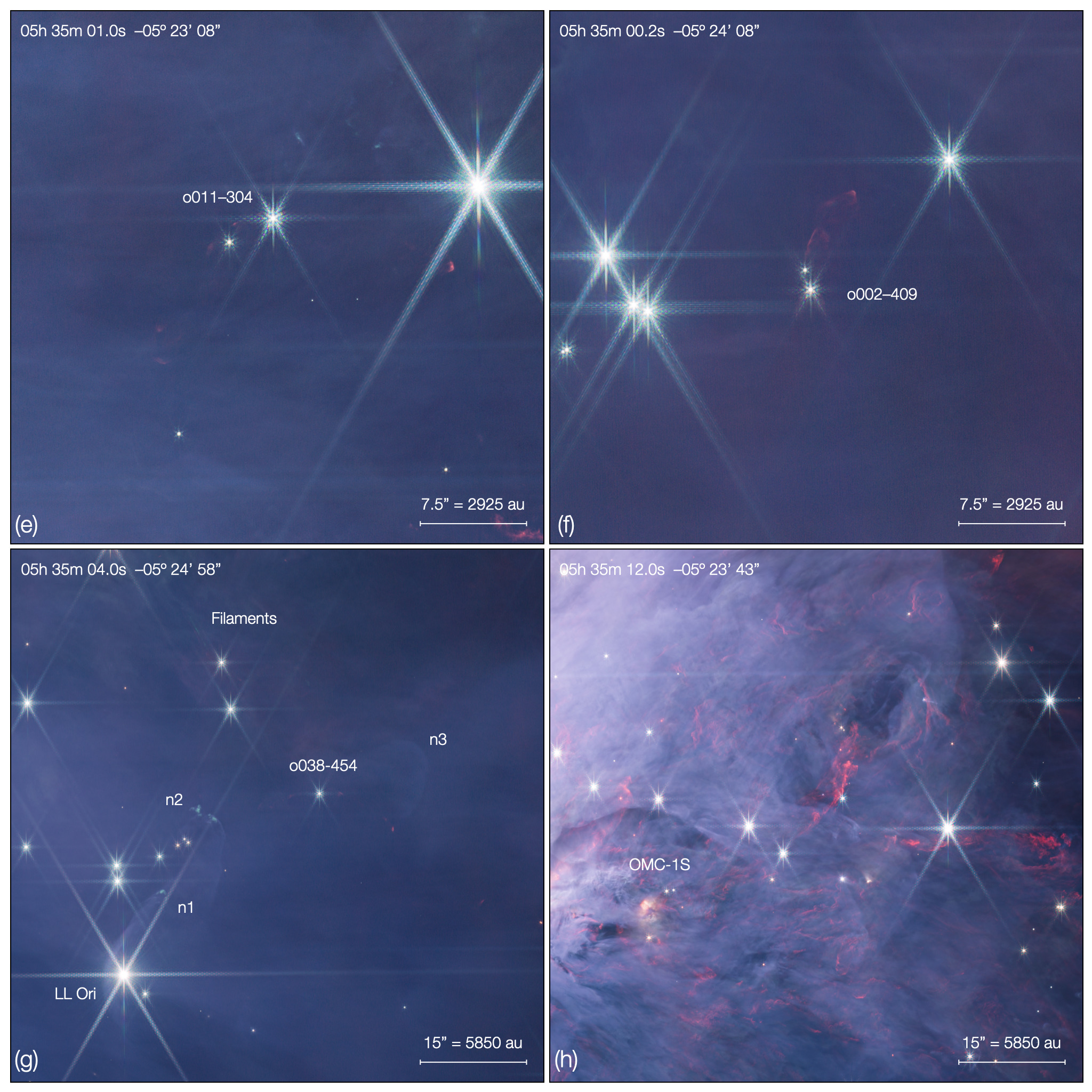}
\caption{%
A selection of outflows from young low-mass stars and binaries in the Orion Nebula taken as
cut-outs from the SW composite. While the central sources are known visible and/or infrared 
stars and binaries, the outflows are predominantly only seen at near-infrared wavelengths and 
have not been previously identified. The panels are: (a) o4538--311; (b) o132--425;
(c) o013--220; (d) o057--305; (e) o011--304; (f) o002--409; (g) o038--454.
Panel (h) shows the region around OMC-1S to 
illustrate the complex of flows emanating from there. Each of the systems is described in the 
text. N is up and E left in each panel: scales are given in each image assuming a 
distance of 390\,pc.
}
\label{fig:outflows}
\end{figure*}

While the massive OMC-1 outflow is most eye-catching in these new JWST images, it is by 
no means the only one in the region, as many young low-mass stars and embedded sources are 
also active. A wide array of optical jets and outflows have been catalogued and studied 
\citep[\eg][]{axon84, bally00, odell01a, odell03, henney07, odellhenney08, odell08} and 
many of them have counterparts in the JWST data, mostly seen in ionised tracers including
Pa-$\alpha$ and \FeII. Making detailed comparisons between the optical and infrared 
emission to study excitation, extinction, and dynamics will necessarily have to be left to
a future dedicated paper. Here though we draw attention to a number of flows that show
shocked H$_2$ emission in the JWST data, although these are just a subset and many
others can be readily identified in the images. We have chosen seven example flows 
associated with apparently low-mass stars or binaries and also show the region surrounding 
OMC-1S and to its west, illustrating the great complexity there due to many overlapping flows.
The flows are shown in Figure~\ref{fig:outflows} and described below: the putative 
source of each flow is labelled using the nomenclature scheme of \citet{odell94} with a prefix
of `o' for outflow. 

Several of these flows are bent or C-shaped, similar to some seen in HST visible 
imaging of irradiated flows in the outskirts of the Orion Nebula \citep{bally01, bally06}. 
It will be interesting to see whether the same proposed bending mechanisms invoked for the 
irradiated jets also apply to outflows seen in molecular material and thus presumably more 
embedded.  At first sight, this may not be the case, as the C-shaped outflows described below do
not bend away from the Trapezium stars.

\begin{description}
\item [\bf o057--305:] This object is identified as a point source in many previous surveys, 
but is clearly seen in the JWST images as a small reddened bipolar reflection nebula, with 
a more extended, bluer bipolar nebula extending to at least 750\,au on either side. 
There is no evidence for a silhouette disk in F187N, but the western inner
nebula is the brighter of the two, indicative of the tilt of the disk. Correspondingly,
a series of H$_2$ knots and bowshocks are seen to the W, clearly curving southwards in
a C-shaped flow. There is another larger knotty bowshock to the SW $\sim 7000$\,au from 
the source. There are two candidates for counterflows on
the E side: one is a small region of H$_2$ roughly equidistant to the SSE, which indicate the
jet bends southward on that side two, but the other candidate is a peculiar umbrella-like 
H$_2$ object to the NE, with a `handle' and semi-circular `canopy'. In that case, the flow would
bend N on the E side. Of course, it is entirely possible that neither object is related to
057--305. A molecular cloud core was detected at the location of the central source by
\citet{shimariji15}. 
\item [\bf o4538--311:] This is a textbook example of a circumstellar disk seen edge-on, with 
two extended, slightly flaring reflection nebulae with hints of reddening towards the disk 
plane. The new JWST images show this in much more detail than the HST discovery images 
\citep{ricci08}. The reflection nebulae are roughly 0.8 arcsec or $\sim 300$\,au in diameter. 
Unlike many similar disks in the Orion Nebula, however, there is no evidence for a 
silhouette disk in F187N simply because it lies far from the core region and the ionised
background is weak there. There are two large outflow lobes signalled by H$_2$ emission
extending $\sim 2000$\,au to the NE and SW, with a brighter, more neutral knot at the end
of the NE lobe, implying emission from other lines there. There is no Pa-$\alpha$ emission 
associated with the outflow, however. The S-shaped mirror symmetry in the two lobes
might be indicative of precession at the source. There are two additional H$_2$ knots (one 
just outside the frame) to the NE, although not along the apparent axis of the main flow. 
The bright horizontal green stripe is a diffraction spike from a nearby star.
\item [\bf o013--220:] This system is a close binary system surrounded by another large 
bipolar nebula extending to at least 2000\,au on either side. It has a complex asymmetric 
structure near the centre, much brighter on one side of the binary than the other. 
Similarly, the faint H$_2$ emission seen inside the reflection nebulae appears to show a 
bent, C-shaped flow, with each side again bending to the S\@. There is a bright H$_2$ 
bowshock 2500\,au out on the W side of the flow. X-ray emission has been detected from 
the central system \citep[COUP155,][]{getman05}.
\item [\bf o011--304]: The central source is a relatively bright M dwarf variable, V2107\,Ori, 
seen in many optical, infrared, and x-ray surveys 
\citep[\eg{} JW202, COUP149,][]{jones88, getman05}. There is no evidence for resolved 
circumstellar material in any of our images, but it easily be swamped by the bright star.
It could also be that the true source lies behind the optical star in the molecular cloud. 
In any case, there is a clear H$_2$ outflow extending at least 5000\,au to the E and W, and 
yet again bent in a C-shape to the S on both sides with an angle between the two lobes 
of $\sim 110$\degree{}. There are knots of ionised emission to the NW, but these are likely 
unrelated. 
\item [\bf o002--409:] The central source is another M-type variable, V2099\,Ori, with a 
fainter companion 1.5 arcsec to the N \citep[source 2290 of][]{dario09}. There is a clear H$_2$ 
flow to the N with two sections extending to $\sim 3000$\,au, again bending in a C-shape, 
but this time to the W\@. From the geometry, it seems clear that the brighter source is 
driving the flow. There is some very faint emission in H$_2$ to the S, but no obvious counterflow.
\item [\bf o132--425:] This is another known low-mass star seen in the optical and infrared 
\citep[MLLA200]{muench02} which is now seen to exhibit extended structure more reminiscent 
of young outflows with cavities expanding away from the source to the NE and SW before
converging again in brighter bowshocks. The flow does not bend like many others in
Orion, but is asymmetric in length, extending 3400\,au to the NE and almost twice that
distance to the SW\@. The SE bowshock is bright and only emitting in F212N; the NE bowshock
shows some hint of green \FeII{} emission as well. 
\item [\bf o038--454:] The driving source here is another bright M-dwarf variable, V375\,Ori, 
also seen in many optical, infrared, and x-ray surveys 
\citep[\eg{} JW236, COUP200,][]{jones88, getman05}. There are H$_2$ knots on either
side extending to at least 4000\,au, with a possible further knot to the SW at double that
distance. Once again, the jet is C-shaped, bending downwards to the S on both sides, 
The well-known windblown bowshock and flow around the bright star LL\,Ori \citep{bally00} is
seen in the same region: the bright high proper motion knots n1 and n2 are seen clearly 
emitting in \FeII, as are knots s1 and s2 (out of the frame). 
The broader ionised flows linked to LL\,Ori are also seen, with n3 lying beyond 038--454. 
Finally, there is an interesting linear filament to the NNE of 038--454 which is emitting 
in \FeII{} (green) with a roughly parallel H$_2$ partner (red) just to its E\@. 
Geometrically at least, these filaments appear unrelated to any of the sources in the field.
\item [\bf OMC-1S flows:] Finally, we include an image of the region around the young
star-forming complex OMC-1S, roughly 90 arcsec S of BN-KL, as an illustration of the complexity
than can arise when there are multiple overlapping flows likely driven by several separate
embedded sources. Many large-scale flows thought to emanate from the region have been 
studied in the visible using multi-epoch HST imaging, including HH202, HH203, HH204, HH269, 
HH507, HH529, and HH530, and some have molecular outflow counterparts \citep{henney07, odell08}.
The existence of shocked molecular hydrogen jets and flows heading W and NW from
OMC-1S was noted by \citet{kaifu00, stanke02} (flows 2--6 and 2--5, respectively, in
the latter paper) and these are revealed by much greater detail now. The NW flow
seems intimately connected to the peculiar dark cavity also seen in HST images and
named HH625 by \citet{odell03}, but there is also H$_2$ emission around the rim of that
cavity and elsewhere in the OMC-1S that might be excited by fluorescence due to the
relatively nearby Trapezium stars, rather than through shocks. The same cavity
features prominently in a further discussion of HH625 below, in 
Section~\ref{sec:darkabsorber}. Discerning which are
moving flows as opposed to externally-illuminated gas will likely require follow-up images
with JWST in the future to establish proper motions and velocities. 
\end{description}

\section{A mystery ``dark absorber''} \label{sec:darkabsorber}
\begin{figure*}[p]
\centering
\includegraphics[width=\textwidth]{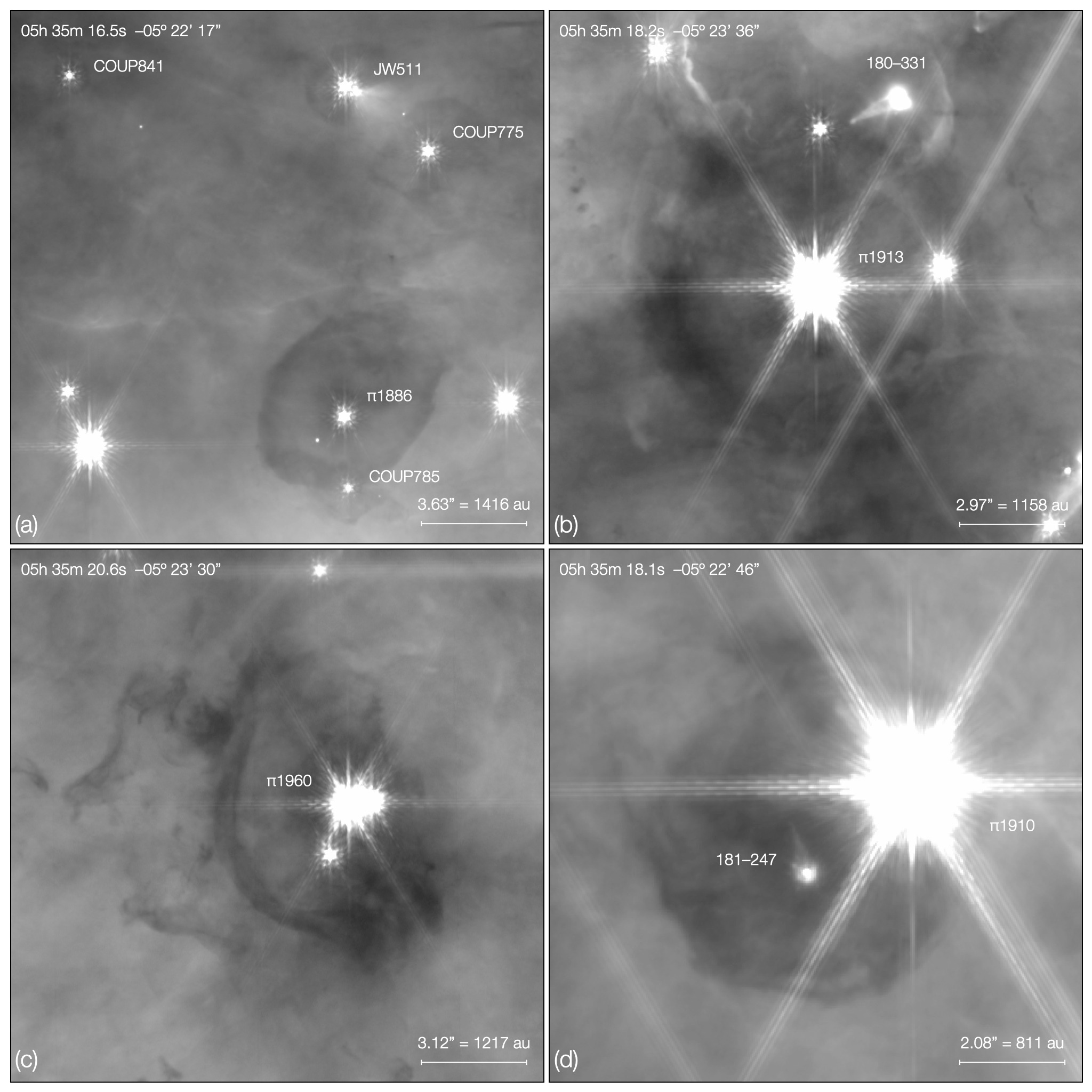} 
\caption{%
A selection of stars with dark ``coffee stains'' seen around them in the F115W mosaic.
The central coordinate for each image is shown along with a scale bar in arcseconds and au, 
the latter assuming a distance of 390\,pc to Orion. N is up and E left in each panel. 
The intensity is displayed logarithmically and the contrast has been enhanced to allow close 
examination of the features, but typically the dark regions are 10--20\% fainter than the 
surrounding nebula. Most of the stars have multiple x-ray, optical, infrared, and radio
identifications, but for simplicity here we use the name from one of the optical surveys
of \citet{parenago54, jones88} and the Chandra Orion Ultradeep project \citep{getman05},
prefixed with $\pi$ (as is traditional), JW, or COUP, respectively.
(a) ``Coffee stains'' of varying sizes are clearly seen centred on five stars (COUP775, COUP785,
JW511 [two stars, one with a reflection nebula], $\pi$1886, COUP841), with other fainter ones 
perhaps visible around three other stars in the field.
(b) A ``bullseye'' with a main dark ring centred on the bright variable star V2325\,Ori or 
$\pi$1913. The proplyd to the NNW is 180--331.
(c) A one-sided ring-type ``coffee stain'' centred on a small group of stars, with extended trails 
of darkness to the east. The brightest star in the group is another variable, 
V1520\,Ori or $\pi$1960.
(d) A ``coffee stain'' that appears to be centred on an ionised proplyd, 181--247. The adjacent
bright star is another variable, MT\,Ori or $\pi$1910, but is quite possibly unrelated.
}
\label{fig:darkstars}
\end{figure*}

While inspecting and cleaning all of the filter individual mosaics, some unexpected
and interesting features were discovered in the F115W data. The first was that some stars
N and E of the Trapezium within the brightest parts of the Huygens Region were seen 
be surrounded by dark ``coffee stains'', broadly circular, relatively large (5--10\arcsec{} 
in diameter), and often fairly sharply defined regions  where the nebular emission was depressed 
in intensity by about 10--20\% relative to the surroundings. 
The dark structures around other stars are more 
arc-like and may have additional tendrils, plus there are many smaller structures around
other stars. Examples are shown in Figure~\ref{fig:darkstars}. 

On seeing these features, thoughts immediately turned to the ``socket stars'' discovered by 
Feibelman (1989), where some stars in star-forming regions including the Orion Nebula 
apparently showed ``empty spaces'' a few arcseconds in size around them in visible wavelength
photographs. Initial follow-up studies \citep[\eg][]{castelaz90} showed that many of the 
socket stars exhibited infrared excess emission suggesting that the wider sockets might
also be dust-related, but subsequent optical and infrared imaging with digital detectors 
revealed the sockets to be chimeras, artefacts of photographic processing \citep{schaefer95},
and they were declared dead \citep{trimble95}. 

While the dark features seen in the JWST F115W images are definitely real, they are not 
associated with the same stars as Feibelman's sockets, so we are not reviving that story 
as such. Nevertheless, the phenomenon deserves further investigation, not least as there
are other important findings related to them: these dark features are not seen in any other 
JWST filter --- it is strictly confined to the F115W filter --- and other ``dark in F115W
only'' structures are seen against the nebular background elsewhere in the F115W mosaic, 
deepening the mystery.

\begin{figure*}[p]
\centering
\includegraphics[width=\textwidth]{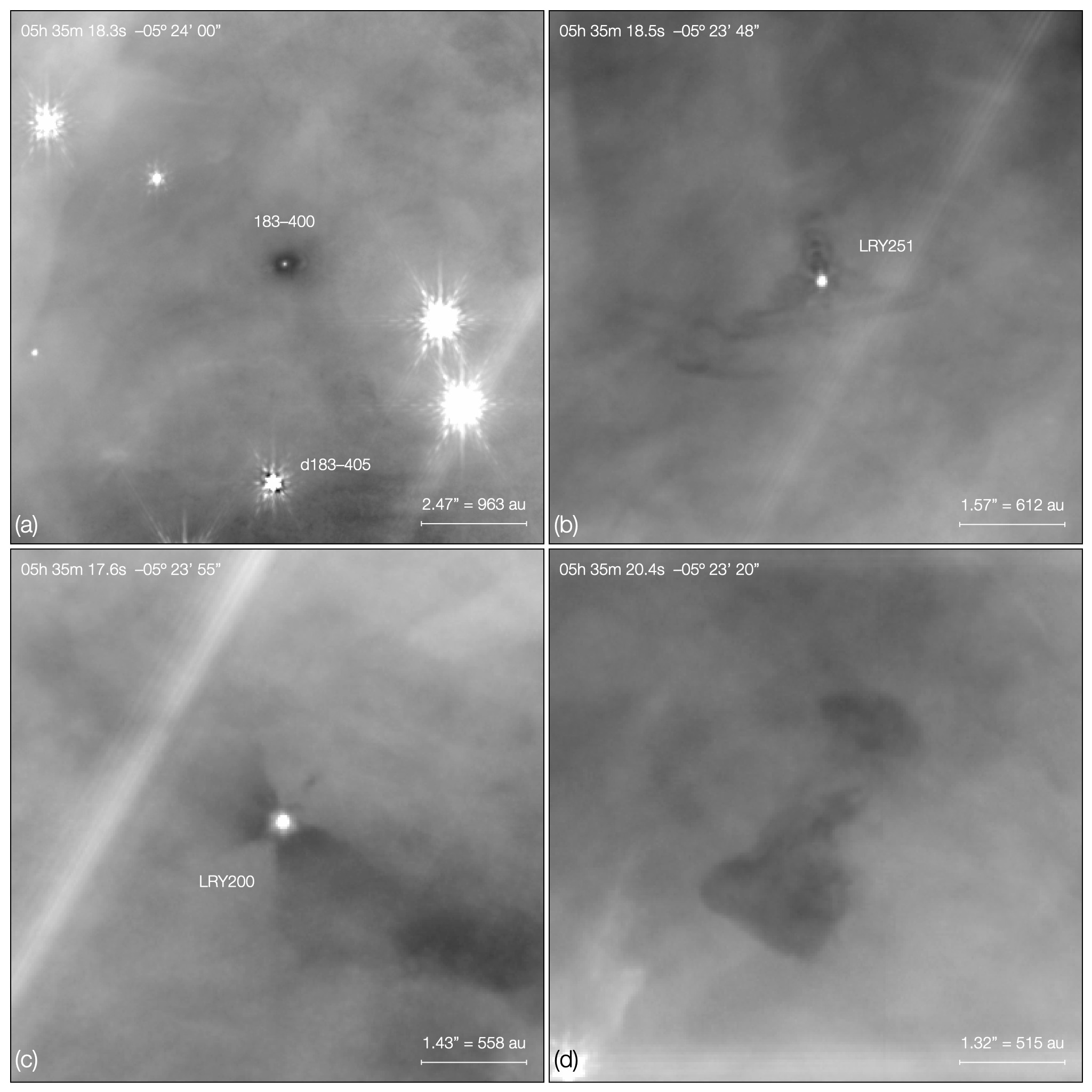} 
compelling \caption{%
A selection of other dark features seen in the F115W mosaic.
The central coordinate for each image is shown along with a scale bar in arcseconds and au, 
the latter assuming a distance of 390\,pc to Orion. N is up and E left in each panel. 
The intensity is displayed logarithmically and the contrast has been enhanced to allow close 
examination of the features, but typically the dark regions are 10--20\% fainter than the 
surrounding nebula. Diffraction spikes from nearby bright stars are seen crossing two of
the images. 
(a) A concentrated dark shadow around a faint source, similar in appearance to a silhouette
disk but only seen in F115W\@. Its coordinate-based name is 183--400; the well-known true 
silhouette disk 183--405 is seen in the same picture. 
(b) The ``Stupa''; the central star is LRY251 \citep{luhman00}.
(c) The ``Bat''; the central star is LRY200 \citep{luhman00}.
(d) An isolated dark stain without any associated point sources.
}
\label{fig:darkfeatures}
\end{figure*}

First, there are a few examples of much more concentrated dark shadows around faint stars,
with Figure~\ref{fig:darkfeatures}(a) showing a particularly striking one. Initially, this 
object was mistaken for one of the classical silhouette circumstellar disks in the group just 
to the SE of the Trapezium, namely d183--405 \citep{mccaughrean96}. But the latter lies 
further south (marked in the image) and this new object is only dark in F115W: in all other 
filters, the faint central source is in emission and slightly nebulous, and there is no 
suggestion of the larger dark shadow.

Second, some of dark shadows exhibit quite peculiar shapes around stars, \eg{} the objects we 
have christened the ``Stupa'' (Figure~\ref{fig:darkfeatures}(b)) and the ``Bat'' 
(Figure~\ref{fig:darkfeatures}(c)). The former appears as a series of wide, almost linear 
structures on one side of the central star and a series of narrower bubble-like structures 
on the other. The latter appears as a kind of ``anti-bipolar nebula'' with two almost 
symmetric dark `wings' and a `head', but with no associated outflow at any wavelength.
Third, in addition to wisps around many stars, there are quite a few locations where blobs of 
darkness are seen against the \HII{} region without any obvious association with point local 
sources, as if shards of molecular clouds are being seen against the nebula 
(\eg{} Figure~\ref{fig:darkfeatures}(d)). 

\begin{figure*}[p]
\centering
\includegraphics[width=\textwidth]{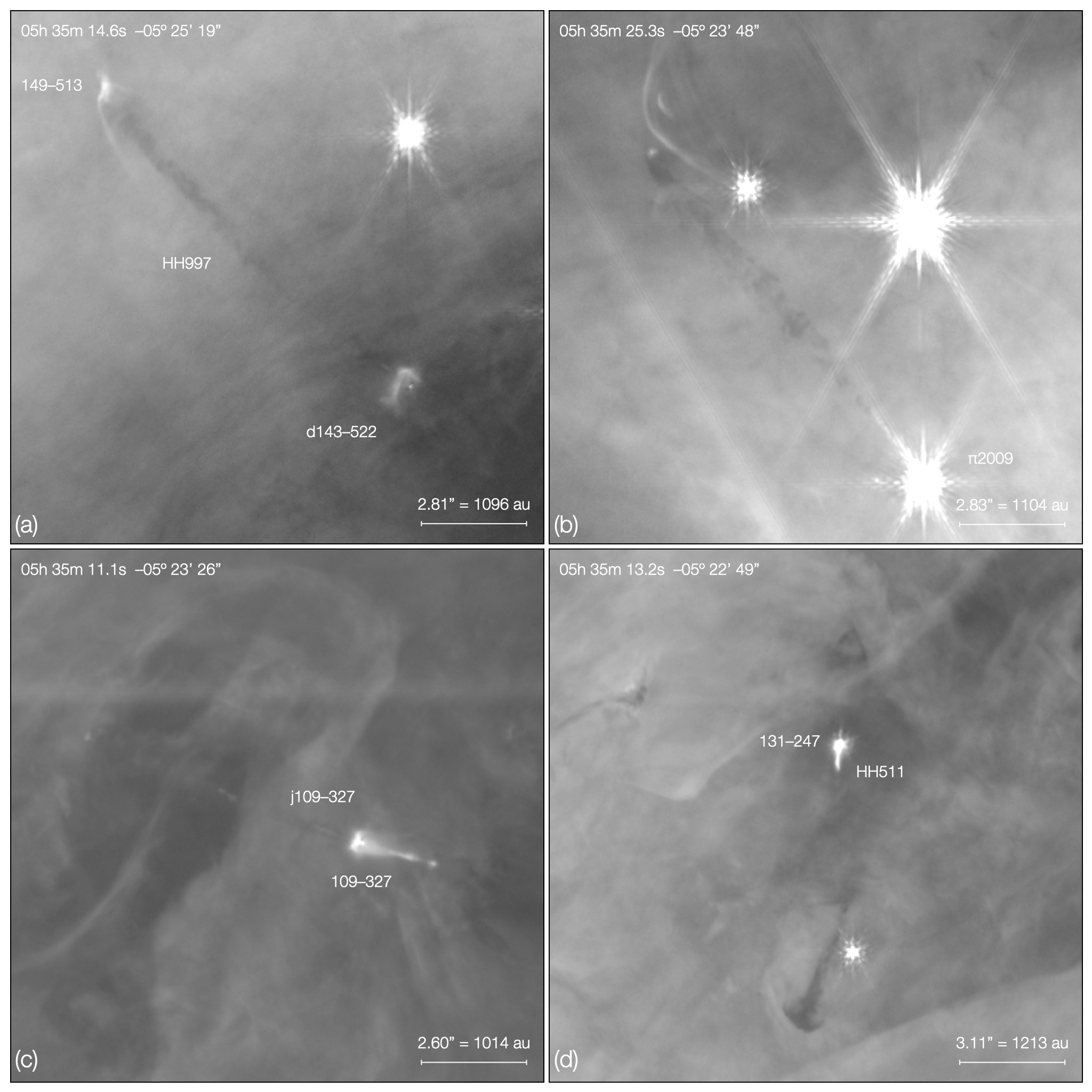} 
\caption{
A selection of ``dark outflows'' seen in the F115W mosaic. The central coordinate for each image is 
shown along with a scale bar in arcseconds and au, the latter assuming a distance of 390\,pc. 
N is up and E left in each panel. The intensity is displayed logarithmically and 
the contrast has been enhanced to allow close examination of the features, but typically the 
dark regions are 10--20\% fainter than the surrounding nebula. Diffraction spikes from nearby 
bright stars are seen crossing two of the images and in each case the source likely driving
the outflow is marked. Descriptions of each object are given in the text, where Dark Outflow~1,
2, 3, and 4 correspond to panels (a), (b), (c), and (d) here, drawing on images 
at other JWST wavelengths in some cases.
}
\label{fig:darkoutflows}
\end{figure*}

Finally and perhaps most compelling of all, there are number of jets and outflows in the region 
where part is seen as a dark shadow against the background nebulosity in F115W, with a turbulent 
appearance down the main outflow axis before terminating in a bright shock, or where the 
jet seen in emission is surrounded by a dark ``sheath''. Figure~\ref{fig:darkoutflows} shows
four examples, as follows:
\begin{description}
\item[\bf Dark Outflow 1:] 
A complex object showing a whole set of characteristic features. The source is a star
in the middle of a near edge-on circumstellar disk seen as a silhouette 
\citep[d143--522,][]{bally00} in most wavelengths, but with a bright cocoon of H$_2$
emission in F212N\@. The whole system is surrounded by an ionised rim 
oriented towards \thetatwoa{} as seen by \citep{bally00}, but no mention is made there
of the faint jet emerging to the NE as seen here. The flow was subsequently identified by 
\citep{odellhenney08} and named HH997. There is a dark ``sheath'' of absorption 
around the initial jet and further along, the outflow appears only as a dark, 
turbulent shadow against the background nebulosity, before ending in a bright ionised knot
and bowshock (149--513). The emitting parts of the outflow are seen at other JWST wavelengths, 
but not the dark shadow. There are no obvious signs of an outflow in the opposite direction 
to the SW, but the region is confused with lots of ionised emission associated with the Bright 
Bar.  In passing, it is worth noting that another silhouette disk, d141--520, lies 5 arcsec
to the NW of this object, but is completely swamped by its central star in the JWST data, 
leaving only evidence for the surrounding ionised proplyd in F187N\@. Just beyond that to
the NW is the large outflow complex HH516 \citep{bally00} seen in the JWST data in
both ionised hydrogen and \FeII{} emission. 
\item[\bf Dark Outflow 2:] 
A similar dark outflow terminating in a bright ionised knot at the NE end, although the 
bright bowshocks there are likely related to a separate source. There an ionised jet closer 
to the source, presumed to be the bright star to the SW \citep[$\pi$2009,][]{parenago54}.
The jet is seen clearly in F187N, but less so here in the SW composite and F115W images, as 
the jet lies close to one of the star's diffraction spikes. Again, there is no trace of a 
counterflow to the southwest.
\item[\bf Dark Outflow 3:]
Another complex system with a star, silhouette disk and bright jet, this time bipolar. 
The source star is proplyd 109--327 and the jet emerging to the E is HH510 
\citep[aka j109--327,][]{bally00}. 
To the WSW, there is a bright ionised cocoon which appears to focus into a 
pair of bright ionised knots with another possible diffuse knot further along, outside the frame,
thus suggesting that the overall outflow is bipolar, counter to the claim by \citet{bally00}
that this is a one-sided system. To the ENE, the bright HH510 jet is initially surrounded by 
a dark sheath in the F115W image, seemingly superimposed on the bright rim of an
expanding cavity seen at visible wavelengths and given the name HH625 \citep{odell03, odell08}. 
As the jet continues to the ENE across the cavity, the ionised knots s1 and s2 of \citet{bally00} 
are also seen.
\item[\bf Dark Outflow 4:] 
A star with two ionised spikes, brighter to the S and fainter to the N, appears to be
surrounded by a darker region. The southern spike was seen in H$\alpha$ by \citet{bally00}
and identified as a one-sided microjet which they named j131--247 and HH511. There is no 
further sign of bright emission in either direction, but both sides of the flow appear to 
terminate in strange dark shapes, a pointed arrow to the N and an elongated shadow and 
hammerhead further to the S\@. 
\end{description}

Again, in all of these cases, the dark feature is only seen in the F115W filter in our JWST 
survey and we have been unable to find any other images (\eg{} HST imaging at visible
wavelengths) where the same features show. That immediately suggests two things: first, it 
is unlikely that the feature is due to dust absorption. Even though there can be strong
features in the scattering phase function of dust \citep{baes22}, it is very unlikely
that dust could yield absorption in F115W but in no other wavelength, including the adjacent
F140M filter. Second, the fact that the F115W filter is quite wide suggests that absorption 
of broad continuum nebulosity by lines in atomic or molecular gas would have to remarkably 
effective to yield a $\sim 10$\% decrease in the flux in the band. In addition, the feature 
is not readily apparent in archival J-band images of the Orion Nebula, suggesting that the 
feature is being imprinted between the blue cut-off of the F115W filter at 1.013\micron{} and 
the blue cut-off of J-band filters at $\sim 1.15$\micron.  

One obvious candidate for emission and possible absorption in that wavelength range is the
singly-excited \HeI{} ($2^3{\rm S}$) triplet at 10830\AA{} or 1.083\micron. Helium is common in
\HII{} regions and the 1.083\micron{} line is seen in extensive emission in the Orion 
Nebula \citep[\eg][]{takami02}. Equally, helium absorption towards Orion has been known for a 
long time, first detected in the 3888\AA{} line by \citet{wilson37} and \citet{adams44}, and 
later just blueward of the 1.083\micron{} line by \citet{boyce66}, who who ascribed it to 
interstellar or ``circumnebular'' material. \HeI{} emission and absorption towards the bright
stars in Orion has since been studied in detail \citep[\eg][]{oudmaijer97}. More generally, 
\HeI{} absorption is used as a key diagnostic in various astrophysical scenarios, including the 
study of solar chromospheric activity such as flares \citep[\eg][]{huang20} and as a tracer 
of winds from young stars \citep[\eg][]{edwards03}.

At first sight then, self-absorption of \HeI{} emission at 1.083\micron{} might fit the bill
well as a way of explaining why the various dark features discussed above are seen only in 
the F115W filter, although the actual physical arrangements of the various emitting and
absorbing helium layers may vary from object to object and in the case of the dark outflows,
perhaps velocity shifts come into play: substantial self-absorption would only be seen if
the velocity in the flow matched that of the background emitting region, \idest{} if the flow
was predominantly in the plane of the sky.

Another possibility however may be \SiI, which also has a series of strong absorption lines
in the relevant region, as seen in supernovae, for example \citep[\eg][]{graham86}. There is 
plenty of silicon in young star-forming regions, \HII{} regions, PDR's, and jets and outflows, 
whether in atomic and ionic form \citep[\eg][]{abel05}, molecular \citep[\eg{} SiO,][]{zapata06}, 
or in dust as silicates \citep[\eg][]{whittet22}, but the question is whether it can be found 
in the right form to serve as a line absorber and whether the lines
can absorb enough background flux. A rough estimate based on the number of \SiI{} lines in 
the F115W filter, their likely widths and depths suggests not, if the background is continuum 
emission, but if the bulk of the emission comes from a few lines at close enough wavelengths,
then it may be possible. For example, one of the stronger \SiI{} lines in this wavelength
region lies at 1.0827\micron, within $\sim 80$\kmpers{} of the \HeI{} line. Absorption of
\HeI{} emission from the \HII{} region by silicon-rich material in an outflow could perhaps
then work, albeit would probably not work for the more `static' features. 

For the outflows at least, an intriguing thought arises if \HeI{} is responsible for both the
background illumination and the absorption. Namely, that this may offer the possibility of 
determining the mass in the outflows directly. Other observations of jets and outflows rely 
on detecting emission from excited, ionised, or dissociated atoms and molecules, and the
fraction of the total amount of material that may currently be shocked excited or swept up,
Therefore they likely only trace an uncertain fraction of the total mass in a flow.
On the other hand, absorption by \HeI{} may be a much less biased tracer of mass and
mass-loss rate in outflows. Combined with radial velocities and proper motions, a full
3D map of the motions, masses, mass-loss rate, and feedback impact on the environment
may be possible. 

\section{Shells around \thetatwoc} \label{sec:thetatwoc}
The B5V star \thetatwoc{} (aka HD37062, Brun 760, $\pi$2085), is the most easterly of the 
three massive stars below the Bright Bar. 
The HST Orion Treasury Survey visible-wavelength images of the star show some some extended 
nebulosity around it \citep{robberto13}, while subsequent HST near-infrared images
hint at some structure \citep{robberto20}\footnote{%
See also the colour composite made by Judy Schmidt from the HST infrared data: \\
\url{https://www.flickr.com/photos/geckzilla/49325834001}
},
and earlier near-infrared imaging polarimetry had showed this to be a 
reflection nebula \citep{tamura06}. 

\begin{figure*}[p]
\centering
\includegraphics[width=\textwidth]{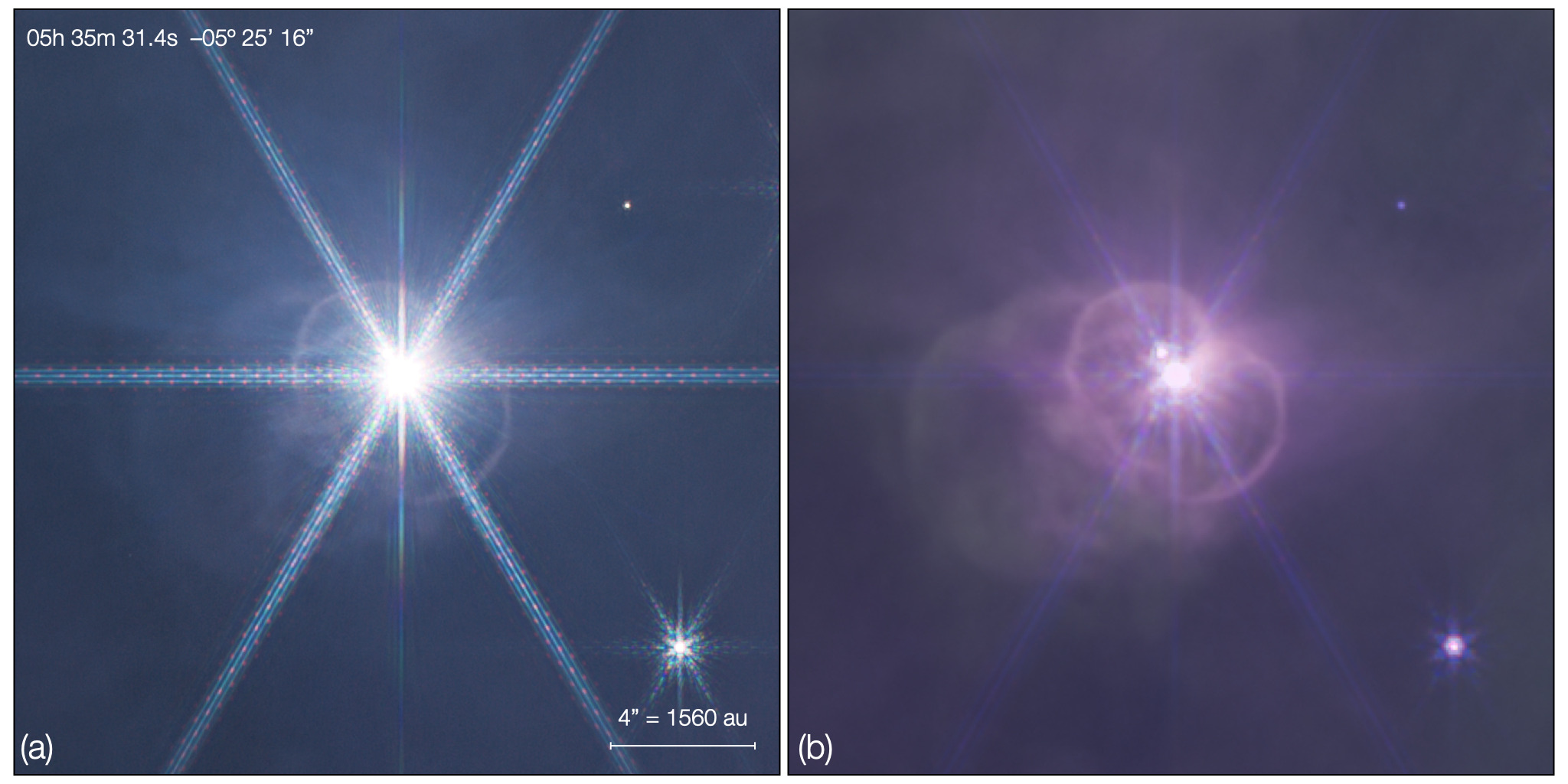} 
\caption{%
The immediate surroundings of the B5V star \thetatwoc, below the Bright Bar to the ESE of the 
Trapezium. Panel (a) shows a cut-out from the SW composite, while panel (b) shows the 
equivalent section from the LW composite. N is up and E left in each panel: the
scale indicated assumes a distance of 390\,pc.
}
\label{fig:thetatwoc}
\end{figure*}

However, as seen in Figure~\ref{fig:thetatwoc},
the higher spatial resolution and sensitivity of the JWST images show that in
addition to a diffuse, blue nebulosity around the star, there is at least one and perhaps
multiple shells of red emission within a $\sim 3$ arcsec or $\sim 1200$\,au radius around
the star. There is also extended structure to the E of the star seen most clearly in the LW
composite with a green-ish tint, thus perhaps a mix of dust and PAH emission. 

It is possible that these shells are linked to episodic ejection of material from 
\thetatwoc. The star is known to have a close spectroscopic companion with a period of
$\sim 13$ days \citep{corporon99}, while a wider companion at a projected separation of 
$\sim 16$\,au was discovered through infrared interferometry \citep{gravity18}. 
The JWST data reveal another component at $\sim 300$\,au to the NNE\@. A proper motion 
study of the shells may reveal their expansion velocity and potentially a link to one of 
the orbital timescales in the system. 

\section{Galaxies} \label{sec:galaxies}
\begin{figure*}[p]
\centering
\includegraphics[height=22cm]{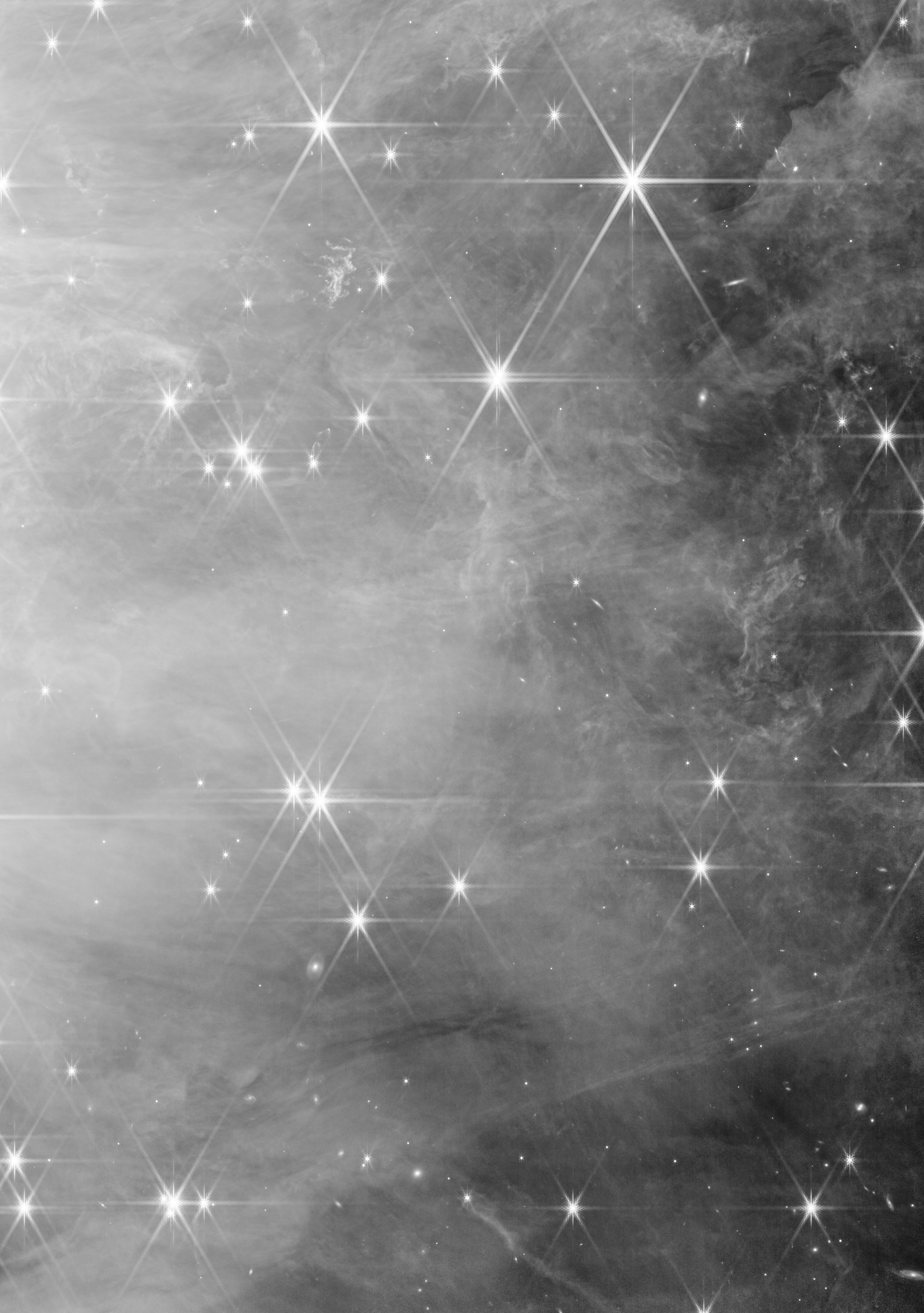} 
\caption{%
A $3.1\times 4.3$ arcmin section of the F277W mosaic, where the lower-right (SW) corner of 
this image is the lower-right corner of the full mosaic. The intensity and contrast have been
adjusted to show that there is significant structured nebulosity all the way to the edge
of the various images, something that is difficult to represent well in the full dynamic
range colour mosaics. In addition to the nebulosity, the many stars of the Trapezium Cluster 
and their strong diffraction spikes, some jets, outflows, and circumstellar disks, a substantial 
number of galaxies are also visible, many of them showing clear spiral structure.
N is up, E left.
}
\label{fig:galaxies}
\end{figure*}

Finally, although the Orion Nebula is associated with a significant extinction thanks to the 
OMC-1 molecular cloud behind it, it is not optically thick at infrared wavelengths across the 
whole area of our survey. As the nebula is fairly far out of the 
galactic plane at $(l,b)\sim (209\degree, -19.5\degree)$, there is not a substantial population
of background field stars, reducing contamination of our cluster sample as described in
Section~\ref{sec:cluster}. On the other hand, towards the W edge and the 
NE corner, the extinction drops to low enough levels that a significant number 
of background galaxies can be seen through and beyond the nebulosity as shown in 
Figure~\ref{fig:galaxies}. 

Many are directly identifiable as large spirals and ellipticals, and through careful examination 
of the images and by searching for obviously extended sources in our aperture photometry, we 
have identified around 200 candidate galaxies in total in the survey. That includes 
a candidate gravitational lens system in the NE corner of the survey, although confirming that 
would need deeper imaging and spectroscopy. 

From a philosophical angle, these distant background galaxies give additional meaning to our
images, placing the nearest region of massive star formation in our Milky Way galaxy in the
wider context of star formation across the Universe and cosmic time. However, beyond being 
just an observational curiosity, it may be possible to use the
colours of the galaxies to make a first order estimate of the extinction along various lines
of sight through the outer Orion Nebula, and the galaxies could also be used as fixed 
points in an astrometric reference frame aiming to detect proper motions in the Trapezium
Cluster, perhaps in search of a signal of overall expansion. This would follow from
previous visible wavelength proper motion studies of the ONC stellar population
\citep[\eg][]{jones88, vanaltena88, platais20} and provide a complement to
{\it Gaia\/}-linked studies \citep{kuhn19}.

\section{Conclusions} \label{sec:conclusions}
While the description of the features and new phenomena seen in these JWST images of the
inner Orion Nebula and Trapezium Cluster has been almost exclusively qualitative in
this paper, it is clear that they lay the ground for important quantitative work in 
more dedicated papers: the first of these focusses on the statistical properties of the 
planetary-mass objects and Jupiter-mass binary objects or JuMBOs (Pearson \& McCaughrean
2023, submitted), and others are in preparation. Equally, the data are now public in
the MAST archive, so freely available for others to conduct their own studies and analyses. 
It is also likely that significant future JWST observational programmes in Orion will 
follow, targeting interesting objects with imaging at longer wavelengths and/or 
spectroscopy across the whole
range: again, the first of these will focus on R$\sim$100 NIRSpec MSA prism spectroscopy 
of the PMOs and JuMBOs in Cycle 2. The potential for future surveys similar to this one
but extending further out, perhaps covering the $30\times 30$ arcmin field of the
HST Treasury Survey or focussing more to the north along the ridge of embedded star formation
between OMC-1 and OMC-2. If conducted in the broad filter set used here, thus covering a
similarly wide range of scientific goals, such programmes will be expensive in observing time, 
but potentially very rewarding.

The discovery of the JuMBOs, the F115W ``dark absorber'', and the H$_2$ spikes ahead
of the OMC-1 outflow fingers in our data were all unexpected. JWST has already begun to 
change the way we see the Orion Nebula and Trapezium Cluster, in much the same way that the 
Hubble Space Telescope did more than thirty years ago, and the 
invention of telescopes and photography did much earlier. More broadly, the observatory is
providing us with remarkable new insights into star and planet formation on all scales, from 
outflows from low-mass protostars such as HH211 \citep{ray23}, the photochemistry of
young externally irradiated disks \citep{berne23}, the structure of debris disks
around young stars such as Fomalhaut \citep{gaspar23}, the properties of galactic star-forming 
regions such as Carina and $\rho$\,Oph, also extending to the SMC \citep{jones23},
and of course also to high redshift, as the first stars and galaxies form. 
Given the powerful imaging and spectroscopic infrared capabilities of JWST, combined
with a hopefully much longer mission lifetime than anticipated at launch, 
it is seems likely that there is much more to come.

\vspace{2cm}

\begin{acknowledgements}
The time used to make these JWST observations come from the Guaranteed Time Observation 
allocation made to MJM upon selection as one of two ESA Interdisciplinary Scientists on the 
JWST Science Working Group (SWG) in response to NASA AO-01-OSS-05 issued in 2001. This 
followed MJM's tenure as a member of the Next Generation Space Telescope (NGST) Interim 
Science Working Group (ISWG), the NGST Ad-hoc Science Working Group (ASWG), and the ESA 
NGST Science Study Team (SST), in reverse order. None of this would have happened but for 
the invitation by Bob Fosbury to join the ESA SST in 1998 and MJM remains eternally grateful 
for that. Equally, he is grateful to his PhD supervisor, Ian McLean, who offered him the
opportunity to be involved in IRCAM, the very first common-user infrared camera for astronomy,
and its first observations of Orion in 1986. 
He also thanks Peter Jakobsen and Pierre Ferruit, the ESA project scientists 
for NGST/JWST down the many years, and Koos Cornelisse, Peter Jensen, and Peter Rumler, the 
ESA project managers, along with all of their teams and the wider international family of 
thousands of engineers, scientists, and others across North America and Europe who 
conceived, designed, built, launched, and commissioned the astonishing JWST, and to those 
who operate it today. Fellow members of the JWST SWG are especially thanked for their 
friendship and dedication to this project over more than two decades of travel, meetings, 
telecons, and shared lows and highs: it has been the privilege of a lifetime. 

Specific thanks for this observing programme are due to the STScI instrument scientists 
and programme reviewers Elizabeth Nance, Massimo Robberto, and Mario Gennaro, and also 
Tony Roman for all his help 
with scheduling and execution. On the science side, he would also like to thank Bob O'Dell 
and John Bally for their collaborations on the HST imaging of the Orion Nebula almost thirty 
years ago, and would also like to remember John Stauffer, for his friendship and collaboration 
on Orion right from the start of HST and more widely. Hans Zinnecker is thanked for his scientific 
and personal friendship down the years, as well as insights contributing to this work. Tom
Haworth and John Bally are thanked for their insightful thoughts and key contributions to
the discussion of the F115W ``dark absorber''; Isabelle Baraffe is thanked for providing
the latest evolutionary models for young planetary mass objects to sub-1\Mjup{} masses;
and Matthew Bate is thanked for his insights on the PMOs and JuMBOs. 

SGP acknowledges support through the ESA research fellowship programme at ESA's ESTEC, 
and would like to thank Victor See for helpful discussions and Katja Fahrion for 
valuable insights on the JWST calibration pipeline.

The data presented in this paper were obtained with the Near Infrared Camera (NIRCam) 
on the NASA/ESA/CSA James Webb Space Telescope, as part of Cycle 1 GTO programme 1256, 
They are available on the Barbara A. Mikulski Archive for Space Telescopes (MAST): 
\url{http://dx.doi.org/10.17909/vjys-x251}.

This work has made use of data from the European Space Agency (ESA) mission
{\it Gaia} (\url{https://www.cosmos.esa.int/gaia}), processed by the {\it Gaia}
Data Processing and Analysis Consortium (DPAC,
\url{https://www.cosmos.esa.int/web/gaia/dpac/consortium}). Funding for the DPAC
has been provided by national institutions, in particular the institutions
participating in the {\it Gaia} Multilateral Agreement.

This research has made use of the Spanish Virtual Observatory 
(\url{https://svo.cab.inta-csic.es}) project funded by 
MCIN/AEI/10.13039/501100011033/ through grant PID2020-112949GB-I00

This research made use of {\tt Photutils}, an {\tt Astropy} package for detection and 
photometry of astronomical sources \citep{bradley23}. 

For certain parts of the image processing, Community IRAF V2.17 was used
(\url{https://iraf-community.github.io}) and MJM is grateful to those who keep this seminal 
piece of software alive and running, not least because there are still things it can do 
better than {\tt Astropy}. And just for old time's sake. 
\end{acknowledgements}

\newpage

\bibliographystyle{aa}
\bibliography{orion_jwst_overview_2col}

\end{document}